\documentclass[twoside,twocolumn,9pt]{article}
\usepackage{extsizes}
\usepackage[super,sort&compress,comma]{natbib} 
\usepackage[version=3]{mhchem}
\usepackage[left=1.5cm, right=1.5cm, top=1.785cm, bottom=2.0cm]{geometry}
\usepackage{balance}
\usepackage{times,mathptmx}
\usepackage{sectsty}
\usepackage{graphicx}
\usepackage{subcaption}
\usepackage{lastpage}
\usepackage[format=plain,justification=justified,singlelinecheck=false,font={stretch=1.125,small,sf},labelfont=bf,labelsep=space]{caption}
\usepackage{float}
\usepackage{fancyhdr}
\usepackage{fnpos}
\usepackage[english]{babel}
\addto{\captionsenglish}{%
  
}
\usepackage{array}
\usepackage{droidsans}
\usepackage{charter}
\usepackage[T1]{fontenc}
\usepackage[usenames,dvipsnames]{xcolor}
\usepackage{setspace}
\usepackage[compact]{titlesec}
\usepackage{hyperref}
\usepackage{placeins}

\makeatletter

\usepackage{xcite}
\usepackage{xr}
\makeatletter
\newcommand*{\addFileDependency}[1]{
  \typeout{(#1)}
  \@addtofilelist{#1}
  \IfFileExists{#1}{}{\typeout{No file #1.}}
}
\makeatother

\newcommand*{\myexternaldocument}[1]{%
    \externaldocument{#1}%
    \addFileDependency{#1.tex}%
    \addFileDependency{#1.aux}%
}

\myexternaldocument{Supplementary_info}

\usepackage{bm} 
\usepackage{verbatim}  

\usepackage{amsmath}

\usepackage{textcomp}

\definecolor{cream}{RGB}{222,217,201}

\begin{document}

\pagestyle{fancy}
\thispagestyle{plain}
\fancypagestyle{plain}{

\renewcommand{\headrulewidth}{0pt}
}

\makeFNbottom
\makeatletter
\renewcommand\LARGE{\@setfontsize\LARGE{15pt}{17}}
\renewcommand\Large{\@setfontsize\Large{12pt}{14}}
\renewcommand\large{\@setfontsize\large{10pt}{12}}
\renewcommand\footnotesize{\@setfontsize\footnotesize{7pt}{10}}
\makeatother

\renewcommand{\thefootnote}{\fnsymbol{footnote}}
\renewcommand\footnoterule{\vspace*{1pt}%
\color{cream}\hrule width 3.5in height 0.4pt \color{black}\vspace*{5pt}} 
\setcounter{secnumdepth}{5}

\makeatletter 
\renewcommand\@biblabel[1]{#1}            
\renewcommand\@makefntext[1]%
{\noindent\makebox[0pt][r]{\@thefnmark\,}#1}
\makeatother 
\renewcommand{\figurename}{\small{Fig.}~}
\sectionfont{\sffamily\Large}
\subsectionfont{\normalsize}
\subsubsectionfont{\bf}
\setstretch{1.125} 
\setlength{\skip\footins}{0.8cm}
\setlength{\footnotesep}{0.25cm}
\setlength{\jot}{10pt}
\titlespacing*{\section}{0pt}{4pt}{4pt}
\titlespacing*{\subsection}{0pt}{15pt}{1pt}

\fancyfoot{}
\fancyfoot[RO]{\footnotesize{\sffamily{1--\pageref{LastPage} ~\textbar  \hspace{2pt}\thepage}}}
\fancyfoot[LE]{\footnotesize{\sffamily{\thepage~\textbar\hspace{3.45cm} 1--\pageref{LastPage}}}}
\fancyhead{}
\renewcommand{\headrulewidth}{0pt} 
\renewcommand{\footrulewidth}{0pt}
\setlength{\arrayrulewidth}{1pt}
\setlength{\columnsep}{6.5mm}
\setlength\bibsep{1pt}

\makeatletter 
\newlength{\figrulesep} 
\setlength{\figrulesep}{0.5\textfloatsep} 

\newcommand{\topfigrule}{\vspace*{-1pt}%
\noindent{\color{cream}\rule[-\figrulesep]{\columnwidth}{1.5pt}} }

\newcommand{\botfigrule}{\vspace*{-2pt}%
\noindent{\color{cream}\rule[\figrulesep]{\columnwidth}{1.5pt}} }

\newcommand{\dblfigrule}{\vspace*{-1pt}%
\noindent{\color{cream}\rule[-\figrulesep]{\textwidth}{1.5pt}} }

\makeatother

\twocolumn[
  \begin{@twocolumnfalse}
\vspace{3cm}
\sffamily
\begin{tabular}{m{4.5cm} p{13.5cm} }

 & \noindent\LARGE{\textbf{IMPRESSION - Prediction of NMR Parameters for 3-dimensional chemical structures using Machine Learning with near quantum chemical accuracy $^\dag$}} \\
\vspace{0.3cm} & \vspace{0.3cm} \\

 & \noindent\large{Will Gerrard, Lars A Bratholm, Martin Packer, Adrian J Mulholland, David R Glowacki*, Craig P Butts*} \\

 & \noindent\normalsize{The IMPRESSION (Intelligent Machine PREdiction of Shift and Scalar information Of Nuclei) machine learning system provides an efficient and accurate method for the prediction of NMR parameters from 3-dimensional molecular structures. Here we demonstrate that machine learning predictions of NMR parameters, trained on quantum chemical computed values, can be as accurate as, but computationally much more efficient (tens of milliseconds per molecular structure) than, quantum chemical calculations (hours/days per molecular structure) starting from the same 3-dimensional structure. Training the machine learning system on quantum chemical predictions, rather than experimental data, circumvents the need for the existence of large, structurally diverse, error-free experimental databases and makes IMPRESSION applicable to solving 3-dimensional problems such as molecular conformation and stereoisomerism.} \\

\end{tabular}

 \end{@twocolumnfalse} \vspace{0.6cm}
  ]

\renewcommand*\rmdefault{bch}\normalfont\upshape
\rmfamily
\section*{}
\vspace{-1cm}


\footnotetext{\textit{University of Bristol, Bristol}}



\section{Introduction}
NMR spectroscopy remains the pre-eminent analytical technique for elucidating molecular structure in solution, with the prediction and interpretation of $^1\mathrm{H}$ and $^{13}\mathrm{C}$ chemical shifts and scalar coupling constants playing a key role. 
The prediction of these parameters, especially in studies of 3-dimensional molecular structure, are increasingly moving towards quantitative comparison between computed values for proposed chemical structures and experiment. 
In such comparisons, the use of fast and accurate NMR prediction methods is crucial.

Fast empirical predictions of chemical shifts for 2-dimensional chemical structures have been used for decades, with the additivity rules exemplified by Pretsch\cite{pretsch2013tables} and HOSE-code\cite{bremser1978hose} variants forming the basis of many analyses. 
However their applicability is limited by being based on 2-dimensional structures and cannot readily deal with 3-dimensional conformational or stereochemical analysis. 
Some modifications to treating 3-dimensional structures have been made by e.g.\ flat-but-stereochemically-aware HOSE codes\cite{kuhn2019stereo} or single conformer models of experimental systems\cite{aires2002prediction, buevich2019enhancing, meiler2002using} but the improvements in 3-dimensional accuracy are limited  as conformation and flexibility must necessarily be accounted for completely to achieve maximum accuracy. 
Multiple-bond $^1\mathrm{H}$-$^1\mathrm{H}$ coupling constants are more directly linked to 3-dimensional structure, however generically applicable Karplus-style empirical relationships, such as the widely used equation reported by Haasnoot \textit{et al}\cite{haasnoot1980relationship}, suffer from lower accuracy when confronted with complex chemical functionality while equations designed for specific sub-structures, e.g.\ carbohydrates\cite{coxon2009developments}, are not applicable to the whole of chemical space. Finally, many NMR parameters, for example 1-bond $^1\mathrm{H}$-$^{13}\mathrm{C}$ scalar coupling constants, $^1\mathrm{\textit{J}_{CH}}$, which are sensitive to both chemical connectivity \textit{and} 3-dimensional structure are rarely used in isotropic studies precisely because there are no general fast predictive methods for $^1\mathrm{\textit{J}_{CH}}$. \\
\begin{figure*}[h]
    \centering
    \includegraphics[width=\textwidth]{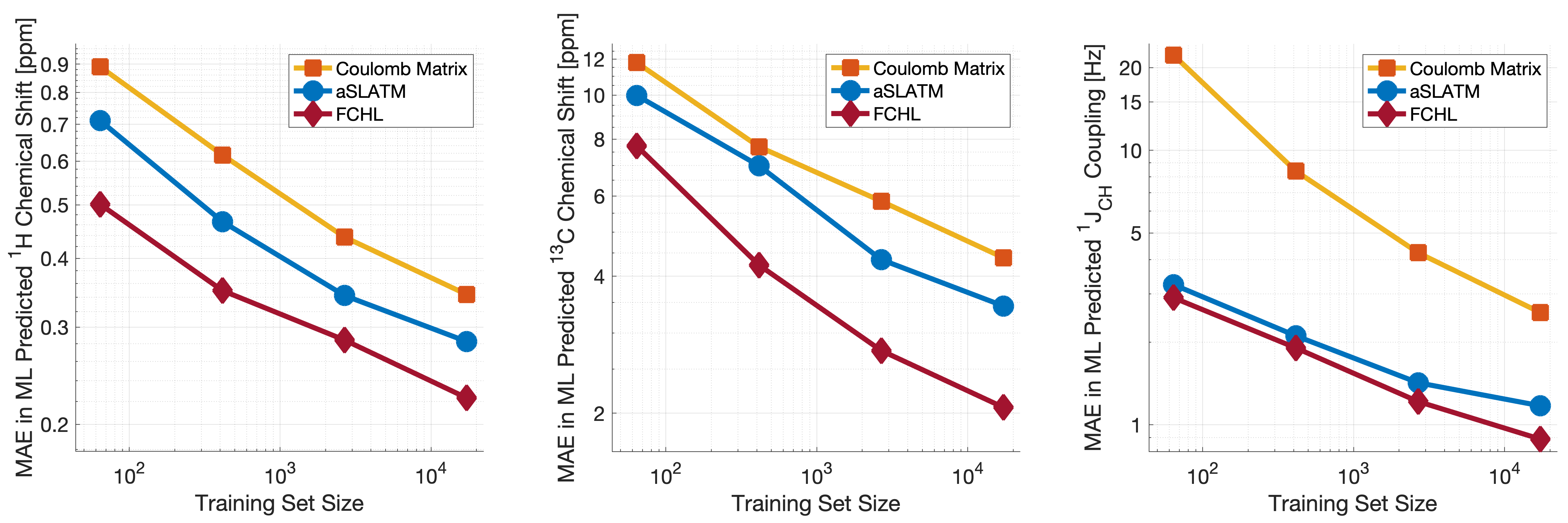}
    \caption{Log-log plot of training set size vs the mean absolute error between ML predictions and DFT of the test set for $\delta ^1\mathrm{H}$ (left), $\delta ^{13}\mathrm{C}$ (centre) and $^1\mathrm{\textit{J}_{CH}}$ couplings (right). Results are shown for the Coulomb Matrix, aSLATM and FCHL kernel similarity measures.}
    \label{fig:loglog}
\end{figure*}
For all of these reasons, the accurate prediction of NMR parameters in modern 3-dimensional structure determinations relies increasingly on the use of quantum chemical calculations, typically based on Density Functional Theory (DFT)\cite{navarro2017state,lodewyk2011computational,steinmann_cs,larsen_procs}. 
Optimal DFT methods can be accurate to within 1-2\%, e.g.\ $^1\mathrm{\textit{J}_{CH}}$  predicted with <4Hz accuracy to experiment\cite{helgaker2008quantum,maximoff2005assessment,san2013computational} (on values that range from roughly 100-250Hz) and <0.2/<2ppm\cite{grimblat2015beyond,semenov2019dft} (on ranges of \textasciitilde10/\textasciitilde200ppm) for $\delta ^1\mathrm{H}$ and $\delta ^{13}\mathrm{C}$ chemical shifts respectively. The substantial downside of DFT is the significant computation time required when using  
methods that can provide sufficient accuracy in NMR predictions. Accurate DFT-based predictions of chemical shift and scalar couplings typically take hours to days of CPU time for a single rigid molecule of even relatively low (\textasciitilde500) molecular mass. The largest proportion of this CPU time is occupied by the NMR computations, especially when computing scalar coupling constants. Naturally, in cases where multiple conformers or isomers must be considered (and thus predictions for multiple structures are required) this becomes days to months of computation for a single study.

Machine learning methods offer a solution to the time-demands of DFT NMR predictions, achieving them in seconds rather than hours or days. Such machines, trained on experimental data, for $^1\mathrm{H}$ and $^{13}\mathrm{C}$ chemical shifts based on 2-dimensional structures are well-established\cite{ACDpredictor,mestrepredictor,castillo2016ask,brandolini2006nmrpredict}. 
These systems are trained on hundreds of thousands of validated experimental chemical shifts arising from tens of thousands of chemical structures. Training such machines for prediction of scalar couplings is more challenging because accurate and validated experimental databases do not exist on this scale (e.g.\ $^1\mathrm{\textit{J}_{CH}}$ values) and they can be critically dependent on 3-dimensional structure (e.g.\ $^3\mathrm{\textit{J}_{HH/CH}}$ values). 
On the other hand, a machine could be trained using large datasets of DFT-computed NMR parameters, such as chemical shifts and scalar couplings, derived from 3-dimensional structures.
Such large DFT-derived datasets can be generated systematically with minimal effort and are not limited to offering accuracy only for structures that are similar to previously experimentally determined molecules. 
With a large enough training database, such a machine would be expected to approach the accuracy of DFT calculation of NMR parameters for 3-dimensional structure analysis, but with several orders of magnitude reduction in time for the NMR predictions. 
This approach was recently reported for solid-state chemical shift predictions by Paruzzo \textit{et  al} (SHIFTML\cite{paruzzo2018chemical}), where the computational demand of DFT calculations on extended lattices are high and comparable to those needed for multi-conformer calculations on solution-state systems.\\
\indent
In this paper we describe the development of our first generation of solution-state NMR prediction machines - IMPRESSION (Intelligent Machine PREdiction of Shift and Scalar Information of Nuclei), trained on DFT-predicted values rather than relying on scarce or error-prone experimental data. We have chosen to demonstrate the versatility of machine learning of NMR parameters using both $^1\mathrm{H}$ and $^{13}\mathrm{C}$ chemical shifts and $^1\mathrm{\textit{J}_{CH}}$ couplings. We include scalar couplings in addition to chemical shift, as the former are less amenable to machine learning based on experimental data, and $^1\mathrm{\textit{J}_{CH}}$ precisely because it has been demonstrated to be valuable for elucidating both 2-dimensional connectivity and 3-dimensional structure \cite{buevich2019enhancing,venkata2014potential} but requires DFT to predict/interpret for most cases. Providing a fast and accurate predictive tool for $^1\mathrm{\textit{J}_{CH}}$ will be especially valuable and could encourage wider acceptance of this and other accessible NMR parameters in structure determinations. We demonstrate that IMPRESSION can predict all these NMR parameters for organic molecules, including 3-dimensional discrimination, with up to DFT accuracy but several orders of magnitude faster and can be applied to experimental data with comparable outcomes to DFT.  
\section{Results and discussion} \label{sec:results}
\begin{figure*}[h]
    \centering
    \includegraphics[width=\textwidth]{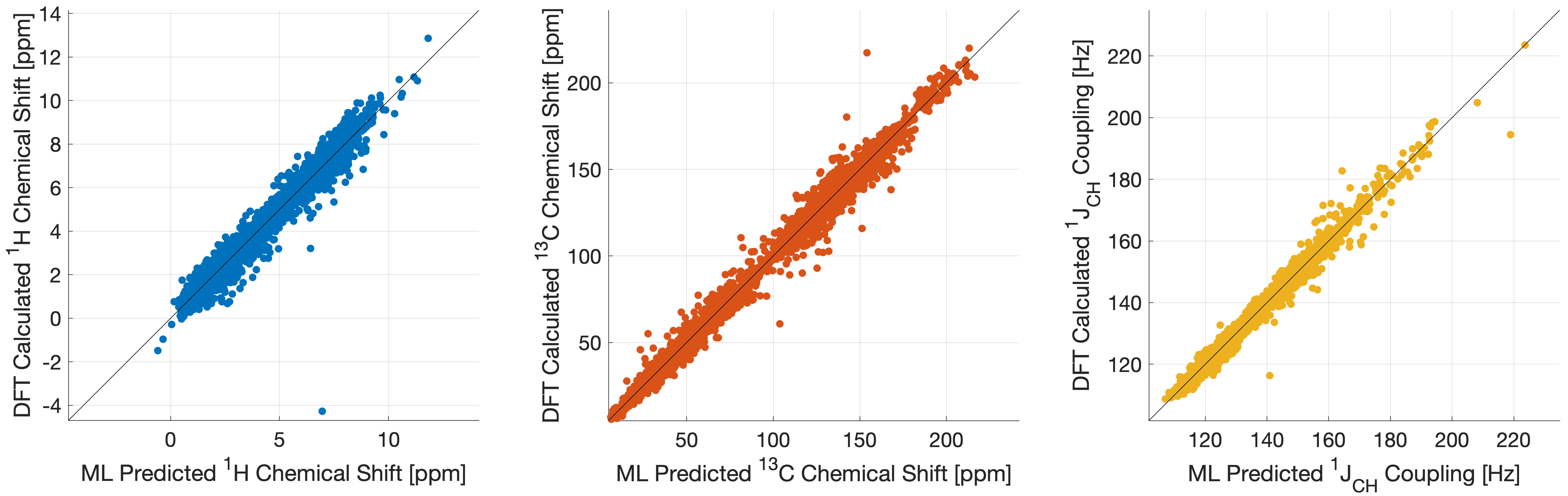}
    \caption{IMPRESSION machine learning predictions compared to DFT computed NMR parameters for $\delta ^1\mathrm{H}$ (left), $\delta ^{13}\mathrm{C}$ (centre) and $^1\mathrm{\textit{J}_{CH}}$ couplings (right) without variance filtering.}
    \label{fig:test_scatter}
\end{figure*}

\subsection{Dataset production and framework}
 In order to train and test IMPRESSION, we developed a dataset of NMR parameters ($\delta ^1\mathrm{H}$, $\delta ^{13}\mathrm{C}$, $^1\mathrm{\textit{J}_{CH}}$), computed using DFT in the Gaussian09 software package \cite{frisch2009gaussian}. 
 While more demanding computational methods could be considered \cite{teale2013benchmarking}, their computational cost would be extortionate with minimal improvement in outcomes for the training and testing datasets described. 
 Instead we found that using $mPW1PW91$/6-311g(d,p) for optimisation and $\omega\text{b}97\text{xd}$/6-311g(d,p)\cite{adamo1998exchange,mclean1980contracted,krishnan1980self-consistent,chai2008systematic,deng2006calculation} for computing the NMR parameters was computationally efficient and sufficiently accurate for comparison to experimental values across a range of NMR parameters. 
 In the geometry optimisations a tight optimisation criteria and ultrafine integral grids were used to minimise molecular orientation affecting geometries and energies (see reference \cite{wilson2019errors} and references therein for a discussion of this). 
 The NMR parameters were calculated using gauge independent atomic orbitals with uncontracted basis sets to improve descriptions of the core orbitals\cite{deng2006calculation} and calculation of all components of the scalar couplings (Fermi contact, spin dipole, diamagnetic spin orbit, paramagnetic spin orbit). 
 The calculated magnetic shielding tensors were converted into chemical shifts using the linear scaling method and reference compounds reported by Tantillo \textit{et al}\cite{lodewyk2011computational,CHESHIRE}.    
  A training set of 882 structures 
(17,222 
$^1\mathrm{\textit{J}_{CH}}$ ; 
18,383 
 $\delta ^1\mathrm{H}$;
 17,081 
 $\delta ^{13}\mathrm{C}$ values/environments) were selected by an adaptive sampling (active learning) procedure\cite{qbc, behler_active_learning, roitberg_active_learning} from a superset of 75,382 chemical structures comprising only C, H, N, O and F atoms in the Cambridge Structural Database\cite{csd} (accessed 7/9/2018). 
The adaptive sampling procedure trains an initial IMPRESSION machine from 100 chemical structures and then uses this machine to predict the parameters for all remaining structures in the superset to measure their variance in a 5-fold cross validation (\textit{i.e.} how much a given parameter changes when predicted from 5 separate machines each trained on a different 80\%  subset  of the current training set). 
The 100 structures in the superset which show the highest variance are then added to the training dataset and the cycle is iterated (see supplementary information for further details).
Adaptive sampling therefore adds the 100 structures at each training iteration which IMPRESSION is the most uncertain about.
In doing so, each added structure provides the maximum benefit to the machine and substantially reduces the overall computational cost required to reach a given accuracy. The test set, against which the quality of the IMPRESSION predictions is independently tested, was comprised of a further 410 chemical structures (7788 $^1\mathrm{\textit{J}_{CH}}$ ; 7832 $\delta ^1\mathrm{H}$; 7522 $\delta ^{13}\mathrm{C}$ environments) harvested from the CSD-500 dataset recently reported by Paruzzo et al\cite{paruzzo2018chemical}.\\
\indent
IMPRESSION uses a Kernel Ridge Regression\cite{saunders1998ridge} (KRR) framework to learn the $^1\mathrm{\textit{J}_{CH}}$ scalar couplings and ${}^{13}\mathrm{C}$ and ${}^{1}\mathrm{H}$ chemical shifts of molecular structures. KRR was successfully used by Paruzzo \textit{et al} to develop SHIFTML\cite{paruzzo2018chemical}. Neural networks have also been used to predict chemical shifts in small molecules from experimental data \cite{meiler2002using, nn1, nn2}, however we found no clear advantages in using feed forward neural networks in this work as the accuracy was comparable to KRR for the datasets used, with the kernel methods being much faster to train with the given training set size. 
In order to encode the similarity between chemical environments of each molecular structure we tested three approaches previously described - Coulomb matrices\cite{aCM}, aSLATM\cite{huang2017dna} and FCHL\cite{faber2018alchemical}, all available from the QML python package\cite{qml}. We refer the reader to section \ref{sec:krr} in the supplementary information and the respective papers describing each representation for more details. 
All of these kernel similarity measures compare \textit{atomic} environments, so in the case of $^1\mathrm{\textit{J}_{CH}}$, we used the product of the separately calculated kernel similarities for the $^1\mathrm{H}$ and $^{13}\mathrm{C}$ nuclei as this performed better than either atomic environment alone.
The KRR procedure is further described in the supplementary information (section \ref{sec:krr}).\\
\indent
Both aSLATM and FCHL were found to outperform Couloumb matrices (Figure \ref{fig:loglog}), which is expected as Coulomb matrices only include 2-body interactions, while aSLATM and FCHL both include three-body interactions as well. As FCHL provided the best performance for all three parameters and was substantially more computationally efficient than aSLATM, it was used in the final development of the full IMPRESSION machine.

\noindent
\subsection{Performance relative to DFT}
During training, the machine performance for prediction of all NMR parameters ($\delta ^1\mathrm{H}$, $\delta ^{13}\mathrm{C}$, $^1\mathrm{\textit{J}_{CH}}$) improved steadily with increasing training set size, as illustrated in the learning curves (Figure \ref{fig:loglog}). This indicates that the accuracy of the machine can be further improved by adding additional training data, however the absolute gains become marginal beyond the dataset size used here with a ten-fold increase in training set size approximately halving the average error between IMPRESSION and DFT.
After training on the full set of 882 chemical structures, 
IMPRESSION predictions achieved mean absolute errors (MAE) of 0.23ppm/2.45ppm/0.87Hz for $\delta ^1\mathrm{H}$/$\delta ^{13}\mathrm{C}$/$^1\mathrm{\textit{J}_{CH}}$/ predictions and root mean squared error (RMSE) of 0.35ppm/3.88ppm/1.39Hz against the independent test set (Figure 2).

Notably however, a very small number of predictions for the test set were much less reliable. For example, 186 (\texttildelow 2.3\%) of the $\delta ^1\mathrm{H}$ values had errors >1ppm between IMPRESSION and DFT, with a maximum error (MaxE) of 11.22ppm. 
Similar outcomes were observed for the other parameters with 187 $\delta ^{13}\mathrm{C}$ values (\texttildelow 2.5\%) with errors >10ppm (MaxE = 63.33ppm) and 14 (\texttildelow 0.2\%) of the 7788 $^1\mathrm{\textit{J}_{CH}}$ predicted $^1\mathrm{\textit{J}_{CH}}$ values having errors of >10Hz (MaxE = 24.63Hz). 
\begin{figure*}[h]
   \centering
   \captionsetup[subfigure]{labelformat=empty}
\begin{subfigure}{0.33\linewidth}
   \includegraphics[width=\linewidth]{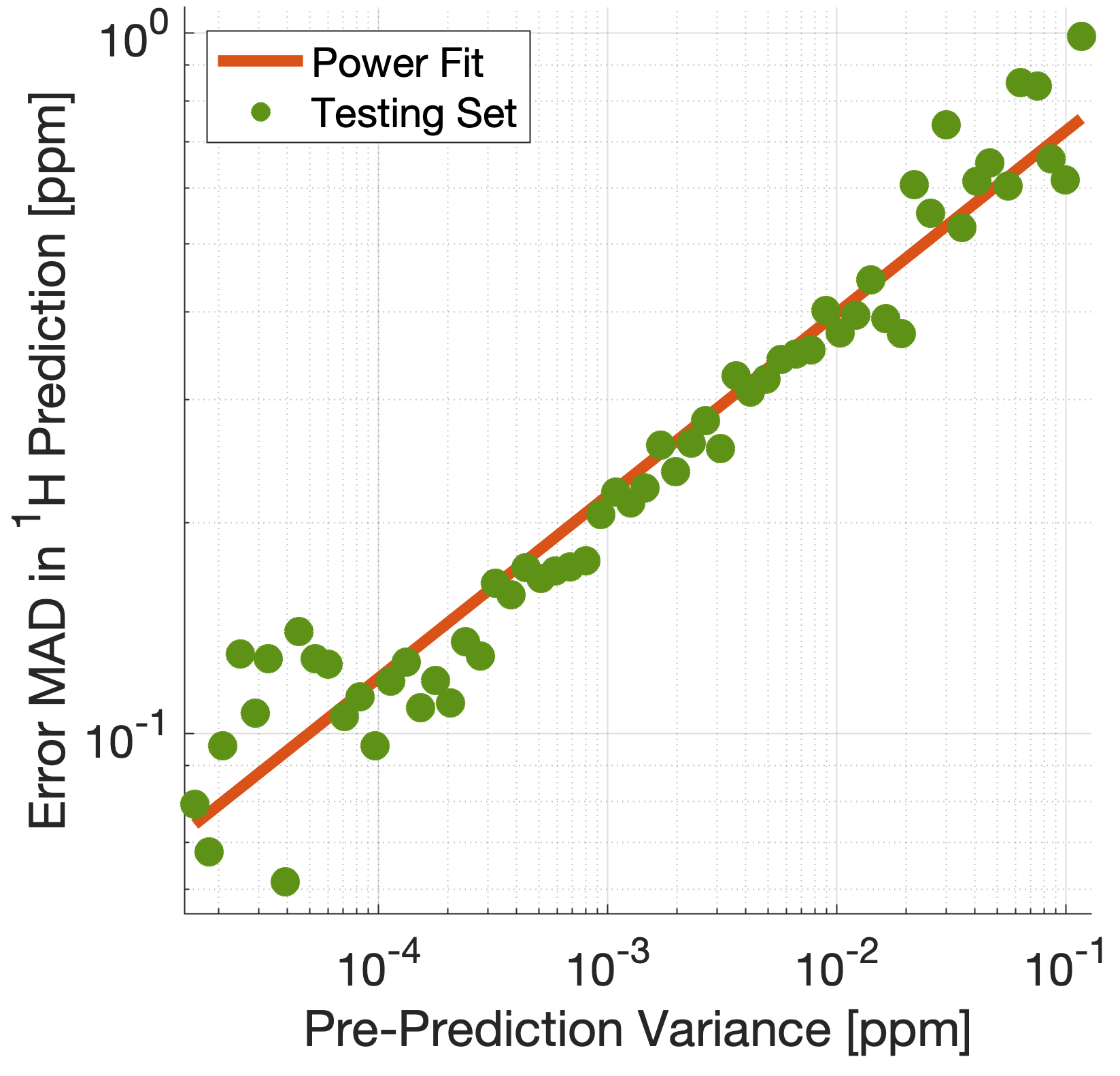}
\end{subfigure}
\begin{subfigure}{0.33\linewidth}
   \includegraphics[width=\linewidth]{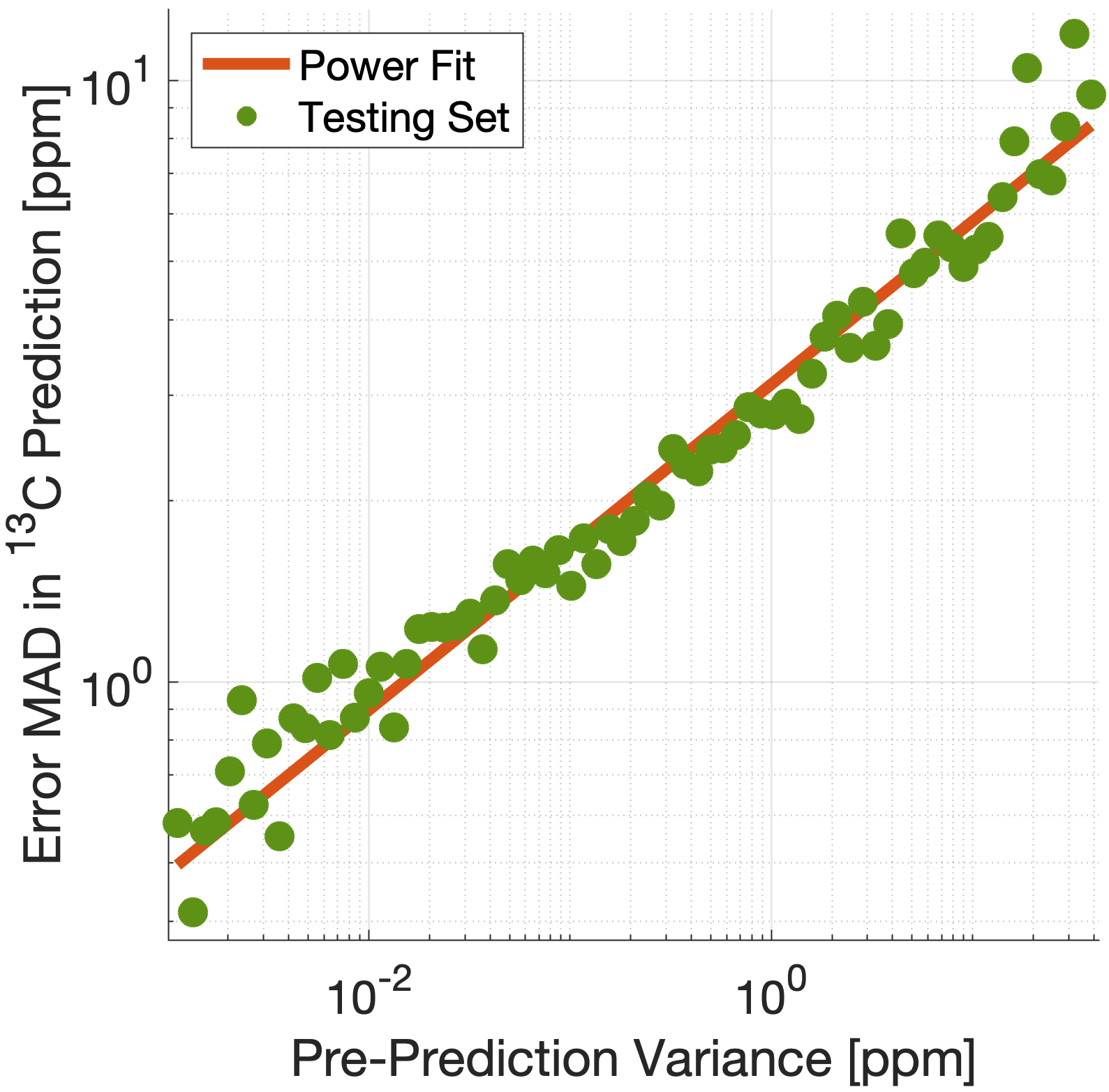}
\end{subfigure}
\begin{subfigure}{0.33\linewidth}
   \includegraphics[width=\linewidth]{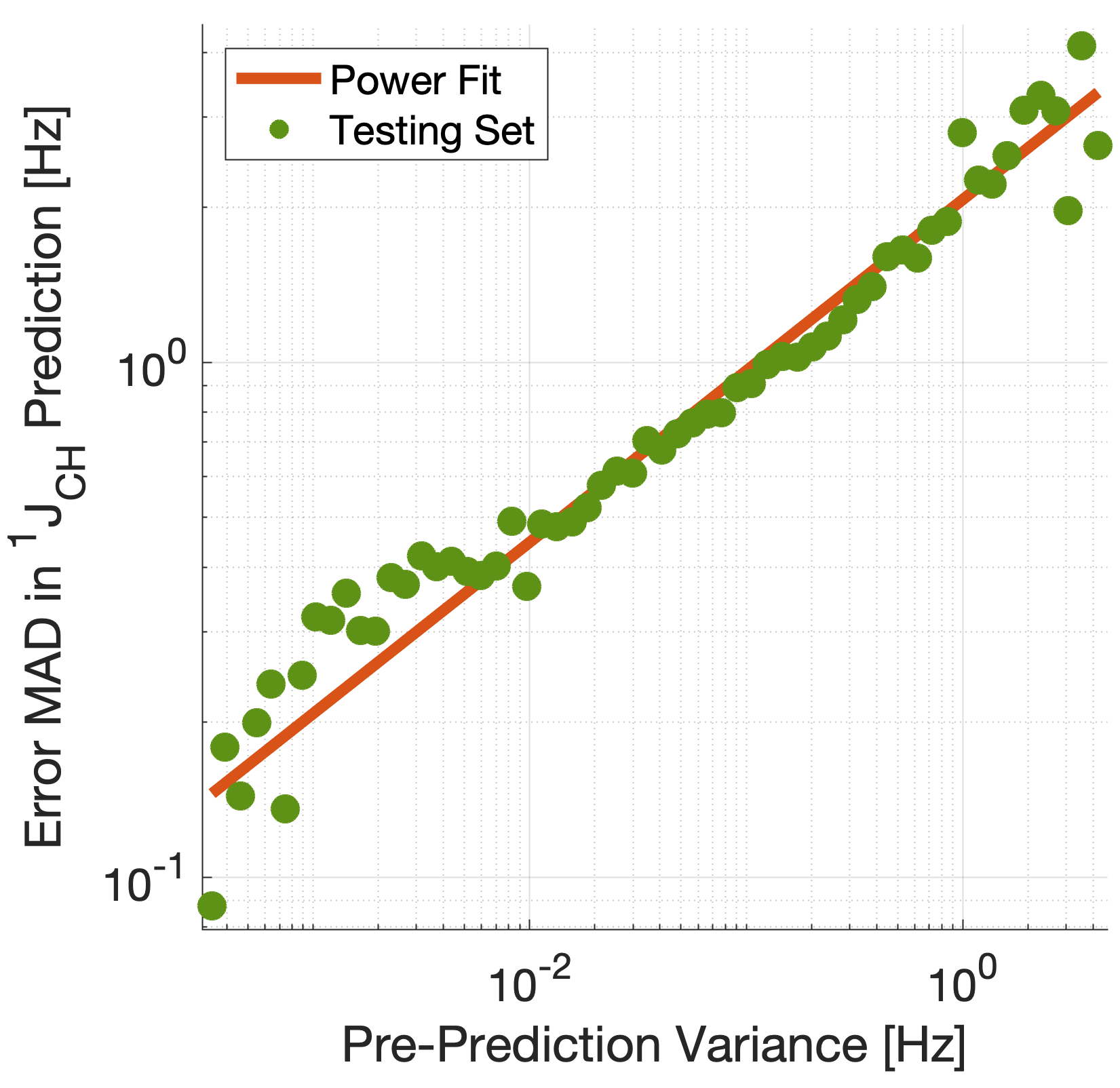}
\end{subfigure}
\\
\begin{subfigure}{0.33\linewidth}
   \includegraphics[width=\linewidth]{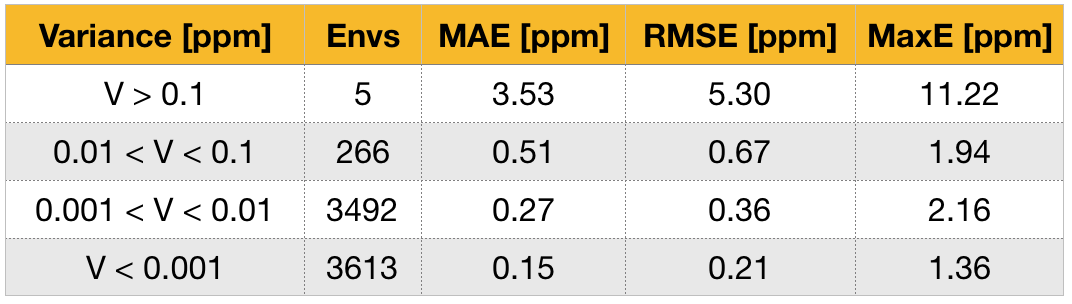}
\end{subfigure}
\begin{subfigure}{0.33\linewidth}
   \includegraphics[width=\linewidth]{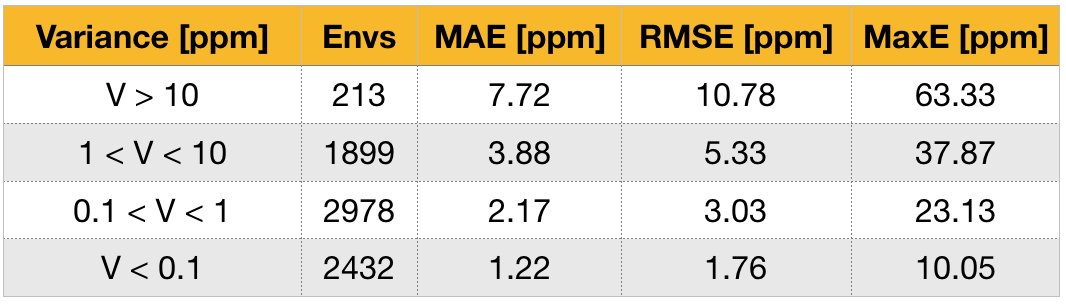}
\end{subfigure}
\begin{subfigure}{0.33\linewidth}
   \includegraphics[width=\linewidth]{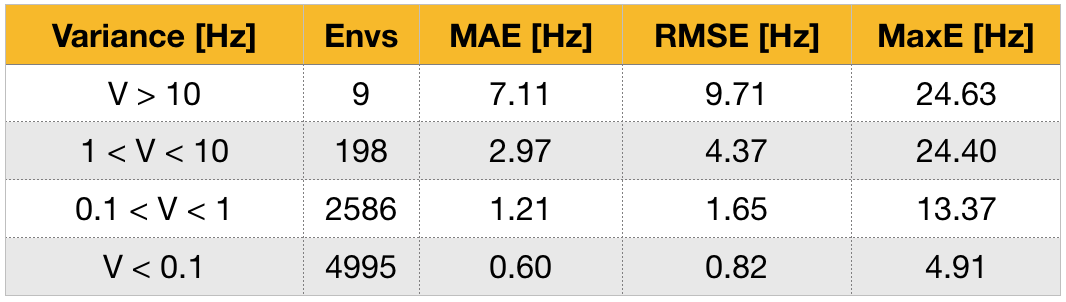}
\end{subfigure}
\caption{Top: correlation between pre-prediction variance and prediction error between DFT and IMPRESSION for $\delta ^1\mathrm{H}$ (left), $\delta ^{13}\mathrm{C}$ (centre) and $^1\mathrm{\textit{J}_{CH}}$ couplings (right) on the test set. The prediction errors were binned by variance and an average error (MAE) was produced for each bin. Bottom: error metrics for different variance ranges.}
\label{fig:variance_func}
\end{figure*}
 Diagrams of the structures containing the five most significant outliers for each NMR parameter are shown in figures \ref{fig:big_errs_hcs}, \ref{fig:big_errs_ccs}, and \ref{fig:big_errs_1jch} in the SI. Examination of the  chemical environments of the most significant outliers show that they arise from unusual functional groups such as those containing sp-hybridised atoms, or unusual 3-dimensional environments such as atoms near pi-systems of aromatic rings.
These outliers suggest that, as desired, the machine learning system is indeed very sensitive to the 3-dimensional relationships of the atoms in the structure. However this same sensitivity also makes IMPRESSION less accurate for chemical environments which are not very similar to environments across the 882 molecular structures used to train IMPRESSION. \\
Crucially, we are able to \textit{a priori} identify poorly described environments using the same variance-based approach used to generate the training set. By assessing the variance in the prediction of a given NMR parameter across a 5-fold cross-validation, we can quantify our confidence in each individual prediction since environments which are poorly described by the chemical structures in the training set will have high variance in this cross-validation. There is indeed a clear correlation of variance against prediction error for the independent test set (Figure \ref{fig:variance_func}). 
The tables in Figure \ref{fig:variance_func} suggest that the bulk of the environments are predicted very accurately, and that the high variance environments are the dominant source of the large outliers.\\
\indent
In principle, removing IMPRESSION-predicted values which show high variances in cross-validation should provide a "pre-prediction variance filter" that will substantially improve the quality of, and thus the confidence in, IMPRESSION predictions.
Selecting an appropriate variance cut-off for each NMR parameter is then simply a balance between desired prediction quality and the number of predictions which will be excluded by that cut-off. 
Reports of DFT accuracy with respect to experiment for $\delta ^1\mathrm{H}$ and $\delta ^{13}\mathrm{C}$ chemical shift predictions vary significantly, but typically in the range of 0.2-0.4ppm/2-4ppm, with the best reported accuracies down to <0.2/<2ppm\cite{grimblat2015beyond,semenov2019dft} in optimal cases. 
Similarly, Buevich \textit{et al} recently highlighted that current best-in-class DFT methods predict $^1\mathrm{\textit{J}_{CH}}$ experimental values with accuracies of 2-4Hz, when presenting an optimised workflow for calculating $^1\mathrm{\textit{J}_{CH}}$ values which achieved an RMSE of 1.61Hz.\\
\indent
We therefore identified variance cut-offs for IMPRESSION predictions that provide a good comprimise between accuracy and excluded values for the test set, which were found to be 1Hz for $^1\mathrm{\textit{J}_{CH}}$, 0.1ppm for $\delta ^1\mathrm{H}$ and 5ppm for $\delta ^{13}\mathrm{C}$. 
Applying these pre-variance filter values improves the fits between IMPRESSION and DFT to levels that are comparable with literature reports for MAE/RMSE of DFT vs experiment (MaxE is rarely reported for large experimental validations, but the reader can find comparators from our experimental validations described below in section 2.3). 
For $\delta ^1\mathrm{H}$ the 0.1ppm filter excludes 5 environments (<0.1\%) and improves the fit to MAE = 0.23ppm, RMSE = 0.32ppm; MaxE = 2.16ppm. 
For $\delta ^{13}\mathrm{C}$ a 5ppm filter provided a good fit MAE = 2.17ppm; RMSE = 3.25ppm; MaxE = 37.87ppm) while excluding 538 (\texttildelow 7.2\%) of the environments. 
For $^1\mathrm{\textit{J}_{CH}}$ a 1Hz filter improved the fit to MAE = 0.81Hz, RMSE = 1.17Hz; MaxE = 13.37Hz while discarding only 207 (<3\%) of the environments. \\
\begin{figure*}[h]
    \centering
    \includegraphics[width=\textwidth]{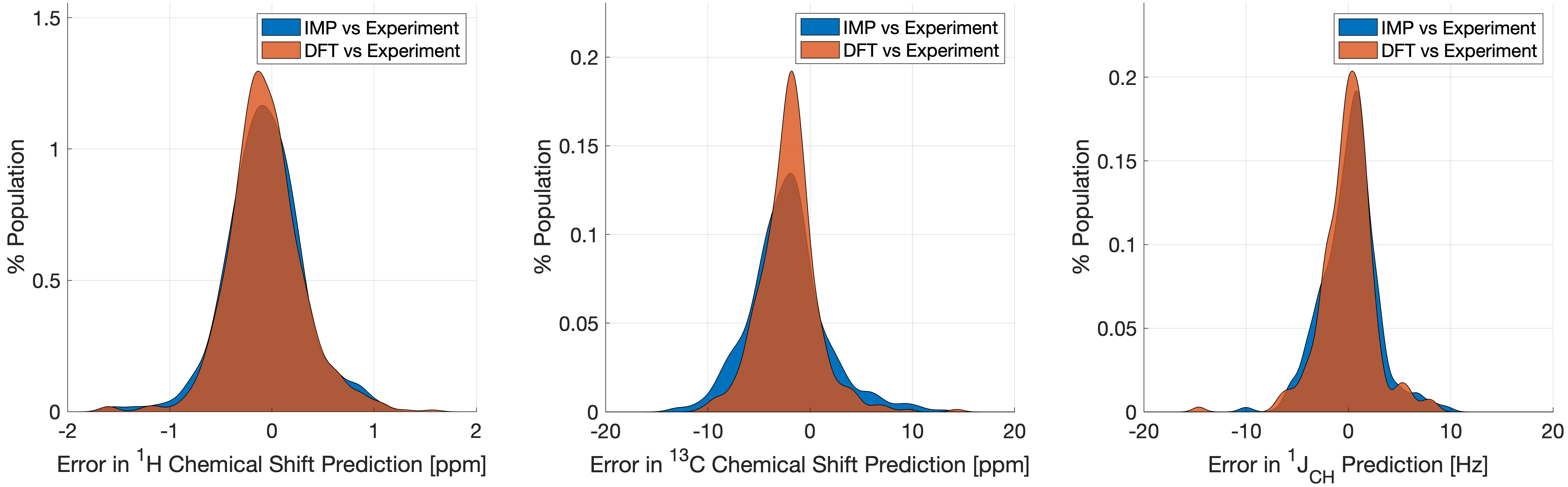}
    \caption{Distribution of errors for machine learning NMR predictions and DFT calculations when compared to the relevant experimental validation dataset for $ \delta ^1\mathrm{H}$ (left), $\delta ^{13}\mathrm{C}$ (centre) and $^1\mathrm{\textit{J}_{CH}}$ couplings (right). Variance filters applied to IMPRESSION predictions: $ \delta ^1\mathrm{H}$ = 0.1ppm (0 of 734 environments removed), $\delta ^{13}\mathrm{C}$ = 5ppm (24 of 457 environments removed), $^1\mathrm{\textit{J}_{CH}}$ = 1Hz (143 of 608 environments removed).}
    \label{fig:overlap}
\end{figure*}
\indent
As highlighted by the learning curves, further improvement to the machine predictions of DFT NMR results can be made by increasing size of the DFT-derived training dataset by around an order of magnitude. However at this stage variance-filtered IMPRESSION compares well enough with respect to DFT that it was taken forward. It should also be noted at this point that IMPRESSION only accelerates NMR prediction, it does not accelerate the 3D structure generation by DFT (which can still take hours/days). This overall time, \textit{i.e.}3D structure generation + NMR prediction, could be reduced further by using 3D structures derived from molecular mechanics rather than DFT. While not the key focus here, the use of molecular mechanics structures as inputs to a re-trained IMPRESSION machine was explored. While practical, this resulted in a \texttildelow 30-50\% increase in the average prediction errors for $\delta ^1\mathrm{H}$ and $^1\mathrm{\textit{J}_{CH}}$ presumably arising from a mismatch between the detail of molecular mechanics geometries and those used to calculate the DFT NMR parameters (see Section S2 in the SI for details). Interestingly, $\delta ^{13}\mathrm{C}$ predictions were relatively insensitive to this change, perhaps reflecting better description of carbon environments by molecular mechanics forcefields. This is an exciting avenue to explore further, but to focus the discussion here on the ability of IMPRESSION to reproduce DFT NMR predictions, the subsequent experimental comparisons are based on the IMPRESSION machine trained on the same DFT-geometries used for the DFT NMR predictions. 

\begin{figure*}
    \centering
   \captionsetup[subfigure]{labelformat=empty}
\begin{subfigure}{0.38\linewidth}
   \includegraphics[width=\linewidth]{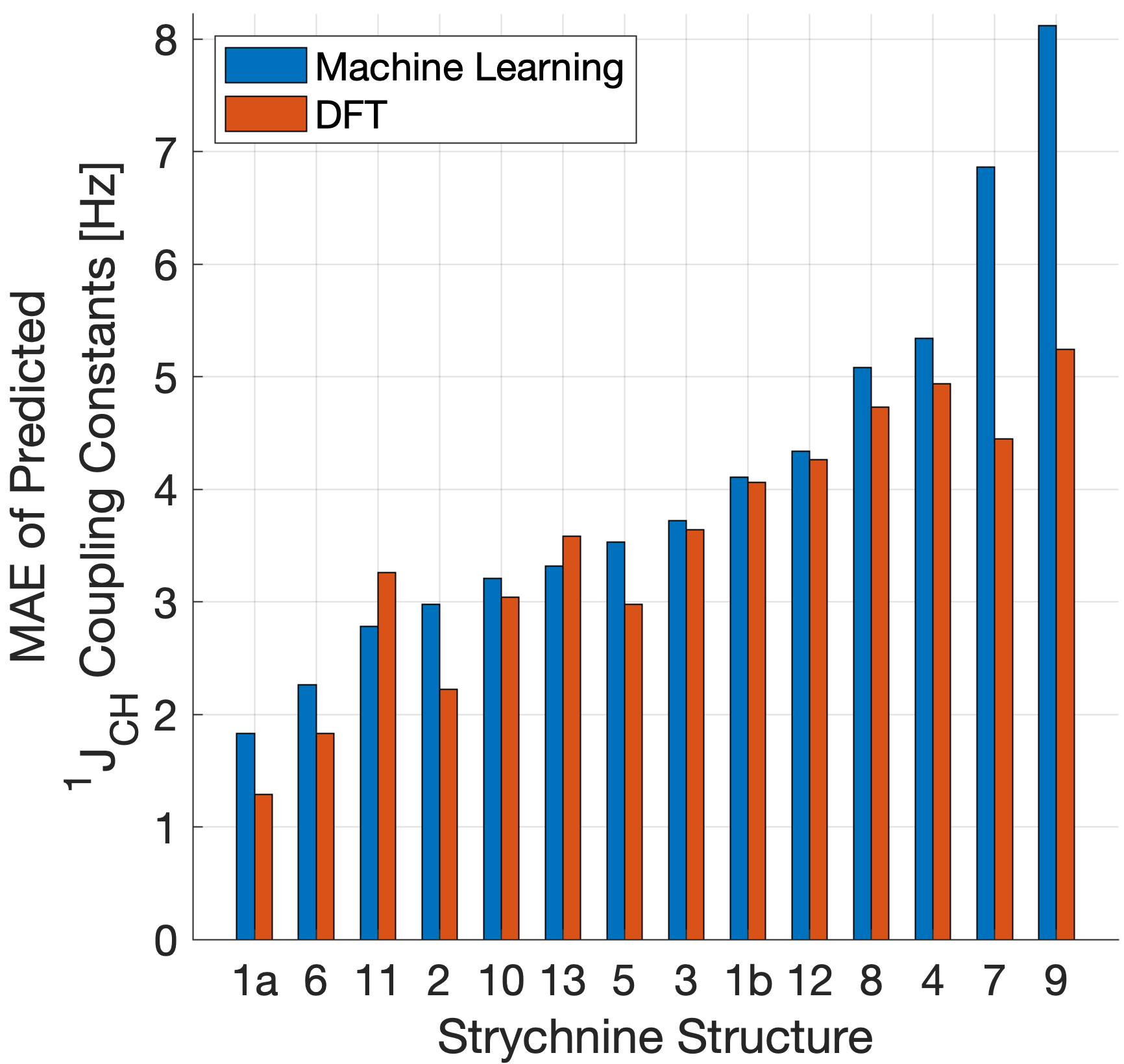}
\end{subfigure}
\quad
\begin{subfigure}{0.19\linewidth}
   \includegraphics[width=\linewidth]{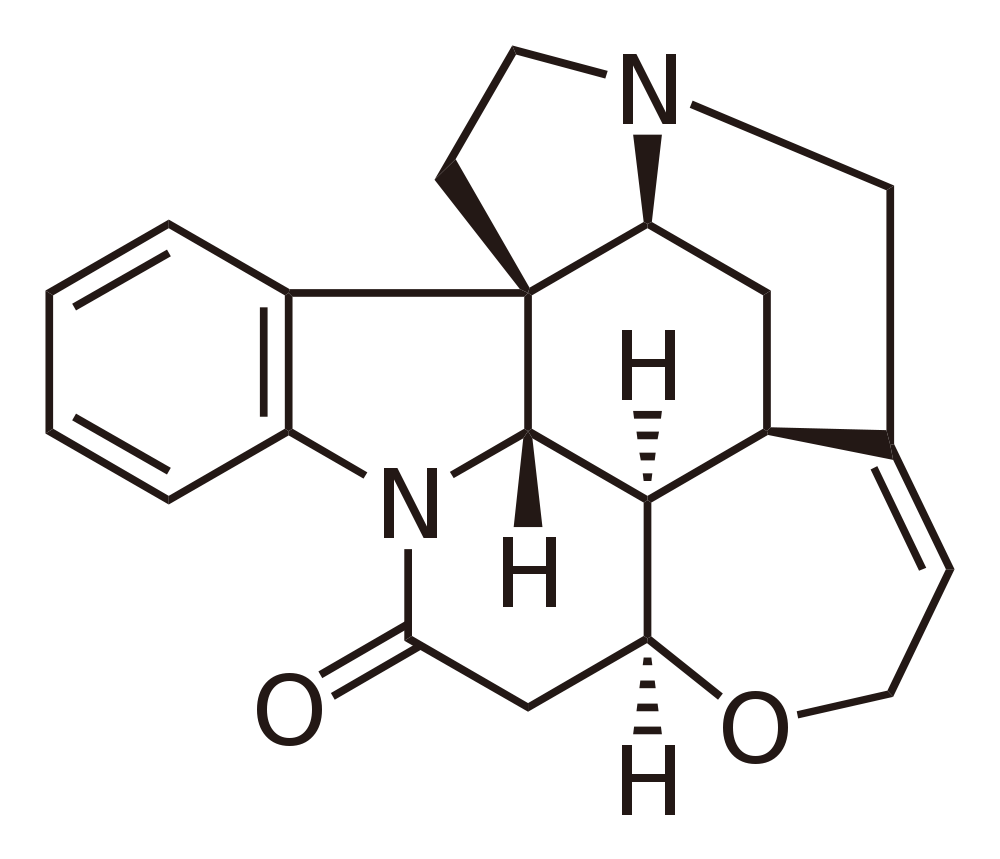}
\end{subfigure}
\quad
\begin{subfigure}{0.38\linewidth}
   \includegraphics[width=\linewidth]{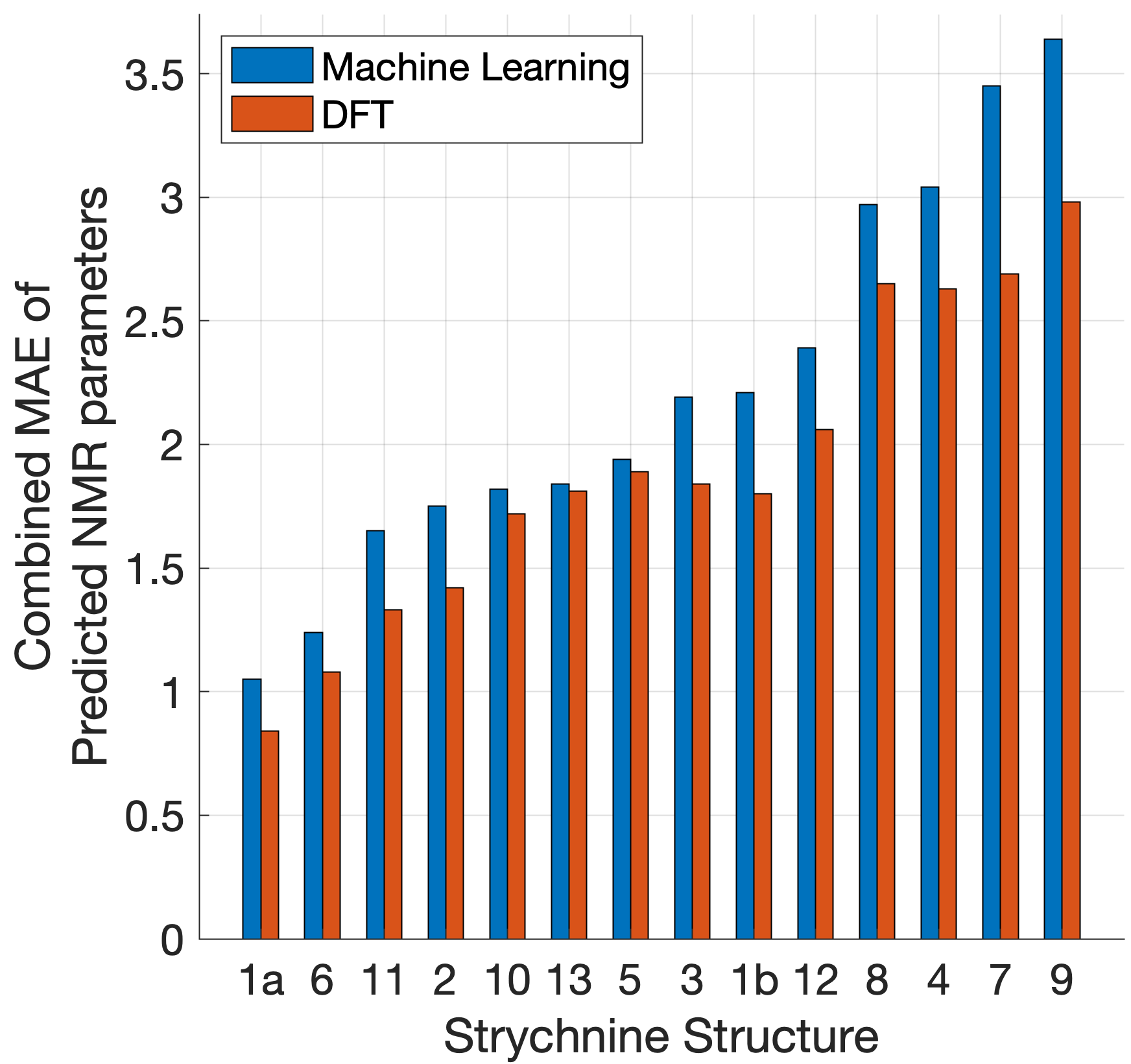}
\end{subfigure}
    \caption{Errors from comparison of NMR experimental data of the natural product strychnine (centre) to IMPRESSION (blue) and DFT (red) predictions for 13 diastereomers of strychnine, including two conformers for the natural product 1: the lowest energy 1a (>97\% populated) and the next lowest energy 1b (<3\% populated). 
    The left hand plot shows MAE for $^1\mathrm{\textit{J}_{CH}}$  while the right hand plot shows the geometric mean error for all NMR parameters ($ \delta ^1\mathrm{H}$, $\delta ^{13}\mathrm{C}$ and $^1\mathrm{\textit{J}_{CH}}$) combined. Variance filters applied to predictions: $ \delta ^1\mathrm{H}$ = 0.1ppm, $\delta ^{13}\mathrm{C}$ = 5ppm, $^1\mathrm{\textit{J}_{CH}}$ = 1Hz.}
    \label{fig:strychnine}
\end{figure*}

\subsection{Performance relative to experiment}
Naturally, a key test of IMPRESSION is its ability to reproduce DFT predictions of experimental values of relevant compounds. To test this for $^1\mathrm{\textit{J}_{CH}}$, a validation set of
608 experimental $^1\mathrm{\textit{J}_{CH}}$ values were taken from structures collated by Venkata \textit{et al}\cite{venkata2014potential} which contain C, H, N, O and F elements only. Firstly, we checked the ability of our $\omega\text{b}97\text{xd}$/6-311g(d,p) DFT method itself to reproduce these experimental results. It should be noted in the subsequent analysis that all DFT and IMPRESSION predictions were based on the single conformers that Venkata \textit{et al} reported for each compound. While not making the predictions entirely experimentally relevant, it allows direct comparison between DFT and IMPRESSION NMR predictions for this data.  
Calculating the 608 couplings with $\omega\text{b}97\text{xd}$/6-311g(d,p) took 156 CPU hours and initially gave a relatively poor fit to experiment (MAE = 10.92Hz) but with a systematic offset from the experimental data by an average of -10.91Hz. 
Adding this systematic offset to the DFT-predicted values provided a good fit between DFT and experiment (MAE = 2.16Hz; RMSE = 3.33Hz; MaxE = 20.05Hz) and this was used for all subsequent comparisons to experiment based on this DFT method. 
As IMPRESSION is trained on DFT data computed with this same $\omega\text{b}97\text{xd}$/6-311g(d,p) method and both methods use only single conformer predictions for each molecule, then these statistics represent a practical limit for the accuracy that we might expect from IMPRESSION on this experimental data.\\
\indent
IMPRESSION took only 60 CPU seconds to predict the full set of 612 $^1\mathrm{\textit{J}_{CH}}$ values but with some substantial outliers (MAE = 4.52Hz; RMSE = 10.49Hz; MaxE = 120.3Hz). 
Applying the 1Hz variance filter gave: MAE = 2.01Hz, RMSE = 2.69Hz, MaxE = 10.01Hz (removing 143 values) which was essentially identical accuracy to that obtained from the DFT method for these same filtered environments: MAE = 1.83Hz, RMSD = 2.60Hz, MaxE = 14.63Hz. 
An overlay of the error distributions for DFT and the 1Hz variance-filtered IMPRESSION vs the experimental values (Figure \ref{fig:overlap}) demonstrates the comparability between machine learning and DFT for $^1\mathrm{\textit{J}_{CH}}$ predictions. 
This represents quite excellent performance of the machine for reproducing experimental data in just a few seconds, with quality for the majority of environments as good as the best MAEs (1.5-4Hz) described by Buevich \textit{et al} as typical for DFT methods, with <25\% of the values being tagged as unreliable by the variance filter. Of course, if a slight loss in prediction quality is acceptable for a given study, then more  predicted values could be retained by using a slightly looser variance-filter.\\
\indent
Similar accuracy could be obtained for IMPRESSION predictions of 734 $^1\mathrm{H}$ chemical shifts for 36 structures reported by Smith and Goodman\cite{smith2010assigning} in their DP4 dataset (again, single conformers were used for both DFT and IMPRESSION predictions). IMPRESSION predictions gave MAE = 0.29ppm, RMSD = 0.38ppm, MaxE = 1.59ppm, variance filter 0.1ppm but in this case no environments were lost to the variance filter and provided essentially the same outcomes as the $\omega\text{b}97\text{xd}$/6-311g(d,p) DFT method on the same single conformer structures (MAE = 0.28ppm, RMSE 0.37ppm, MaxE 1.62ppm, see Figure \ref{fig:overlap} for an overlay of errors). 
The IMPRESSION predictions for $\delta ^{13}\mathrm{C}$ using the 5ppm variance filter identified during training and testing of the machine compared slightly less well to the DP4 experimental dataset (MAE = 3.44ppm, RMSE = 4.30ppm, MaxE = 13.06ppm, removing 24 environments) than DFT (MAE = 2.78ppm, RMSE = 3.48ppm, MaxE = 14.33ppm). 
A tighter 1ppm variance filter for the  $\delta ^{13}\mathrm{C}$ predictions was examined, but gave only a slight improvement in prediction quality MAE = 3.20ppm, RMSE = 4.00ppm, MaxE = 13.03ppm while removing 120 out of the 458 carbon environments. 

At every stage in this study we found that the  IMPRESSION $\delta ^{13}\mathrm{C}$ predictions have a wider distribution of errors than the other NMR parameters when compared to the quality of the DFT from which they are trained. This is unsurprising given that the structural environments of $^{13}\mathrm{C}$ nuclei in molecules are inherently more complex than 1H given the higher valency and thus more complex bonding environments and geometries, so in future development, larger training datasets focussed on optimising $\delta ^{13}\mathrm{C}$ predictions will be beneficial. 

\FloatBarrier
\subsection{3-Dimensional structure discrimination}

A demanding test of IMPRESSION is in its ability to predict and discriminate experimental NMR data for stereoisomeric compounds i.e. those that differ only in their 3-dimensional structure, but not connectivity. Even though IMPRESSION has not been explicitly trained to deal with multiple conformers/isomers of any one compound, 3-dimensional variation is implicit  within  the  varied  chemical  structural space of the adaptively sampled training set. Buevich \textit{et al} recently demonstrated \cite{buevich2019enhancing} that DFT prediction of $^1\mathrm{\textit{J}_{CH}}$ values can successfully discriminate the naturally occurring structure \textbf{1} of the polycyclic alkaloid strychnine (Figure \ref{fig:strychnine}, centre) from 12 other diastereomers (see supplementary information section \ref{sec:strych_SI} for the structures) based on comparison with the experimental $^1\mathrm{\textit{J}_{CH}}$ values of the natural product.
Pleasingly, the same test conducted with IMPRESSION-predicted $^1\mathrm{\textit{J}_{CH}}$ values (blue bars in Figure \ref{fig:strychnine}, left) also correctly identifies the natural product diastereomer \textbf{1a} as having the smallest error (MAE = 1.87Hz; RMSE = 2.50Hz; MaxE = 6.19Hz). 
The error for the correct structure is \textasciitilde30\% lower than the diastereomer with the second lowest error \textbf{6} (MAE = 2.48Hz; RMSE = 3.38Hz; MaxE = 8.42Hz) and this is very similar to the discrimination offered by $\omega\text{b}97\text{xd}$/6-311g(d,p) (red bars in Figure \ref{fig:strychnine}). 
Indeed IMPRESSION could also distinguish between the 3-dimensional structures of \textbf{1a}, the lowest energy conformer of the natural product (97\% population in solution), and \textbf{1b} which is the second lowest energy conformer (3\% population in solution)\cite{butts2011high}. So while the absolute accuracy of IMPRESSION for predicting $^1\mathrm{\textit{J}_{CH}}$ values for strychnine (MAE = 1.87Hz) is slightly lower than that obtained from the DFT method (MAE = 1.31Hz), its discriminating power between structural isomers is nearly the same. \\
\indent
Combining IMPRESSION predictions for $^1\mathrm{\textit{J}_{CH}}$ with  $^1\mathrm{H}$ and $^{13}\mathrm{C}$ chemical shifts also provides correct identification of the naturally occurring structure, but IMPRESSION and DFT now both see structure \textbf{2} as the next best candidate (Figure \ref{fig:strychnine}, right). This is due to the experimental $\delta ^1\mathrm{H}$ values having better agreement with the predictions for diastereomer \textbf{2} than \textbf{1a} for DFT and also IMPRESSION. While this is obviously problematic for structure elucidation purposes, it clearly arises because of a deficiency in the DFT prediction of $^1\mathrm{H}$ chemical shifts, which is then faithfully reproduced by IMPRESSION. For the individual MAE values across all three parameters see supplementary information section \ref{sec:strych_SI}.\\
\indent
Similarly, we found that IMPRESSION predictions can be used to correctly assign the diastereotopic protons in strychnine. IMPRESSION and DFT predictions of $^1\mathrm{\textit{J}_{CH}}$ for the diastereotopic protons in strychnine were consistently in line with each other (details can be found in Section \ref{sec:strych_protons} of the SI) and for the three methylene groups where there is a significant difference (>>2Hz) in experimental $^1\mathrm{\textit{J}_{CH}}$ values both methods correctly assign these protons (Figure \ref{fig:protons1}). \\
\indent
Finally, we validated IMPRESSION chemical shift predictions for natural product structures. We conducted DFT and IMPRESSION predictions on structures from a recent report which suggested structural reassignments for oxirane-containing natural products on the basis of DU8+ DFT calculations\cite{kutateladze2018addressing}. 
To avoid complications with incorrect DFT prediction of conformer energies leading to poor population averaging of NMR parameters from the constituent conformers, we limited the validation to 'rigid' structures in the report that contained only one dominant conformer after conformational searching. 
Pleasingly, while our results did not always agree with the DU8+ analysis, IMPRESSION  was just as effective as our underlying  $\omega\text{b}97\text{xd}$/6-311g(d,p) DFT method in discriminating each original and revised chemical structure (see section \ref{sec:csdiscrimination} in the supplementary information for more details). Once again this confirms that IMPRESSION is capable of making predictions that are of comparable quality to it's underlying DFT method $\omega\text{b}97\text{xd}$/6-311g(d,p), and thus any improvements in the DFT method used to train IMPRESSION will be subsequently expressed in the quality of IMPRESSION predictions.

\FloatBarrier
\section{Conclusions}
In summary, this first generation IMPRESSION machine, trained on DFT-computed NMR parameters derived from a set of 3-dimensional structures is capable of reproducing DFT-predicted NMR parameters for a range of experimentally relevant systems with high accuracy but in a fraction of the time. Accurate and generalised prediction of NMR parameters for 3-dimensional applications has not been addressed by previous machine learning systems but the confidence provided by the variance-filtered IMPRESSION results makes this tool essentially as robust for 3-dimensional applications to experimental systems as DFT. At this stage, the two primary sources of error in IMPRESSION predictions of experimental data are errors in the underlying DFT method on which it is trained (of which there can be several\cite{iron2017evaluation, bagno2006toward, laskowski2013assessment}) and the range of chemical space covered by the current IMPRESSION training set. We are working to improve both of these factors, as well as extending the predictions to multiple-bond scalar couplings for future generations of IMPRESSION, along with developing a more rigorous statistical treatment of the predicted values taking into account the pre-prediction variance. 

\section*{Conflicts of interest}
There are no conflicts to declare.

\section*{Acknowledgements}
This work was carried out using the computational facilities of the Advanced Computing Research Centre, University of Bristol - http://www.bristol.ac.uk/acrc/. We thank Dr Peter Howe (Syngenta, UK) for useful discussions regarding the experimental $^1\mathrm{\textit{J}_{CH}}$ dataset used. WG thanks the EPSRC National Productivity Investment Fund (NPIF) for Doctoral Studentship funding.
LAB thanks the Alan Turing Institute under the EPSRC grant EP/N510129/1. DRG acknowledges funding from the Royal Society as a University Research Fellow, and also from EPSRC grant EP/M022129/1.AJM thanks EPSRC for funding (EP/M022609/1, CCP-BioSim). 
LAB and DRG acknowledge support of this work through EPSRC grant EP/P021123/1.
We further acknowledge the use of the following software: BayesianOptimization\cite{bayes}, Open Babel \cite{obabel}, Pybel \cite{pybel}, NumPy \cite{numpy}, OpenMP \cite{openmp}, F2PY \cite{f2py}.


\balance


\bibliography{rsc} 

\providecommand*{\mcitethebibliography}{\thebibliography}
\csname @ifundefined\endcsname{endmcitethebibliography}
{\let\endmcitethebibliography\endthebibliography}{}
\begin{mcitethebibliography}{55}
\providecommand*{\natexlab}[1]{#1}
\providecommand*{\mciteSetBstSublistMode}[1]{}
\providecommand*{\mciteSetBstMaxWidthForm}[2]{}
\providecommand*{\mciteBstWouldAddEndPuncttrue}
  {\def\EndOfBibitem{\unskip.}}
\providecommand*{\mciteBstWouldAddEndPunctfalse}
  {\let\EndOfBibitem\relax}
\providecommand*{\mciteSetBstMidEndSepPunct}[3]{}
\providecommand*{\mciteSetBstSublistLabelBeginEnd}[3]{}
\providecommand*{\EndOfBibitem}{}
\mciteSetBstSublistMode{f}
\mciteSetBstMaxWidthForm{subitem}
{(\emph{\alph{mcitesubitemcount}})}
\mciteSetBstSublistLabelBeginEnd{\mcitemaxwidthsubitemform\space}
{\relax}{\relax}

\bibitem[Pretsch \emph{et~al.}(2013)Pretsch, Clerc, Seibl, and
  Simon]{pretsch2013tables}
E.~Pretsch, T.~Clerc, J.~Seibl and W.~Simon, \emph{Tables of spectral data for
  structure determination of organic compounds}, Springer Science \& Business
  Media, 2013\relax
\mciteBstWouldAddEndPuncttrue
\mciteSetBstMidEndSepPunct{\mcitedefaultmidpunct}
{\mcitedefaultendpunct}{\mcitedefaultseppunct}\relax
\EndOfBibitem
\bibitem[Bremser(1978)]{bremser1978hose}
W.~Bremser, \emph{Anal. Chim. Acta}, 1978, \textbf{103}, 355--365\relax
\mciteBstWouldAddEndPuncttrue
\mciteSetBstMidEndSepPunct{\mcitedefaultmidpunct}
{\mcitedefaultendpunct}{\mcitedefaultseppunct}\relax
\EndOfBibitem
\bibitem[Kuhn and Johnson(2019)]{kuhn2019stereo}
S.~Kuhn and S.~R. Johnson, \emph{ACS Omega}, 2019, \textbf{4}, 7323--7329\relax
\mciteBstWouldAddEndPuncttrue
\mciteSetBstMidEndSepPunct{\mcitedefaultmidpunct}
{\mcitedefaultendpunct}{\mcitedefaultseppunct}\relax
\EndOfBibitem
\bibitem[Aires-de Sousa \emph{et~al.}(2002)Aires-de Sousa, Hemmer, and
  Gasteiger]{aires2002prediction}
J.~Aires-de Sousa, M.~C. Hemmer and J.~Gasteiger, \emph{Anal. Chem.}, 2002,
  \textbf{74}, 80--90\relax
\mciteBstWouldAddEndPuncttrue
\mciteSetBstMidEndSepPunct{\mcitedefaultmidpunct}
{\mcitedefaultendpunct}{\mcitedefaultseppunct}\relax
\EndOfBibitem
\bibitem[Buevich \emph{et~al.}(2019)Buevich, Saur{\'\i}, Parella, De~Tommasi,
  Bifulco, Williamson, and Martin]{buevich2019enhancing}
A.~V. Buevich, J.~Saur{\'\i}, T.~Parella, N.~De~Tommasi, G.~Bifulco, R.~T.
  Williamson and G.~E. Martin, \emph{Chem. Commun.}, 2019, \textbf{55},
  5781--5784\relax
\mciteBstWouldAddEndPuncttrue
\mciteSetBstMidEndSepPunct{\mcitedefaultmidpunct}
{\mcitedefaultendpunct}{\mcitedefaultseppunct}\relax
\EndOfBibitem
\bibitem[Meiler \emph{et~al.}(2002)Meiler, Maier, Will, and
  Meusinger]{meiler2002using}
J.~Meiler, W.~Maier, M.~Will and R.~Meusinger, \emph{J. Mag. Reson.}, 2002,
  \textbf{157}, 242--252\relax
\mciteBstWouldAddEndPuncttrue
\mciteSetBstMidEndSepPunct{\mcitedefaultmidpunct}
{\mcitedefaultendpunct}{\mcitedefaultseppunct}\relax
\EndOfBibitem
\bibitem[Haasnoot \emph{et~al.}(1980)Haasnoot, de~Leeuw, and
  Altona]{haasnoot1980relationship}
C.~Haasnoot, F.~A. de~Leeuw and C.~Altona, \emph{Tetrahedron}, 1980,
  \textbf{36}, 2783--2792\relax
\mciteBstWouldAddEndPuncttrue
\mciteSetBstMidEndSepPunct{\mcitedefaultmidpunct}
{\mcitedefaultendpunct}{\mcitedefaultseppunct}\relax
\EndOfBibitem
\bibitem[Coxon(2009)]{coxon2009developments}
B.~Coxon, \emph{Adv. Carbohydr. Chem. Biochem.}, 2009, \textbf{62},
  17--82\relax
\mciteBstWouldAddEndPuncttrue
\mciteSetBstMidEndSepPunct{\mcitedefaultmidpunct}
{\mcitedefaultendpunct}{\mcitedefaultseppunct}\relax
\EndOfBibitem
\bibitem[Navarro-V{\'a}zquez(2017)]{navarro2017state}
A.~Navarro-V{\'a}zquez, \emph{Magn. Reson. Chem.}, 2017, \textbf{55},
  29--32\relax
\mciteBstWouldAddEndPuncttrue
\mciteSetBstMidEndSepPunct{\mcitedefaultmidpunct}
{\mcitedefaultendpunct}{\mcitedefaultseppunct}\relax
\EndOfBibitem
\bibitem[Lodewyk \emph{et~al.}(2011)Lodewyk, Siebert, and
  Tantillo]{lodewyk2011computational}
M.~W. Lodewyk, M.~R. Siebert and D.~J. Tantillo, \emph{Chem. Rev.}, 2011,
  \textbf{112}, 1839--1862\relax
\mciteBstWouldAddEndPuncttrue
\mciteSetBstMidEndSepPunct{\mcitedefaultmidpunct}
{\mcitedefaultendpunct}{\mcitedefaultseppunct}\relax
\EndOfBibitem
\bibitem[Steinmann \emph{et~al.}(2017)Steinmann, Bratholm, Olsen, and
  Kongsted]{steinmann_cs}
C.~Steinmann, L.~A. Bratholm, J.~M.~H. Olsen and J.~Kongsted, \emph{J. Chem.
  Theory Comput.}, 2017, \textbf{13}, 525--536\relax
\mciteBstWouldAddEndPuncttrue
\mciteSetBstMidEndSepPunct{\mcitedefaultmidpunct}
{\mcitedefaultendpunct}{\mcitedefaultseppunct}\relax
\EndOfBibitem
\bibitem[Larsen \emph{et~al.}(2015)Larsen, Bratholm, Christensen, Channir, and
  Jensen]{larsen_procs}
A.~S. Larsen, L.~A. Bratholm, A.~S. Christensen, M.~Channir and J.~H. Jensen,
  \emph{PeerJ}, 2015, \textbf{3}, e1344\relax
\mciteBstWouldAddEndPuncttrue
\mciteSetBstMidEndSepPunct{\mcitedefaultmidpunct}
{\mcitedefaultendpunct}{\mcitedefaultseppunct}\relax
\EndOfBibitem
\bibitem[Helgaker \emph{et~al.}(2008)Helgaker, Jaszu{\'n}ski, and
  Pecul]{helgaker2008quantum}
T.~Helgaker, M.~Jaszu{\'n}ski and M.~Pecul, \emph{Prog. Nucl. Magn. Reson.
  Spectrosc.}, 2008, \textbf{4}, 249--268\relax
\mciteBstWouldAddEndPuncttrue
\mciteSetBstMidEndSepPunct{\mcitedefaultmidpunct}
{\mcitedefaultendpunct}{\mcitedefaultseppunct}\relax
\EndOfBibitem
\bibitem[Maximoff \emph{et~al.}(2005)Maximoff, Peralta, Barone, and
  Scuseria]{maximoff2005assessment}
S.~N. Maximoff, J.~E. Peralta, V.~Barone and G.~E. Scuseria, \emph{J. Chem.
  Theory Comput.}, 2005, \textbf{1}, 541--545\relax
\mciteBstWouldAddEndPuncttrue
\mciteSetBstMidEndSepPunct{\mcitedefaultmidpunct}
{\mcitedefaultendpunct}{\mcitedefaultseppunct}\relax
\EndOfBibitem
\bibitem[San \emph{et~al.}(2013)San, de~la Vega~Garc{\'\i}a, Suard{\'\i}az,
  Fern{\'a}ndez-Oliva, P{\'e}rez, Crespo-Otero, and
  Contreras]{san2013computational}
J.~F. San, J.~de~la Vega~Garc{\'\i}a, R.~Suard{\'\i}az, M.~Fern{\'a}ndez-Oliva,
  C.~P{\'e}rez, R.~Crespo-Otero and R.~Contreras, \emph{Magn. Reson. Chem.},
  2013, \textbf{51}, 775--787\relax
\mciteBstWouldAddEndPuncttrue
\mciteSetBstMidEndSepPunct{\mcitedefaultmidpunct}
{\mcitedefaultendpunct}{\mcitedefaultseppunct}\relax
\EndOfBibitem
\bibitem[Grimblat \emph{et~al.}(2015)Grimblat, Zanardi, and
  Sarotti]{grimblat2015beyond}
N.~Grimblat, M.~M. Zanardi and A.~M. Sarotti, \emph{J. Org. Chem.}, 2015,
  \textbf{80}, 12526--12534\relax
\mciteBstWouldAddEndPuncttrue
\mciteSetBstMidEndSepPunct{\mcitedefaultmidpunct}
{\mcitedefaultendpunct}{\mcitedefaultseppunct}\relax
\EndOfBibitem
\bibitem[Semenov and Krivdin(2019)]{semenov2019dft}
V.~A. Semenov and L.~B. Krivdin, \emph{Magn. Reson. Chem.}, 2019\relax
\mciteBstWouldAddEndPuncttrue
\mciteSetBstMidEndSepPunct{\mcitedefaultmidpunct}
{\mcitedefaultendpunct}{\mcitedefaultseppunct}\relax
\EndOfBibitem
\bibitem[ACD()]{ACDpredictor}
\emph{NMR Prediction Software from ACD/Labs},
  \url{https://www.acdlabs.com/products/adh/nmr/nmr_pred/}\relax
\mciteBstWouldAddEndPuncttrue
\mciteSetBstMidEndSepPunct{\mcitedefaultmidpunct}
{\mcitedefaultendpunct}{\mcitedefaultseppunct}\relax
\EndOfBibitem
\bibitem[mes()]{mestrepredictor}
\emph{NMR Prediction Software from Mestrelab},
  \url{https://mestrelab.com/software/mnova/nmr-predict/}\relax
\mciteBstWouldAddEndPuncttrue
\mciteSetBstMidEndSepPunct{\mcitedefaultmidpunct}
{\mcitedefaultendpunct}{\mcitedefaultseppunct}\relax
\EndOfBibitem
\bibitem[Castillo \emph{et~al.}(2016)Castillo, Bernal, Dieden, Patiny, and
  Wist]{castillo2016ask}
A.~M. Castillo, A.~Bernal, R.~Dieden, L.~Patiny and J.~Wist, \emph{J.
  Cheminf.}, 2016, \textbf{8}, 26\relax
\mciteBstWouldAddEndPuncttrue
\mciteSetBstMidEndSepPunct{\mcitedefaultmidpunct}
{\mcitedefaultendpunct}{\mcitedefaultseppunct}\relax
\EndOfBibitem
\bibitem[Brandolini(2006)]{brandolini2006nmrpredict}
A.~J. Brandolini, \emph{NMRPredict Modgraph Consultants, Ltd, 1348 Graham
  Place, Escondido, CA 92129. http://www.modgraph-usa.com}, 2006\relax
\mciteBstWouldAddEndPuncttrue
\mciteSetBstMidEndSepPunct{\mcitedefaultmidpunct}
{\mcitedefaultendpunct}{\mcitedefaultseppunct}\relax
\EndOfBibitem
\bibitem[Paruzzo \emph{et~al.}(2018)Paruzzo, Hofstetter, Musil, De, Ceriotti,
  and Emsley]{paruzzo2018chemical}
F.~M. Paruzzo, A.~Hofstetter, F.~Musil, S.~De, M.~Ceriotti and L.~Emsley,
  \emph{Nat. Commun.}, 2018, \textbf{9}, 4501\relax
\mciteBstWouldAddEndPuncttrue
\mciteSetBstMidEndSepPunct{\mcitedefaultmidpunct}
{\mcitedefaultendpunct}{\mcitedefaultseppunct}\relax
\EndOfBibitem
\bibitem[Venkata \emph{et~al.}(2014)Venkata, Forster, Howe, and
  Steinbeck]{venkata2014potential}
C.~Venkata, M.~J. Forster, P.~W. Howe and C.~Steinbeck, \emph{PLOS ONE}, 2014,
  \textbf{9}, e111576\relax
\mciteBstWouldAddEndPuncttrue
\mciteSetBstMidEndSepPunct{\mcitedefaultmidpunct}
{\mcitedefaultendpunct}{\mcitedefaultseppunct}\relax
\EndOfBibitem
\bibitem[Frisch \emph{et~al.}(2016)Frisch, Trucks, Schlegel, Scuseria, Robb,
  Cheeseman, Scalmani, Barone, Mennucci, Petersson, and others (for the~full
  reference]{frisch2009gaussian}
M.~Frisch, G.~Trucks, H.~Schlegel, G.~Scuseria, M.~Robb, J.~Cheeseman,
  G.~Scalmani, V.~Barone, B.~Mennucci, G.~Petersson and s.~S. S. o. S.~I.
  others (for the~full reference, \emph{Wallingford, CT}, 2016\relax
\mciteBstWouldAddEndPuncttrue
\mciteSetBstMidEndSepPunct{\mcitedefaultmidpunct}
{\mcitedefaultendpunct}{\mcitedefaultseppunct}\relax
\EndOfBibitem
\bibitem[Teale \emph{et~al.}(2013)Teale, Lutn{\ae}s, Helgaker, Tozer, and
  Gauss]{teale2013benchmarking}
A.~M. Teale, O.~B. Lutn{\ae}s, T.~Helgaker, D.~J. Tozer and J.~Gauss, \emph{J.
  Chem. Phys.}, 2013, \textbf{138}, 024111\relax
\mciteBstWouldAddEndPuncttrue
\mciteSetBstMidEndSepPunct{\mcitedefaultmidpunct}
{\mcitedefaultendpunct}{\mcitedefaultseppunct}\relax
\EndOfBibitem
\bibitem[Adamo and Barone(1998)]{adamo1998exchange}
C.~Adamo and V.~Barone, \emph{J. Chem. Phys.}, 1998, \textbf{108},
  664--675\relax
\mciteBstWouldAddEndPuncttrue
\mciteSetBstMidEndSepPunct{\mcitedefaultmidpunct}
{\mcitedefaultendpunct}{\mcitedefaultseppunct}\relax
\EndOfBibitem
\bibitem[McLean and Chandler(1980)]{mclean1980contracted}
A.~McLean and G.~Chandler, \emph{J. Chem. Phys.}, 1980, \textbf{72},
  5639--5648\relax
\mciteBstWouldAddEndPuncttrue
\mciteSetBstMidEndSepPunct{\mcitedefaultmidpunct}
{\mcitedefaultendpunct}{\mcitedefaultseppunct}\relax
\EndOfBibitem
\bibitem[Krishnan \emph{et~al.}(1980)Krishnan, Binkley, Seeger, and
  Pople]{krishnan1980self-consistent}
R.~Krishnan, J.~S. Binkley, R.~Seeger and J.~A. Pople, \emph{J. Chem. Phys.},
  1980, \textbf{72}, 650--654\relax
\mciteBstWouldAddEndPuncttrue
\mciteSetBstMidEndSepPunct{\mcitedefaultmidpunct}
{\mcitedefaultendpunct}{\mcitedefaultseppunct}\relax
\EndOfBibitem
\bibitem[Chai and Head-Gordon(2008)]{chai2008systematic}
J.-D. Chai and M.~Head-Gordon, \emph{J. Chem. Phys.}, 2008, \textbf{128},
  084106\relax
\mciteBstWouldAddEndPuncttrue
\mciteSetBstMidEndSepPunct{\mcitedefaultmidpunct}
{\mcitedefaultendpunct}{\mcitedefaultseppunct}\relax
\EndOfBibitem
\bibitem[Deng \emph{et~al.}(2006)Deng, Cheeseman, and
  Frisch]{deng2006calculation}
W.~Deng, J.~R. Cheeseman and M.~J. Frisch, \emph{J. Chem. Theory Comput.},
  2006, \textbf{2}, 1028--1037\relax
\mciteBstWouldAddEndPuncttrue
\mciteSetBstMidEndSepPunct{\mcitedefaultmidpunct}
{\mcitedefaultendpunct}{\mcitedefaultseppunct}\relax
\EndOfBibitem
\bibitem[Wilson \emph{et~al.}(2019)Wilson, Grootveld, and
  Kamerlin]{wilson2019errors}
P.~B. Wilson, M.~Grootveld and S.~C.~L. Kamerlin, \emph{Magn. Reson. Chem.},
  2019\relax
\mciteBstWouldAddEndPuncttrue
\mciteSetBstMidEndSepPunct{\mcitedefaultmidpunct}
{\mcitedefaultendpunct}{\mcitedefaultseppunct}\relax
\EndOfBibitem
\bibitem[Laskowski \emph{et~al.}(2019 (accessed October 2nd, 2019))Laskowski,
  Blaha, and Tran]{CHESHIRE}
R.~Laskowski, P.~Blaha and F.~Tran, \emph{CHESHIRE Chemical Shift Repository},
  2019 (accessed October 2nd, 2019)\relax
\mciteBstWouldAddEndPuncttrue
\mciteSetBstMidEndSepPunct{\mcitedefaultmidpunct}
{\mcitedefaultendpunct}{\mcitedefaultseppunct}\relax
\EndOfBibitem
\bibitem[Seung \emph{et~al.}(1992)Seung, Opper, and Sompolinsky]{qbc}
H.~S. Seung, M.~Opper and H.~Sompolinsky, Proc. 5th Ann. Work. Comp. Learn.
  Theory, New York, NY, USA, 1992, pp. 287--294\relax
\mciteBstWouldAddEndPuncttrue
\mciteSetBstMidEndSepPunct{\mcitedefaultmidpunct}
{\mcitedefaultendpunct}{\mcitedefaultseppunct}\relax
\EndOfBibitem
\bibitem[Gastegger \emph{et~al.}(2017)Gastegger, Behler, and
  Marquetand]{behler_active_learning}
M.~Gastegger, J.~Behler and P.~Marquetand, \emph{Chem. Sci.}, 2017, \textbf{8},
  6924--6935\relax
\mciteBstWouldAddEndPuncttrue
\mciteSetBstMidEndSepPunct{\mcitedefaultmidpunct}
{\mcitedefaultendpunct}{\mcitedefaultseppunct}\relax
\EndOfBibitem
\bibitem[Smith \emph{et~al.}(2018)Smith, Nebgen, Lubbers, Isayev, and
  Roitberg]{roitberg_active_learning}
J.~S. Smith, B.~Nebgen, N.~Lubbers, O.~Isayev and A.~E. Roitberg, \emph{J.
  Chem. Phys.}, 2018, \textbf{148}, 241733\relax
\mciteBstWouldAddEndPuncttrue
\mciteSetBstMidEndSepPunct{\mcitedefaultmidpunct}
{\mcitedefaultendpunct}{\mcitedefaultseppunct}\relax
\EndOfBibitem
\bibitem[Groom \emph{et~al.}(2016)Groom, Bruno, Lightfoot, and Ward]{csd}
C.~R. Groom, I.~J. Bruno, M.~P. Lightfoot and S.~C. Ward, \emph{Acta
  Crystallogr. B}, 2016, \textbf{72}, 171--179\relax
\mciteBstWouldAddEndPuncttrue
\mciteSetBstMidEndSepPunct{\mcitedefaultmidpunct}
{\mcitedefaultendpunct}{\mcitedefaultseppunct}\relax
\EndOfBibitem
\bibitem[Saunders \emph{et~al.}(1998)Saunders, Gammerman, and
  Vovk]{saunders1998ridge}
C.~Saunders, A.~Gammerman and V.~Vovk, 1998\relax
\mciteBstWouldAddEndPuncttrue
\mciteSetBstMidEndSepPunct{\mcitedefaultmidpunct}
{\mcitedefaultendpunct}{\mcitedefaultseppunct}\relax
\EndOfBibitem
\bibitem[Binev and Aires-de Sousa(2004)]{nn1}
Y.~Binev and J.~Aires-de Sousa, \emph{J. Chem. Inf. Comput. Sci.}, 2004,
  \textbf{44}, 940--945\relax
\mciteBstWouldAddEndPuncttrue
\mciteSetBstMidEndSepPunct{\mcitedefaultmidpunct}
{\mcitedefaultendpunct}{\mcitedefaultseppunct}\relax
\EndOfBibitem
\bibitem[Binev \emph{et~al.}(2007)Binev, Marques, and Aires-de Sousa]{nn2}
Y.~Binev, M.~M. Marques and J.~Aires-de Sousa, \emph{J. Chem. Inf. Model.},
  2007, \textbf{47}, 2089--2097\relax
\mciteBstWouldAddEndPuncttrue
\mciteSetBstMidEndSepPunct{\mcitedefaultmidpunct}
{\mcitedefaultendpunct}{\mcitedefaultseppunct}\relax
\EndOfBibitem
\bibitem[Rupp \emph{et~al.}(2015)Rupp, Ramakrishnan, and Von~Lilienfeld]{aCM}
M.~Rupp, R.~Ramakrishnan and O.~A. Von~Lilienfeld, \emph{J. Phys. Chem. Lett.},
  2015, \textbf{6}, 3309--3313\relax
\mciteBstWouldAddEndPuncttrue
\mciteSetBstMidEndSepPunct{\mcitedefaultmidpunct}
{\mcitedefaultendpunct}{\mcitedefaultseppunct}\relax
\EndOfBibitem
\bibitem[Huang and von Lilienfeld(2017)]{huang2017dna}
B.~Huang and O.~A. von Lilienfeld, \emph{arXiv preprint arXiv:1707.04146},
  2017\relax
\mciteBstWouldAddEndPuncttrue
\mciteSetBstMidEndSepPunct{\mcitedefaultmidpunct}
{\mcitedefaultendpunct}{\mcitedefaultseppunct}\relax
\EndOfBibitem
\bibitem[Faber \emph{et~al.}(2018)Faber, Christensen, Huang, and von
  Lilienfeld]{faber2018alchemical}
F.~A. Faber, A.~S. Christensen, B.~Huang and O.~A. von Lilienfeld, \emph{J.
  Chem. Phys.}, 2018, \textbf{148}, 241717\relax
\mciteBstWouldAddEndPuncttrue
\mciteSetBstMidEndSepPunct{\mcitedefaultmidpunct}
{\mcitedefaultendpunct}{\mcitedefaultseppunct}\relax
\EndOfBibitem
\bibitem[Christensen \emph{et~al.}(2019)Christensen, Bratholm, Amabilino,
  Kromann, Faber, Huang, Tkatchenko, Müller, and von Lilienfeld]{qml}
A.~S. Christensen, L.~A. Bratholm, S.~Amabilino, J.~C. Kromann, F.~A. Faber,
  B.~Huang, A.~Tkatchenko, K.~R. Müller and O.~A. von Lilienfeld, \emph{QML: A
  Python Toolkit for Quantum Machine Learning}, 2019,
  \url{https://github.com/qmlcode/qml}\relax
\mciteBstWouldAddEndPuncttrue
\mciteSetBstMidEndSepPunct{\mcitedefaultmidpunct}
{\mcitedefaultendpunct}{\mcitedefaultseppunct}\relax
\EndOfBibitem
\bibitem[Smith and Goodman(2010)]{smith2010assigning}
S.~G. Smith and J.~M. Goodman, \emph{J. Am. Chem. Soc.}, 2010, \textbf{132},
  12946--12959\relax
\mciteBstWouldAddEndPuncttrue
\mciteSetBstMidEndSepPunct{\mcitedefaultmidpunct}
{\mcitedefaultendpunct}{\mcitedefaultseppunct}\relax
\EndOfBibitem
\bibitem[Butts \emph{et~al.}(2011)Butts, Jones, and Harvey]{butts2011high}
C.~P. Butts, C.~R. Jones and J.~N. Harvey, \emph{Chem. Commun.}, 2011,
  \textbf{47}, 1193--1195\relax
\mciteBstWouldAddEndPuncttrue
\mciteSetBstMidEndSepPunct{\mcitedefaultmidpunct}
{\mcitedefaultendpunct}{\mcitedefaultseppunct}\relax
\EndOfBibitem
\bibitem[Kutateladze \emph{et~al.}(2018)Kutateladze, Kuznetsov, Beloglazkina,
  and Holt]{kutateladze2018addressing}
A.~G. Kutateladze, D.~M. Kuznetsov, A.~A. Beloglazkina and T.~Holt, \emph{J.
  Org. Chem.}, 2018, \textbf{83}, 8341--8352\relax
\mciteBstWouldAddEndPuncttrue
\mciteSetBstMidEndSepPunct{\mcitedefaultmidpunct}
{\mcitedefaultendpunct}{\mcitedefaultseppunct}\relax
\EndOfBibitem
\bibitem[Iron(2017)]{iron2017evaluation}
M.~A. Iron, \emph{J. Chem. Theory Comp.}, 2017, \textbf{13}, 5798--5819\relax
\mciteBstWouldAddEndPuncttrue
\mciteSetBstMidEndSepPunct{\mcitedefaultmidpunct}
{\mcitedefaultendpunct}{\mcitedefaultseppunct}\relax
\EndOfBibitem
\bibitem[Bagno \emph{et~al.}(2006)Bagno, Rastrelli, and
  Saielli]{bagno2006toward}
A.~Bagno, F.~Rastrelli and G.~Saielli, \emph{Chem.: Eur. J.}, 2006,
  \textbf{12}, 5514--5525\relax
\mciteBstWouldAddEndPuncttrue
\mciteSetBstMidEndSepPunct{\mcitedefaultmidpunct}
{\mcitedefaultendpunct}{\mcitedefaultseppunct}\relax
\EndOfBibitem
\bibitem[Laskowski \emph{et~al.}(2013)Laskowski, Blaha, and
  Tran]{laskowski2013assessment}
R.~Laskowski, P.~Blaha and F.~Tran, \emph{Physical Review B}, 2013,
  \textbf{87}, 195130\relax
\mciteBstWouldAddEndPuncttrue
\mciteSetBstMidEndSepPunct{\mcitedefaultmidpunct}
{\mcitedefaultendpunct}{\mcitedefaultseppunct}\relax
\EndOfBibitem
\bibitem[Nogueira(2019)]{bayes}
F.~Nogueira, \emph{A Python implementation of global optimization with gaussian
  processes}, 2019, \url{https://github.com/fmfn/BayesianOptimization}\relax
\mciteBstWouldAddEndPuncttrue
\mciteSetBstMidEndSepPunct{\mcitedefaultmidpunct}
{\mcitedefaultendpunct}{\mcitedefaultseppunct}\relax
\EndOfBibitem
\bibitem[O'Boyle \emph{et~al.}(2011)O'Boyle, Banck, James, Morley,
  Vandermeersch, and Hutchison]{obabel}
N.~M. O'Boyle, M.~Banck, C.~A. James, C.~Morley, T.~Vandermeersch and G.~R.
  Hutchison, \emph{J. Cheminf.}, 2011, \textbf{3}, 33\relax
\mciteBstWouldAddEndPuncttrue
\mciteSetBstMidEndSepPunct{\mcitedefaultmidpunct}
{\mcitedefaultendpunct}{\mcitedefaultseppunct}\relax
\EndOfBibitem
\bibitem[O'Boyle \emph{et~al.}(2008)O'Boyle, Morley, and Hutchison]{pybel}
N.~M. O'Boyle, C.~Morley and G.~R. Hutchison, \emph{Chem. Cent. J.}, 2008,
  \textbf{2}, 5\relax
\mciteBstWouldAddEndPuncttrue
\mciteSetBstMidEndSepPunct{\mcitedefaultmidpunct}
{\mcitedefaultendpunct}{\mcitedefaultseppunct}\relax
\EndOfBibitem
\bibitem[Oliphant(2006)]{numpy}
T.~E. Oliphant, \emph{A guide to NumPy}, Trelgol Publishing USA, 2006,
  vol.~1\relax
\mciteBstWouldAddEndPuncttrue
\mciteSetBstMidEndSepPunct{\mcitedefaultmidpunct}
{\mcitedefaultendpunct}{\mcitedefaultseppunct}\relax
\EndOfBibitem
\bibitem[Dagum and Menon(1998)]{openmp}
L.~Dagum and R.~Menon, \emph{Comput. Sci. Eng.}, 1998,  46--55\relax
\mciteBstWouldAddEndPuncttrue
\mciteSetBstMidEndSepPunct{\mcitedefaultmidpunct}
{\mcitedefaultendpunct}{\mcitedefaultseppunct}\relax
\EndOfBibitem
\bibitem[Peterson(2009)]{f2py}
P.~Peterson, \emph{Int. J. Comput. Sci. Eng.}, 2009, \textbf{4}, 296--305\relax
\mciteBstWouldAddEndPuncttrue
\mciteSetBstMidEndSepPunct{\mcitedefaultmidpunct}
{\mcitedefaultendpunct}{\mcitedefaultseppunct}\relax
\EndOfBibitem
\end{mcitethebibliography}


\providecommand*{\mcitethebibliography}{\thebibliography}
\csname @ifundefined\endcsname{endmcitethebibliography}
{\let\endmcitethebibliography\endthebibliography}{}
\begin{mcitethebibliography}{22}
\providecommand*{\natexlab}[1]{#1}
\providecommand*{\mciteSetBstSublistMode}[1]{}
\providecommand*{\mciteSetBstMaxWidthForm}[2]{}
\providecommand*{\mciteBstWouldAddEndPuncttrue}
  {\def\EndOfBibitem{\unskip.}}
\providecommand*{\mciteBstWouldAddEndPunctfalse}
  {\let\EndOfBibitem\relax}
\providecommand*{\mciteSetBstMidEndSepPunct}[3]{}
\providecommand*{\mciteSetBstSublistLabelBeginEnd}[3]{}
\providecommand*{\EndOfBibitem}{}
\mciteSetBstSublistMode{f}
\mciteSetBstMaxWidthForm{subitem}
{(\emph{\alph{mcitesubitemcount}})}
\mciteSetBstSublistLabelBeginEnd{\mcitemaxwidthsubitemform\space}
{\relax}{\relax}

\bibitem[Saunders \emph{et~al.}(1998)Saunders, Gammerman, and
  Vovk]{saunders1998ridge}
C.~Saunders, A.~Gammerman and V.~Vovk, 1998\relax
\mciteBstWouldAddEndPuncttrue
\mciteSetBstMidEndSepPunct{\mcitedefaultmidpunct}
{\mcitedefaultendpunct}{\mcitedefaultseppunct}\relax
\EndOfBibitem
\bibitem[Rupp \emph{et~al.}(2015)Rupp, Ramakrishnan, and Von~Lilienfeld]{aCM}
M.~Rupp, R.~Ramakrishnan and O.~A. Von~Lilienfeld, \emph{J. Phys. Chem. Lett.},
  2015, \textbf{6}, 3309--3313\relax
\mciteBstWouldAddEndPuncttrue
\mciteSetBstMidEndSepPunct{\mcitedefaultmidpunct}
{\mcitedefaultendpunct}{\mcitedefaultseppunct}\relax
\EndOfBibitem
\bibitem[Huang and von Lilienfeld(2017)]{huang2017dna}
B.~Huang and O.~A. von Lilienfeld, \emph{arXiv preprint arXiv:1707.04146},
  2017\relax
\mciteBstWouldAddEndPuncttrue
\mciteSetBstMidEndSepPunct{\mcitedefaultmidpunct}
{\mcitedefaultendpunct}{\mcitedefaultseppunct}\relax
\EndOfBibitem
\bibitem[Faber \emph{et~al.}(2018)Faber, Christensen, Huang, and von
  Lilienfeld]{faber2018alchemical}
F.~A. Faber, A.~S. Christensen, B.~Huang and O.~A. von Lilienfeld, \emph{J.
  Chem. Phys.}, 2018, \textbf{148}, 241717\relax
\mciteBstWouldAddEndPuncttrue
\mciteSetBstMidEndSepPunct{\mcitedefaultmidpunct}
{\mcitedefaultendpunct}{\mcitedefaultseppunct}\relax
\EndOfBibitem
\bibitem[Christensen \emph{et~al.}(2019)Christensen, Bratholm, Amabilino,
  Kromann, Faber, Huang, Tkatchenko, Müller, and von Lilienfeld]{qml}
A.~S. Christensen, L.~A. Bratholm, S.~Amabilino, J.~C. Kromann, F.~A. Faber,
  B.~Huang, A.~Tkatchenko, K.~R. Müller and O.~A. von Lilienfeld, \emph{QML: A
  Python Toolkit for Quantum Machine Learning}, 2019,
  \url{https://github.com/qmlcode/qml}\relax
\mciteBstWouldAddEndPuncttrue
\mciteSetBstMidEndSepPunct{\mcitedefaultmidpunct}
{\mcitedefaultendpunct}{\mcitedefaultseppunct}\relax
\EndOfBibitem
\bibitem[Seung \emph{et~al.}(1992)Seung, Opper, and Sompolinsky]{qbc}
H.~S. Seung, M.~Opper and H.~Sompolinsky, Proc. 5th Ann. Work. Comp. Learn.
  Theory, New York, NY, USA, 1992, pp. 287--294\relax
\mciteBstWouldAddEndPuncttrue
\mciteSetBstMidEndSepPunct{\mcitedefaultmidpunct}
{\mcitedefaultendpunct}{\mcitedefaultseppunct}\relax
\EndOfBibitem
\bibitem[Gastegger \emph{et~al.}(2017)Gastegger, Behler, and
  Marquetand]{behler_active_learning}
M.~Gastegger, J.~Behler and P.~Marquetand, \emph{Chem. Sci.}, 2017, \textbf{8},
  6924--6935\relax
\mciteBstWouldAddEndPuncttrue
\mciteSetBstMidEndSepPunct{\mcitedefaultmidpunct}
{\mcitedefaultendpunct}{\mcitedefaultseppunct}\relax
\EndOfBibitem
\bibitem[Smith \emph{et~al.}(2018)Smith, Nebgen, Lubbers, Isayev, and
  Roitberg]{roitberg_active_learning}
J.~S. Smith, B.~Nebgen, N.~Lubbers, O.~Isayev and A.~E. Roitberg, \emph{J.
  Chem. Phys.}, 2018, \textbf{148}, 241733\relax
\mciteBstWouldAddEndPuncttrue
\mciteSetBstMidEndSepPunct{\mcitedefaultmidpunct}
{\mcitedefaultendpunct}{\mcitedefaultseppunct}\relax
\EndOfBibitem
\bibitem[Paruzzo \emph{et~al.}(2018)Paruzzo, Hofstetter, Musil, De, Ceriotti,
  and Emsley]{paruzzo2018chemical}
F.~M. Paruzzo, A.~Hofstetter, F.~Musil, S.~De, M.~Ceriotti and L.~Emsley,
  \emph{Nat. Commun.}, 2018, \textbf{9}, 4501\relax
\mciteBstWouldAddEndPuncttrue
\mciteSetBstMidEndSepPunct{\mcitedefaultmidpunct}
{\mcitedefaultendpunct}{\mcitedefaultseppunct}\relax
\EndOfBibitem
\bibitem[Frisch \emph{et~al.}(2016)Frisch, Trucks, Schlegel, Scuseria, Robb,
  Cheeseman, Scalmani, Barone, Mennucci, Petersson, and others (for the~full
  reference]{frisch2009gaussian}
M.~Frisch, G.~Trucks, H.~Schlegel, G.~Scuseria, M.~Robb, J.~Cheeseman,
  G.~Scalmani, V.~Barone, B.~Mennucci, G.~Petersson and s.~S. S. o. S.~I.
  others (for the~full reference, \emph{Wallingford, CT}, 2016\relax
\mciteBstWouldAddEndPuncttrue
\mciteSetBstMidEndSepPunct{\mcitedefaultmidpunct}
{\mcitedefaultendpunct}{\mcitedefaultseppunct}\relax
\EndOfBibitem
\bibitem[Adamo and Barone(1998)]{adamo1998exchange}
C.~Adamo and V.~Barone, \emph{J. Chem. Phys.}, 1998, \textbf{108},
  664--675\relax
\mciteBstWouldAddEndPuncttrue
\mciteSetBstMidEndSepPunct{\mcitedefaultmidpunct}
{\mcitedefaultendpunct}{\mcitedefaultseppunct}\relax
\EndOfBibitem
\bibitem[McLean and Chandler(1980)]{mclean1980contracted}
A.~McLean and G.~Chandler, \emph{J. Chem. Phys.}, 1980, \textbf{72},
  5639--5648\relax
\mciteBstWouldAddEndPuncttrue
\mciteSetBstMidEndSepPunct{\mcitedefaultmidpunct}
{\mcitedefaultendpunct}{\mcitedefaultseppunct}\relax
\EndOfBibitem
\bibitem[Krishnan \emph{et~al.}(1980)Krishnan, Binkley, Seeger, and
  Pople]{krishnan1980self-consistent}
R.~Krishnan, J.~S. Binkley, R.~Seeger and J.~A. Pople, \emph{J. Chem. Phys.},
  1980, \textbf{72}, 650--654\relax
\mciteBstWouldAddEndPuncttrue
\mciteSetBstMidEndSepPunct{\mcitedefaultmidpunct}
{\mcitedefaultendpunct}{\mcitedefaultseppunct}\relax
\EndOfBibitem
\bibitem[Wilson \emph{et~al.}(2019)Wilson, Grootveld, and
  Kamerlin]{wilson2019errors}
P.~B. Wilson, M.~Grootveld and S.~C.~L. Kamerlin, \emph{Magn. Reson. Chem.},
  2019\relax
\mciteBstWouldAddEndPuncttrue
\mciteSetBstMidEndSepPunct{\mcitedefaultmidpunct}
{\mcitedefaultendpunct}{\mcitedefaultseppunct}\relax
\EndOfBibitem
\bibitem[Chai and Head-Gordon(2008)]{chai2008systematic}
J.-D. Chai and M.~Head-Gordon, \emph{J. Chem. Phys.}, 2008, \textbf{128},
  084106\relax
\mciteBstWouldAddEndPuncttrue
\mciteSetBstMidEndSepPunct{\mcitedefaultmidpunct}
{\mcitedefaultendpunct}{\mcitedefaultseppunct}\relax
\EndOfBibitem
\bibitem[Deng \emph{et~al.}(2006)Deng, Cheeseman, and
  Frisch]{deng2006calculation}
W.~Deng, J.~R. Cheeseman and M.~J. Frisch, \emph{J. Chem. Theory Comput.},
  2006, \textbf{2}, 1028--1037\relax
\mciteBstWouldAddEndPuncttrue
\mciteSetBstMidEndSepPunct{\mcitedefaultmidpunct}
{\mcitedefaultendpunct}{\mcitedefaultseppunct}\relax
\EndOfBibitem
\bibitem[Lodewyk \emph{et~al.}(2011)Lodewyk, Siebert, and
  Tantillo]{lodewyk2011computational}
M.~W. Lodewyk, M.~R. Siebert and D.~J. Tantillo, \emph{Chem. Rev.}, 2011,
  \textbf{112}, 1839--1862\relax
\mciteBstWouldAddEndPuncttrue
\mciteSetBstMidEndSepPunct{\mcitedefaultmidpunct}
{\mcitedefaultendpunct}{\mcitedefaultseppunct}\relax
\EndOfBibitem
\bibitem[Laskowski \emph{et~al.}(2019 (accessed October 2nd, 2019))Laskowski,
  Blaha, and Tran]{CHESHIRE}
R.~Laskowski, P.~Blaha and F.~Tran, \emph{CHESHIRE Chemical Shift Repository},
  2019 (accessed October 2nd, 2019)\relax
\mciteBstWouldAddEndPuncttrue
\mciteSetBstMidEndSepPunct{\mcitedefaultmidpunct}
{\mcitedefaultendpunct}{\mcitedefaultseppunct}\relax
\EndOfBibitem
\bibitem[Nogueira(2019)]{bayes}
F.~Nogueira, \emph{A Python implementation of global optimization with gaussian
  processes}, 2019, \url{https://github.com/fmfn/BayesianOptimization}\relax
\mciteBstWouldAddEndPuncttrue
\mciteSetBstMidEndSepPunct{\mcitedefaultmidpunct}
{\mcitedefaultendpunct}{\mcitedefaultseppunct}\relax
\EndOfBibitem
\bibitem[Halgren(1996)]{halgren1996merck}
T.~A. Halgren, \emph{J. Comput. Chem.}, 1996, \textbf{17}, 490--519\relax
\mciteBstWouldAddEndPuncttrue
\mciteSetBstMidEndSepPunct{\mcitedefaultmidpunct}
{\mcitedefaultendpunct}{\mcitedefaultseppunct}\relax
\EndOfBibitem
\bibitem[Kutateladze \emph{et~al.}(2018)Kutateladze, Kuznetsov, Beloglazkina,
  and Holt]{kutateladze2018addressing}
A.~G. Kutateladze, D.~M. Kuznetsov, A.~A. Beloglazkina and T.~Holt, \emph{J.
  Org. Chem.}, 2018, \textbf{83}, 8341--8352\relax
\mciteBstWouldAddEndPuncttrue
\mciteSetBstMidEndSepPunct{\mcitedefaultmidpunct}
{\mcitedefaultendpunct}{\mcitedefaultseppunct}\relax
\EndOfBibitem
\bibitem[Buevich \emph{et~al.}(2019)Buevich, Saur{\'\i}, Parella, De~Tommasi,
  Bifulco, Williamson, and Martin]{buevich2019enhancing}
A.~V. Buevich, J.~Saur{\'\i}, T.~Parella, N.~De~Tommasi, G.~Bifulco, R.~T.
  Williamson and G.~E. Martin, \emph{Chem. Commun.}, 2019, \textbf{55},
  5781--5784\relax
\mciteBstWouldAddEndPuncttrue
\mciteSetBstMidEndSepPunct{\mcitedefaultmidpunct}
{\mcitedefaultendpunct}{\mcitedefaultseppunct}\relax
\EndOfBibitem
\end{mcitethebibliography}
\bibliographystyle{rsc} 

\end{document}


\maketitle
 
\tableofcontents
 
\newpage
\section{Methods}
\label{sec:SIMethods}
\subsection{Kernel ridge regression}
\label{sec:krr}
Kernel Ridge Regression\cite{saunders1998ridge} (KRR) provides a systematic way to map geometric features of a chemical environment (i.e. the chemical identity and  geometry of atoms in the environment surrounding atoms of interest) to a target observable (in this case scalar coupling constants or chemical shifts), effectively interpolating between known data points.
The observable of interest ($y_i$) for a given environment (${\mathbf E}_i$) is estimated as a linear combination of it's similarity to the environments (${\mathbf E}_j$), for which the corresponding observable is known:

\begin{equation} \label{eq:krr}
    y_i^\mathrm{pred} = \sum_j^N \alpha_j k\left({\mathbf E}_i, {\mathbf E}_j\right),
\end{equation}
Here N is the number of chemical environments in the training data set and $k$ is a kernel function that computes the similarity between two environments. 
The kernel function typically takes a value of 1 for identical environments and approaches asymptotically 0 when environments become increasingly different.
The regression parameters $\bm \alpha$ are regression coefficients that can be fitted to the training data by regularized least-squares optimization:
\begin{equation}
    \displaystyle{\minimise_{\bm \alpha} \;\sum_i^N \left(y_i^\mathrm{exp} - y_i^\mathrm{pred}\right)^2 + \lambda \sum_i^N \alpha_i^2},
\end{equation}
where $y_i^\mathrm{pred}$ is given by equation (\ref{eq:krr}). $\lambda$ controls the strength of the $l_{\,2}$-regularization, which is a penalty term to the loss function that favours the regression coefficients to be more uniform and to take smaller values.
This effectively reduces overfitting and if properly tuned can improve transferability to new chemical environments.\\ \\
Several functional forms of the kernel similarity measure has been proposed in recent years.
In this work we compare three different kernel functions.
The atomic Coulomb Matrix\cite{aCM} was one of the early successful vector representations of the chemical environment around an atom and includes two-body interactions (distances) between a given atom and all atoms within a specified cutoff radius.
The Atomic Spectral London Axilrod-Teller-Muto\cite{huang2017dna} (aSLATM) representation is a separate approach that also includes three-body interactions (angles).
Both representations generate a vector (${\bm x}$) per environment, where the kernel similarity can computed with a laplacian kernel:
\begin{equation}
    k\left({\mathbf E}_i, {\mathbf E}_j\right) = \exp{\left(-\frac{\|{\bm x}_i - {\bm x}_j\|_1}{\sigma}\right)},
\end{equation}
where the kernel width $\sigma$ determines how quickly the similarity measure converges towards 0.\\
FCHL \cite{faber2018alchemical} (acronym derived from the authors surnames) also includes three-body terms, but generates the kernel similarity directly, rather than through an intermediate vector representation step.\\ \\
Since the above kernel similarity measure indicates how similar the chemical environment around two {\it atoms} are, we chose to use the product of the kernel similarity between the two hydrogens and the two carbons to represent ${}^1\mathrm{J_{CH}}$ environments:
\begin{equation}
k\left({\mathbf E}_i^\mathrm{CH},{\mathbf E}_j^\mathrm{CH}\right) = k\left({\mathbf E}_i^\mathrm{H},{\mathbf E}_j^\mathrm{H}\right) k\left({\mathbf E}_i^\mathrm{C},{\mathbf E}_j^\mathrm{C}\right),
\end{equation}
where ${\mathbf E}_i^\mathrm{CH}$ is the joint set of chemical environments around the hydrogen and carbon atom involved in the scalar coupling. Alternatively just the kernel similarity between hydrogen atoms could be used, but we found an improvement in performance by also including the carbon similarity. \\
All representations and kernels as well as optimisation of the regression parameters were performed with the QML python library \cite{qml}.

\subsection{Training and test data}
\label{sec:si_train_test_data}
The KRR machine was trained using 
17,222 
coupling environments from 
882 
chemical structures selected by adaptive sampling (active learning)\cite{qbc,behler_active_learning,roitberg_active_learning} from the Cambridge structural database (filtering first for structures that contain only C, H, N, O and F elements, see section \ref{sec:active_learning} for details) then optimising the structures and calculating the DFT NMR parameters (see next paragraph for details). The test set contained an independent set of
7832 
environments from 
410 
chemical structures from the randomly selected CSD-500 test set reported by Emsley et al \cite{paruzzo2018chemical}.  
All DFT calculations were carried out using the Gaussian09 Rev. D software package \cite{frisch2009gaussian} (See section S7 for example input files). 
The 3-dimensional chemical structures were each optimised with $mPW1PW91$\cite{adamo1998exchange}/6-311g(d,p)\cite{mclean1980contracted, krishnan1980self-consistent} using tight optimisation criteria and ultrafine integral grids were used to minimise molecular orientation affecting geometries and energies (see reference \cite{wilson2019errors} and references therein for a discussion of this) and the resulting optimised structures were used to compute NMR parameters with $\omega b97xd$\cite{chai2008systematic}/6-311g(d,p). 
The NMR computations used gauge independent atomic orbitals and were conducted with an uncontracted basis set for coupling calculations \cite{deng2006calculation}, called with the 'mixed' option within the Gaussian09 software. The scalar coupling values obtained from the calculations included all terms calculated: Fermi contact, spin-dipolar, paramagnetic spin-orbit and diamagnetic spin orbit terms are all included in the total nuclear spin-spin coupling produced in the output files. Some DFT structure optimisations failed to converge to an energy minima and these were excluded from the final datasets.

\FloatBarrier
\newpage
\subsection{Correction of DFT NMR parameter predictions for comparison to experiment}
\label{sec:si_param_correction}
\begin{figure}[h]
    \centering
    \subfloat{\includegraphics[width=0.45\linewidth]{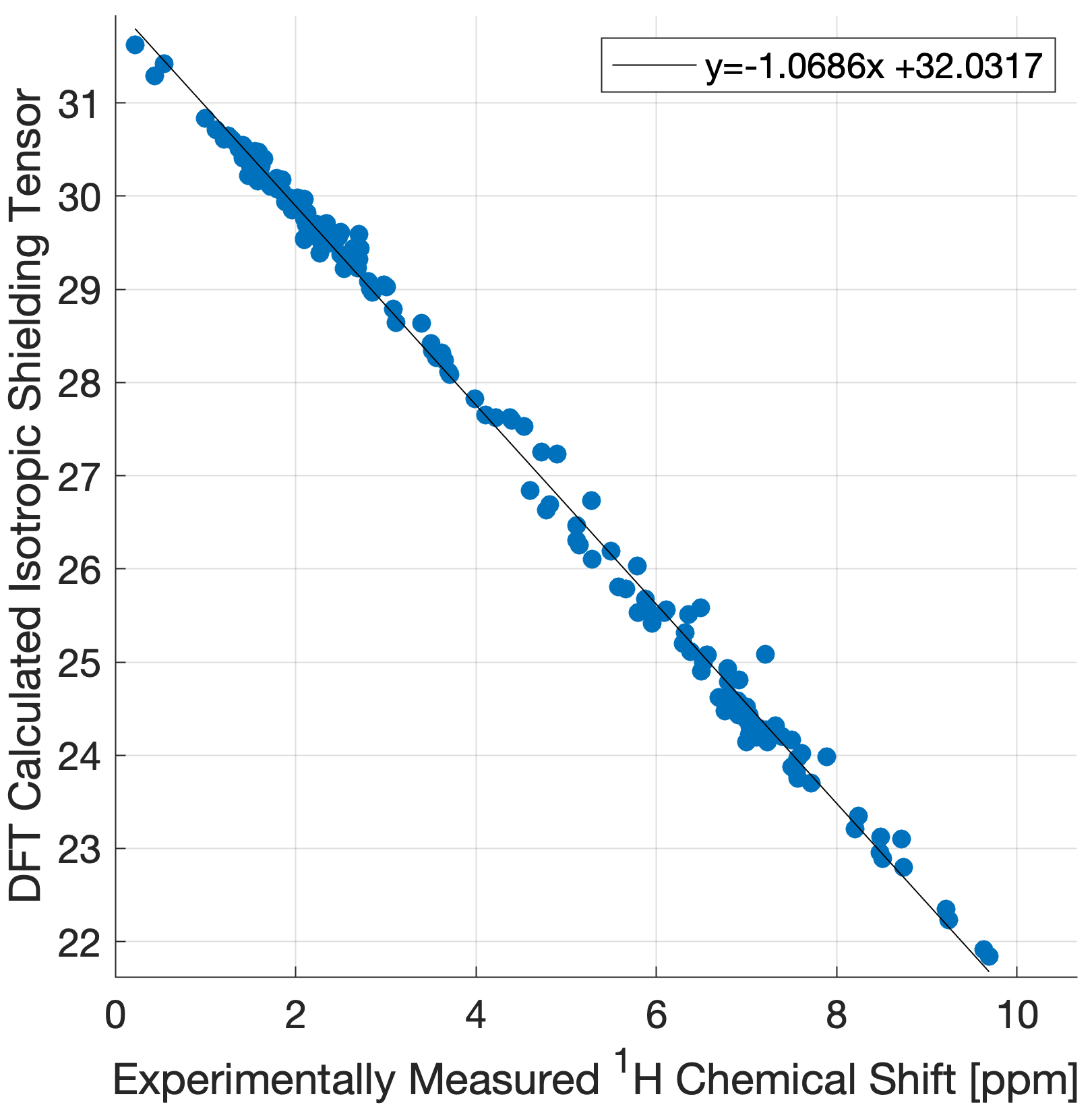}}
    \quad
    \subfloat{\includegraphics[width=0.45\linewidth]{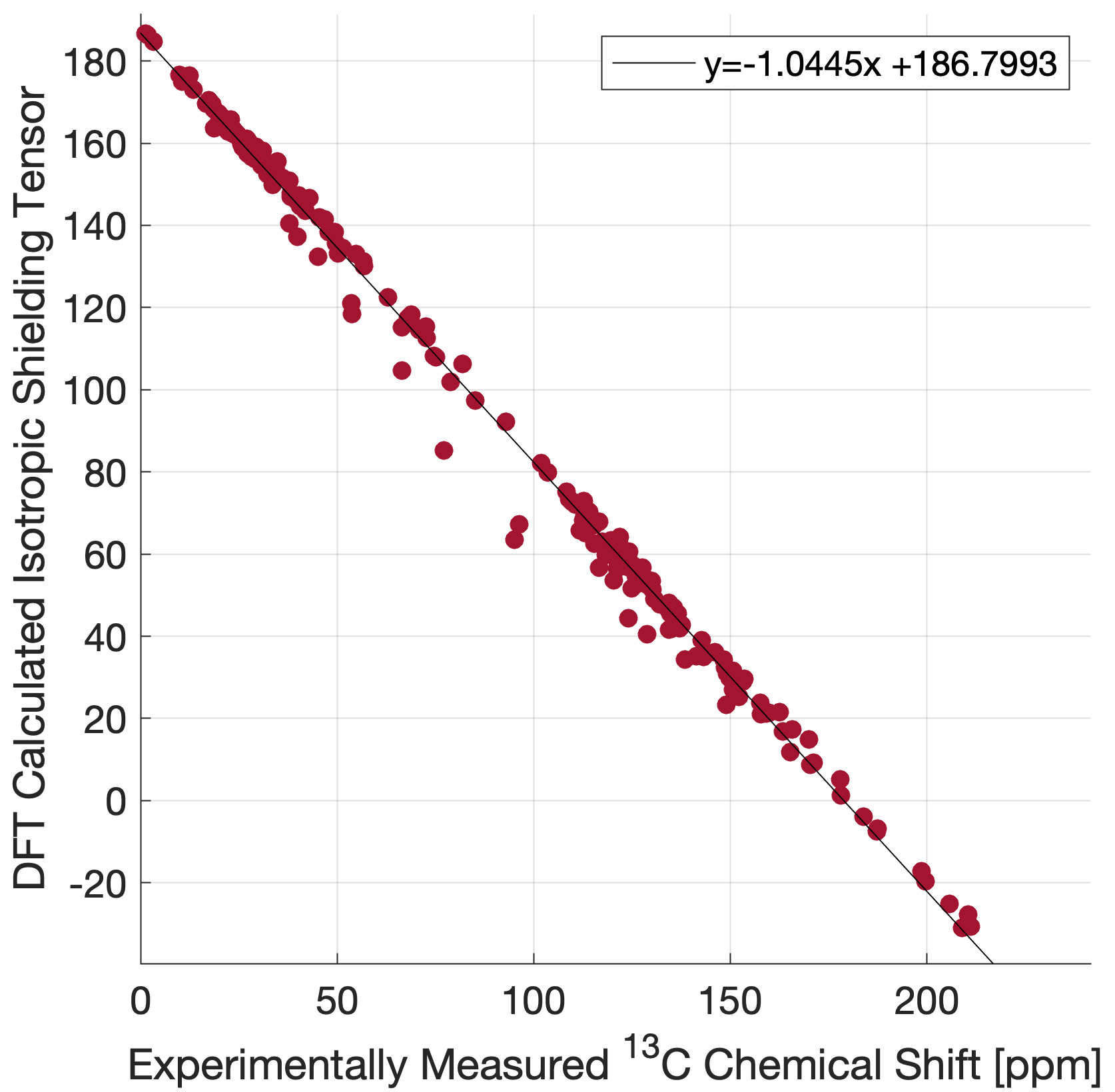}}
    \caption{Calculation of tantillo regression scaling factors for a) $\delta ^1\mathrm{H}$ and b) $\delta ^{13}\mathrm{C}$}
    \label{fig:tantillo}
\end{figure}

The DFT calculated magnetic shielding tensors were converted to chemical shifts using a linear scaling method and reference compounds reported by Tantillo \textit{et al}\cite{lodewyk2011computational,CHESHIRE}. The results of this linear scaling are shown in figure \ref{fig:tantillo}.

For the $^1\mathrm{\textit{J}_{CH}}$ data used in this work, a clear linear offset was found upon comparison of the DFT values to experimentally measured data. As a result, the offset (10.91Hz) was applied to the DFT values in both the training and test datasets.

\subsection{Adaptive sampling} \label{sec:active_learning}
The training set was obtained via an adaptive sampling approach. An initial set of 100 structures were chosen at random from the CSD-500 test set already obtained from the work by Emsley et al\cite{paruzzo2018chemical}.
5 subsets of 80 structures each were then used to train separate models to predict $^1\mathrm{J_{CH}}$ coupling constants, $^1\mathrm{H}$ and $^{13}\mathrm{C}$ chemical shifts for all organic structures in the Cambridge Structural Database containing only H/C/N/O/F atoms.
The variance in the predictions of the five models (pre-prediction variance) is a measure of how confident one can be in a given prediction.
300 structures containing the environments with the highest variance were selected to be added to the training set (100 each based on the $^1\mathrm{J_{CH}}$, $^1\mathrm{H}$ and $^{13}\mathrm{C}$ variance). Structure optimizations and NMR computations were performed for these to build the training set. The initial random set of 100 structures was discarded after the first round, and the process was repeated four times. Some structures failed to optimise in each round and were discarded leading to a training set consisting of 882 structures.

\subsection{Hyper-parameter optimisation}
The following hyper-parameters were optimized for the machine learning procedure: the cutoff radius, the kernel width and the l2-regularisation factor. The optimal combination of these three variables was found through a cross-validated gaussian-process led search using the python module BayesianOptimization\cite{bayes}. The optimal parameters were determined as those with the lowest average mean absolute deviation across a five-fold cross-validation using the training set environments.

\newpage
\FloatBarrier
\section{IMPRESSION performance using Molecular Mechanics geometries}
\label{sec:mm_models}
Whilst the focus of this work is to develop a machine learning method to replace the DFT calculation of NMR parameters, the geometry optimisation used in preparing the structures in all datasets still accounts for 26\% of the total CPU time. The effect of replacing the DFT geometry optimisation step with a molecular mechanics based optimisation was investigated through two methods. 

\subsection{DFT trained model}
\begin{figure*}[h]
    \centering
    \includegraphics[width=\textwidth]{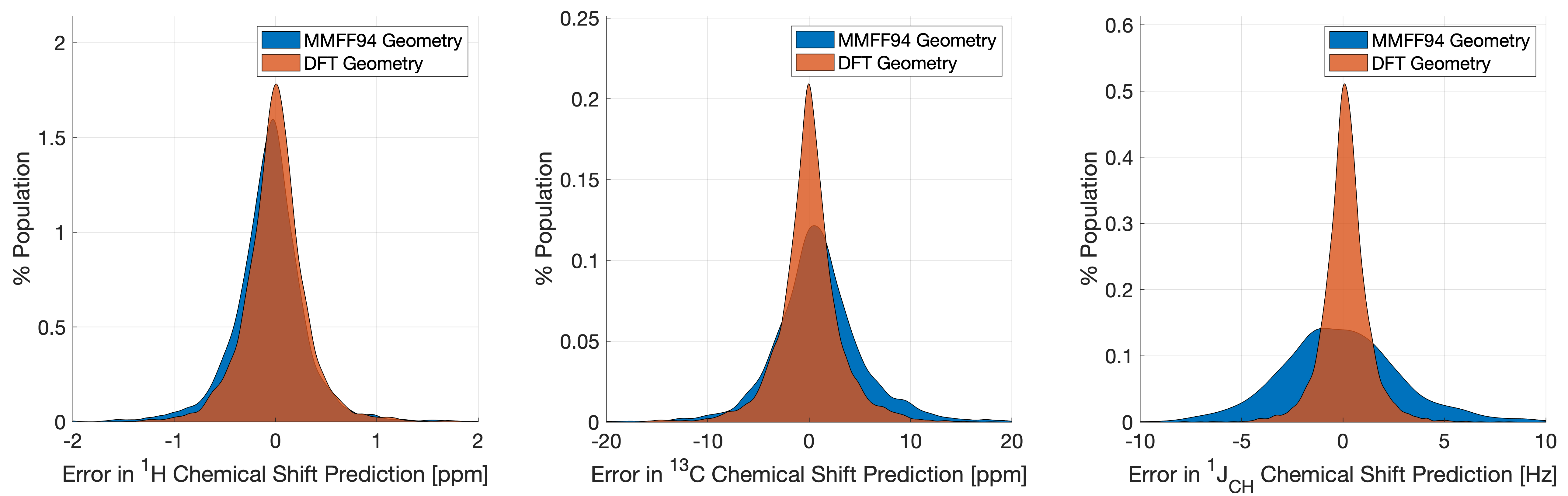}
    \caption{Error distributions for IMPRESSION predictions of molecular mechanics structures, using IMPRESSION models trained using DFT geometries. Variance filters applied: $\delta ^1\mathrm{H}$ = 0.1ppm, $\delta ^{13}\mathrm{C}$ = 5ppm, $^1\mathrm{\textit{J}_{CH}}$ = 1Hz.}
    \label{fig:dftmodel_comp}
\end{figure*}

Firstly, the existing models (trained using DFT optimised geomtries) were used to make predictions on structures optimised through the MMFF94 forcefield\cite{halgren1996merck}. The result was a decrease in accuracy of all three models but especially so for $1J_{CH}$ and $\delta ^{13}\mathrm{C}$. The error distributions in figure \ref{fig:dftmodel_comp} show a reduction in the quality of the predictions on all three parameters.

\begin{center}
\begin{tabular}{|c||c|c|c|c|c|}
\hline
     & MAE & RMSE & MaxE & Variance Cutoff & Envs removed\\
    \hline
    $\delta ^1\mathrm{H}$ & 0.26ppm & 0.38ppm & 5.55ppm & 0.1ppm & 1\\
        \hline
    $\delta ^{13}\mathrm{C}$ & 3.30ppm & 4.63ppm & 37.42ppm & 5ppm & 949\\
    \hline
    $^1\mathrm{\textit{J}_{CH}}$ & 2.30Hz & 3.00Hz & 20.44Hz & 1Hz & 5009\\
    \hline
\end{tabular}
\end{center}

\newpage
\FloatBarrier
\subsection{MMFF94 trained model}
Additionally, new models were trained based on molecular mechanics optimised training structures. 
The entire set of training and testing structures were reoptimised using the MMFF94\cite{halgren1996merck} forcefield. 
These structures were associated with the previously calculated DFT NMR parameters and used to train and test new models. 
The model hyper-parameters were optimised using the same method as the DFT trained models and achieved an accuracy which was up to 50\% worse than the models trained using DFT optimised structures. 

\begin{center}
\begin{tabular}{|c||c|c|c|c|c|}
\hline
     & MAE & RMSE & MaxE & Variance Cutoff & Envs removed\\
    \hline
    $\delta ^1\mathrm{H}$ & 0.28ppm & 0.40ppm & 5.20ppm & 0.1ppm & 3\\
        \hline
    $\delta ^{13}\mathrm{C}$ & 2.31ppm & 3.48ppm & 39.92ppm & 5ppm & 952\\
    \hline
    $^1\mathrm{\textit{J}_{CH}}$ & 1.19Hz & 1.75Hz & 20.40Hz & 5Hz & 143\\
    \hline
\end{tabular}
\end{center}

\begin{figure*}[h]
    \centering
    \includegraphics[width=\textwidth]{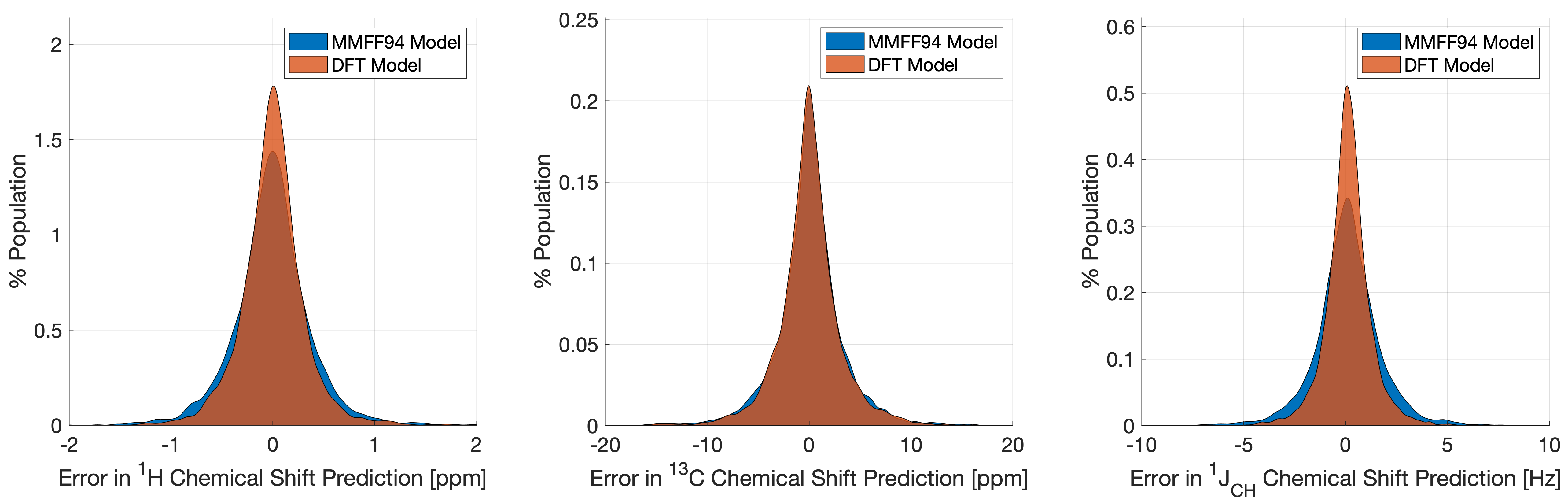}
    \caption{Error distributions for models trained using MMFF94 geometries predicting on structures with MMFF94 geometries, compared to the original DFT models from the main text. Variance filters applied: $\delta ^1\mathrm{H}$ = 0.1ppm, $\delta ^{13}\mathrm{C}$ = 5ppm, $^1\mathrm{\textit{J}_{CH}}$ = 1Hz.}
    \label{fig:mmmodelcomp}
\end{figure*}

\FloatBarrier
\newpage
\subsection{Computational timings}

\begin{figure}[h]
    \centering
    \subfloat[]{\includegraphics[width=0.45\linewidth]{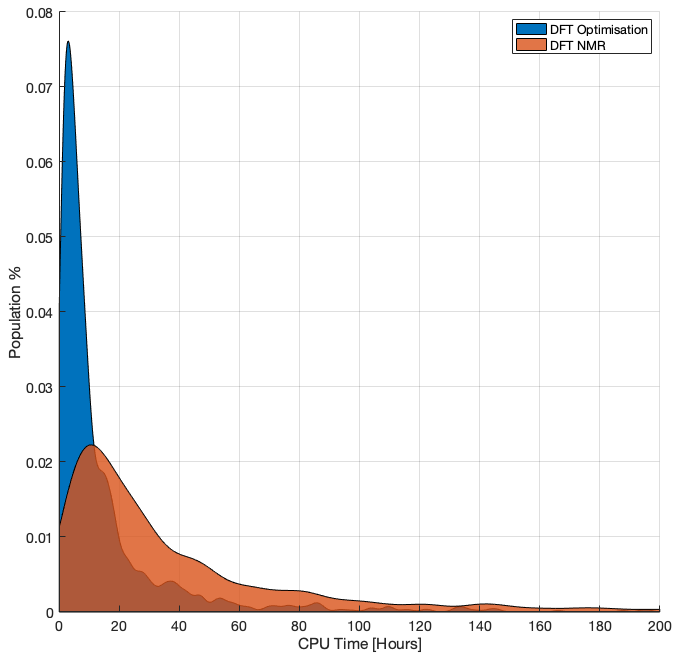}}
    \quad
    \subfloat[]{\includegraphics[width=0.45\linewidth]{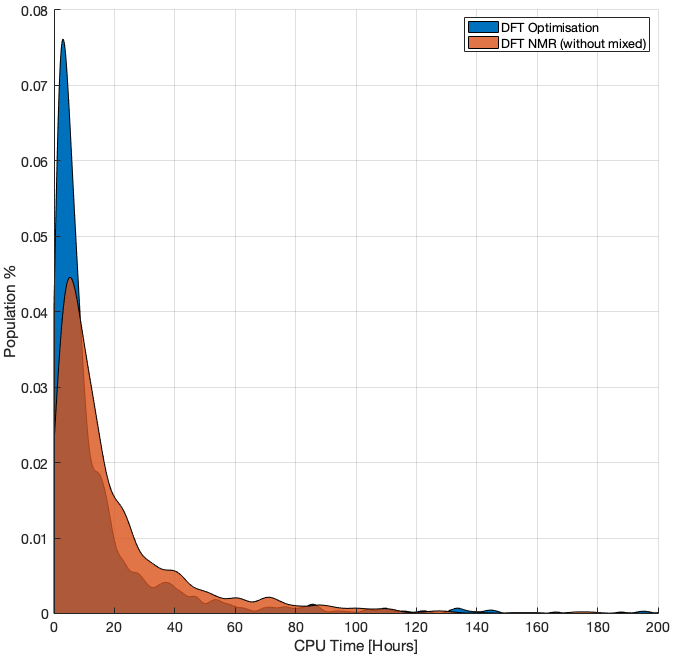}}
    \caption{Distribution of CPU time for DFT calculations on the training dataset. Mean time for optimisation = 15 Hours. a) with mixed option: Mean time for NMR calculation = 42 Hours. b) without mixed option: Mean time for NMR calculation = 22 Hours.}
    \label{fig:cputimecomp}
\end{figure}

To highlight the value of replacing the NMR calculation with a machine learning solution, the distributions of CPU cost for all calculations in producing the training set are included here in figure \ref{fig:cputimecomp}. The 'mixed' option which uses an un-contracted basis set for calculating the fermi contact term is only relevant for coupling calculations so this has been removed from figure \ref{fig:cputimecomp}b. 
The mean CPU time for an optimisation was 15 hours across all 882 structures, whilst the mean CPU time for a DFT NMR calculation was 42 hours (or 22 hours without mixed). The use of a machine learning model to replace the NMR calculation therefore represents a significant time saving.

\FloatBarrier
\newpage
\section{Structure revision examples} \label{sec:csdiscrimination}
To further demonstrate the applicability of the IMPRESSION predictions to structural elucidation problems, 5 examples of proposed natural product structure revisions from the literature were investigated \cite{kutateladze2018addressing}. No $^1J_\mathrm{CH}$ values were reported for these compounds, so we can only make comparisons using the chemical shift models. 

For the 5 compounds, Cartesian coordinates for the original and revised structures were obtained from the literature along with the experimental $\delta ^{1}\mathrm{H}$ and $\delta ^{13}\mathrm{C}$ assignments.
The Cartesian coordinates were optimised, NMR parameters were computed and IMPRESSION predictions were made for each structure. The mean absolute error between the IMPRESSION predictions and experiment were compared to the corresponding MAE between the DFT calculations and the experimental values. Variance cutoffs of 1Hz, 0.1ppm, and 5ppm were used for $^1J_\mathrm{CH}$, $\delta ^1\mathrm{H}$, and $\delta ^{13}\mathrm{C}$ respectively.

\subsection{Geometric mean for diastereomer discrimination}
As we combine different types of data to gather evidence for a given diastereomer, we take the geometric mean of mean absolute errors for each of the parameters:

\begin{equation}
\mathrm{MAE_{combined}} = \sqrt[3]{\mathrm{MAE_{{}^1\mathrm{\textit{J}_{CH}}}MAE_{\delta{}^1\mathrm{H}}MAE_{\delta{}^{13}\mathrm{C}}}}
\end{equation}

or in the case where $^1\mathrm{\textit{J}_{CH}}$ values are not avalable:

\begin{equation}
\mathrm{MAE_{combined}} = \sqrt[2]{\mathrm{MAE_{\delta{}^1\mathrm{H}}MAE_{\delta{}^{13}\mathrm{C}}}}
\end{equation}

\FloatBarrier
\subsection{Crithmifolide}
Comparing the results from our DFT method to that used in the original work, the predictions for the $^1\mathrm{H}$ chemical shifts do not show the same improvement in accuracy between the original and revised structures. In the original work an improvement of 0.08ppm RMSE was reported, whereas comparisons using our DFT method found an increase in MAE of 0.03ppm (and RMSE of 0.01ppm). Pleasingly the IMPRESSION results mirror this discrepancy and match the DFT method on which the model was trained.

The $^{13}\mathrm{C}$ chemical shift results from our DFT method agree with the literature, showing an improvement in fit from the original to the revised structure. The indecisive results from the geometric mean comparison reflect this discrepancy between the two chemical shift comparisons.

\begin{figure}[h]
    \centering
    \includegraphics[width=0.5\linewidth]{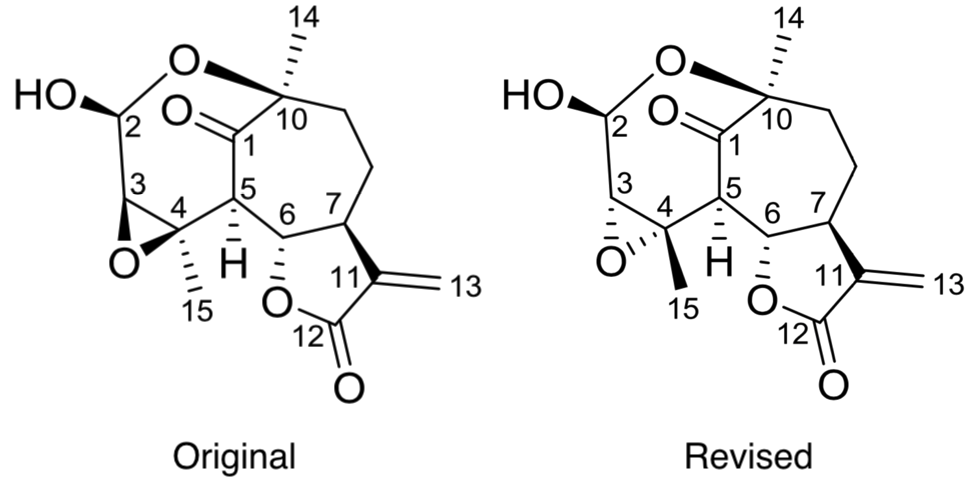}
    \caption{Original and revised structures for Crithmifolide.}
    \label{fig:crithmifolide}
\end{figure}
\begin{figure}[h]
    \centering
    \subfloat[]{\includegraphics[width=0.3\linewidth]{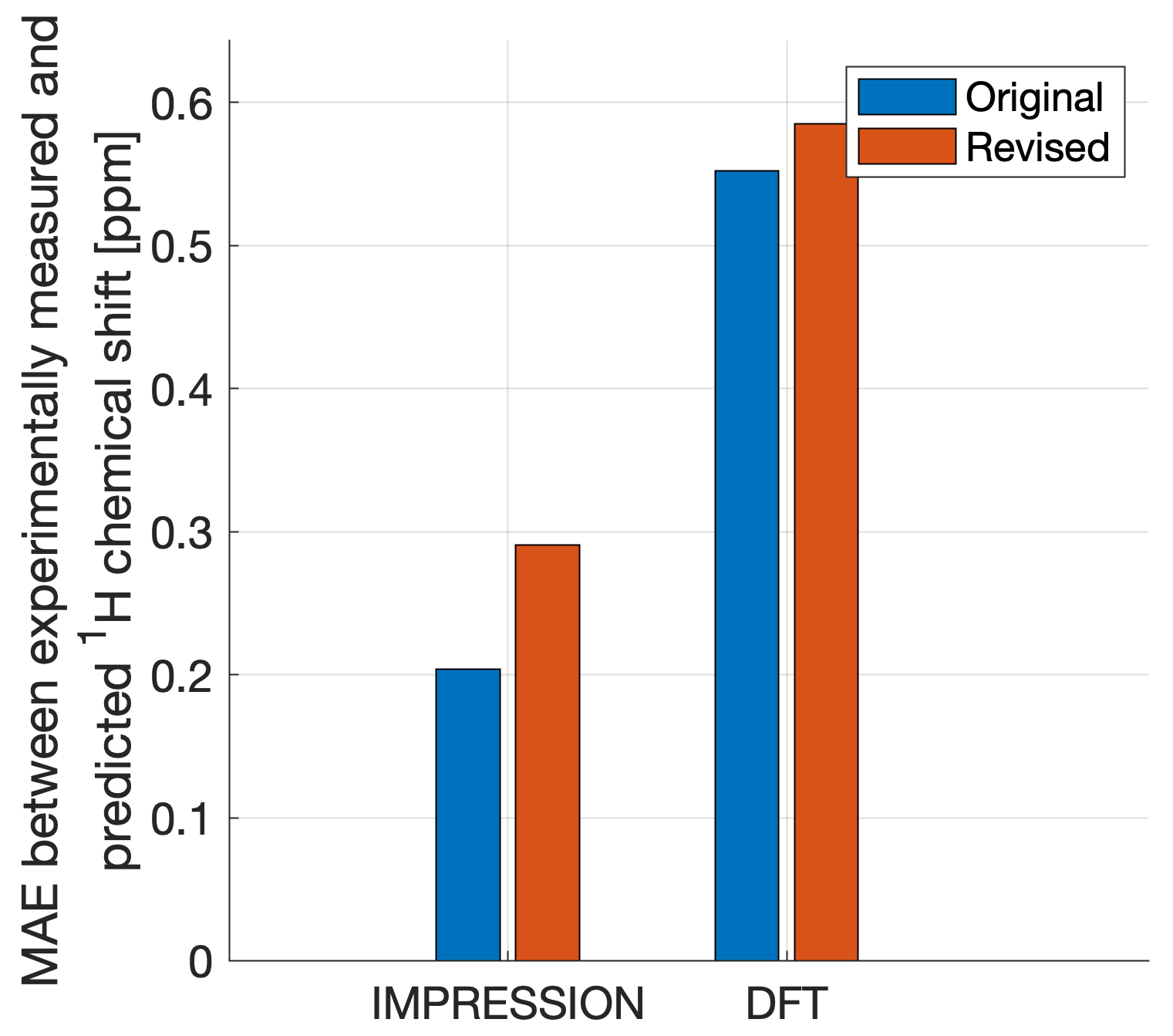}}
    \subfloat[]{\includegraphics[width=0.3\linewidth]{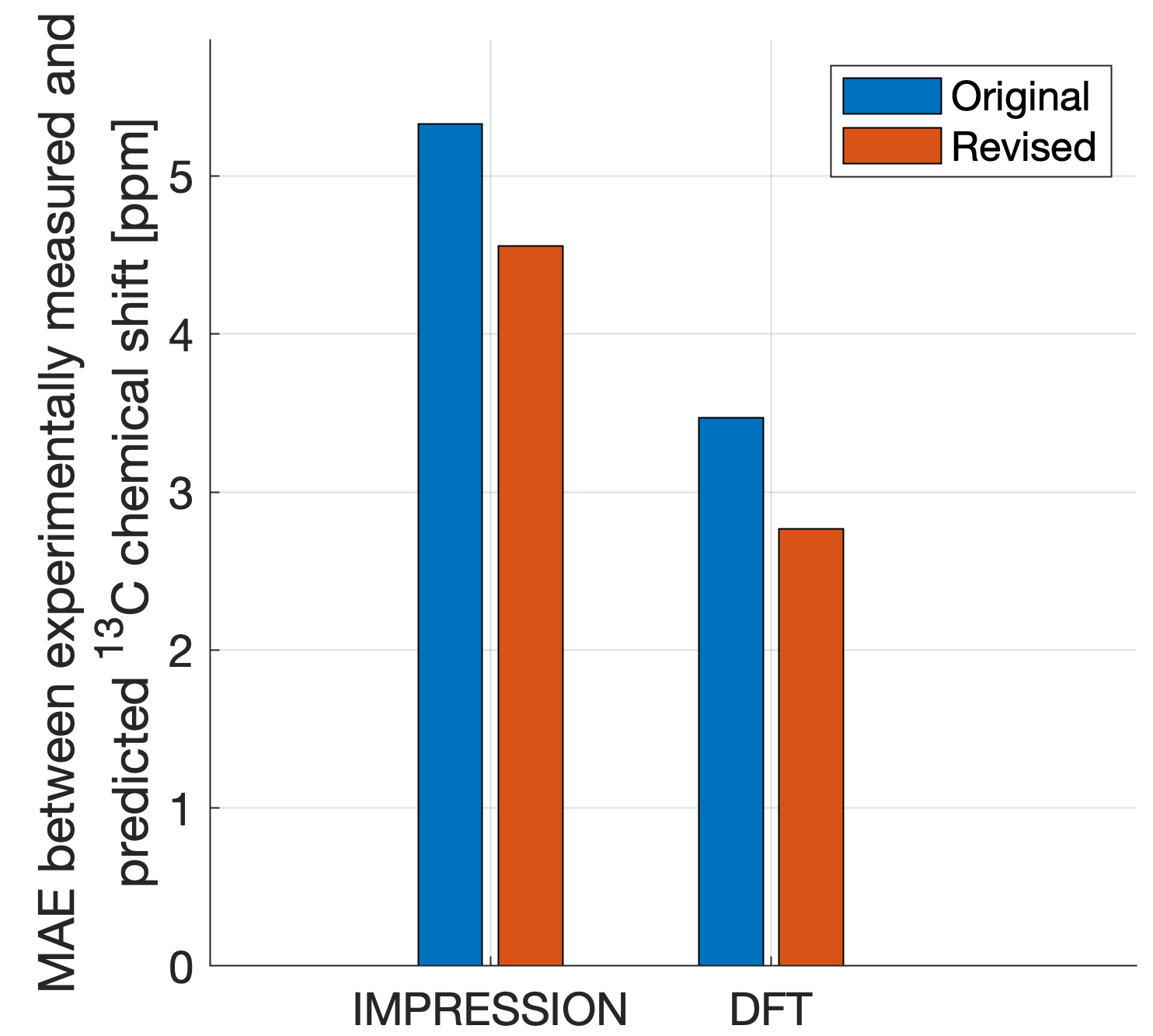}}
    \subfloat[]{\includegraphics[width=0.3\linewidth]{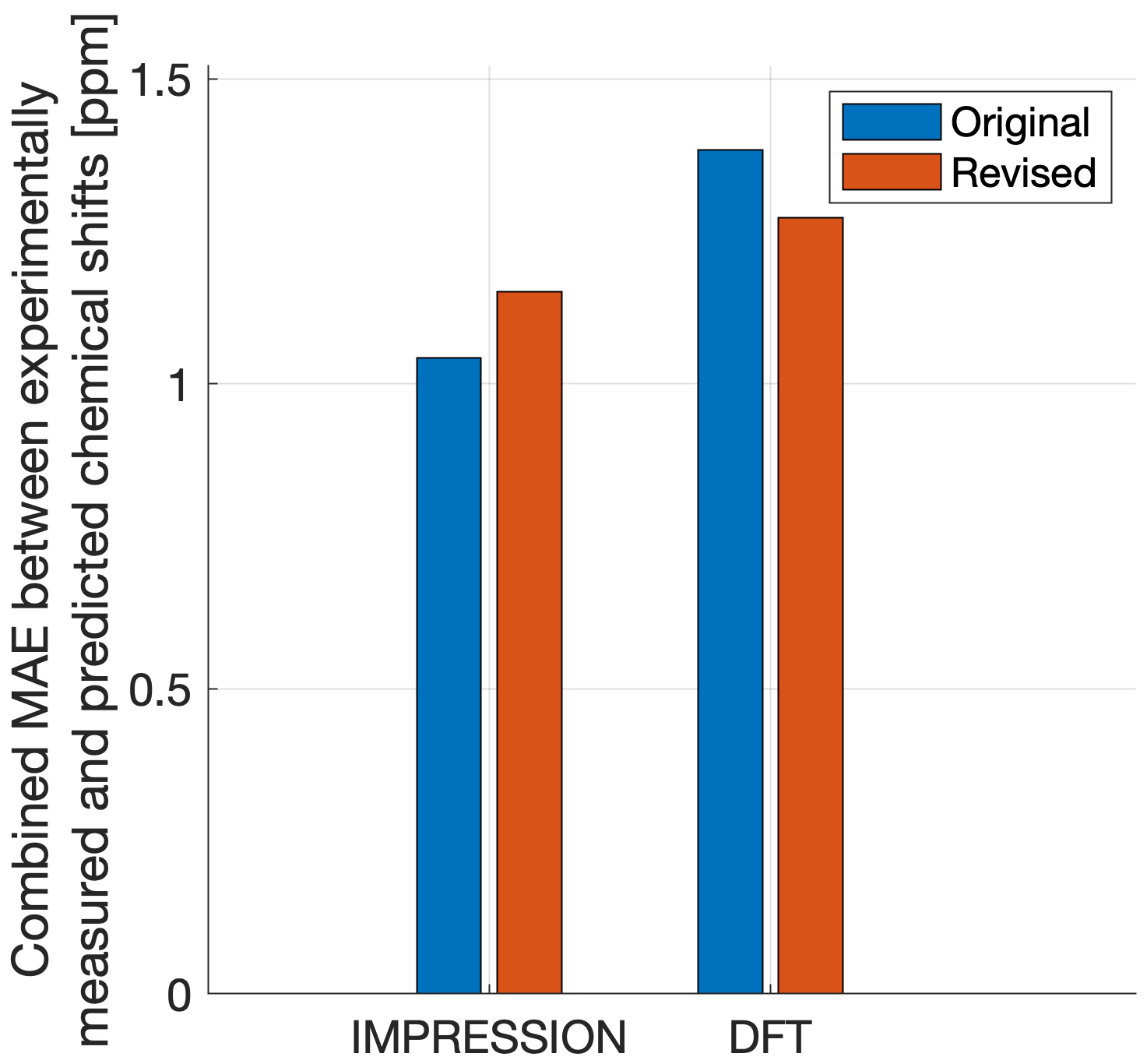}}
    
    \caption{Change in the fit between prediction and experiment for both DFT and IMPRESSION for Crithmifolide. a) $\delta ^1\mathrm{H}$. b) $\delta ^{13}\mathrm{C}$. c) Geometric mean across both parameters.}
    \label{fig:crithmifolide_bars}
\end{figure}

\FloatBarrier
\subsection{Caespitenone}
The results for Caespitenone show good agreement between IMPRESSION, our DFT method, and the previously published results. Large deviations (>5ppm) were reported in the $^{13}\mathrm{C}$ chemical shift results, and the DFT method used in this work reproduces this. The IMPRESSION predictions show the same change in the fit to experiment.

The methods used in this work also showed a significant improvement in fit for the $^1\mathrm{H}$ chemical shifts, resulting in a reduction in error of over 50\% for both our DFT method and IMPRESSION.

\begin{figure}[h]
    \centering
    \includegraphics[width=0.5\linewidth]{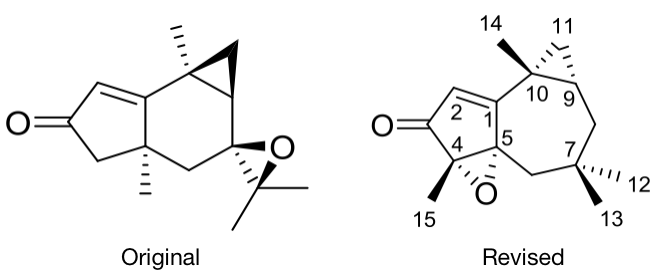}
    \caption{Original and revised structures for Caespitenone}
    \label{fig:caespitenone}
\end{figure}
\begin{figure}[h]
    \centering
    \subfloat[]{\includegraphics[width=0.3\linewidth]{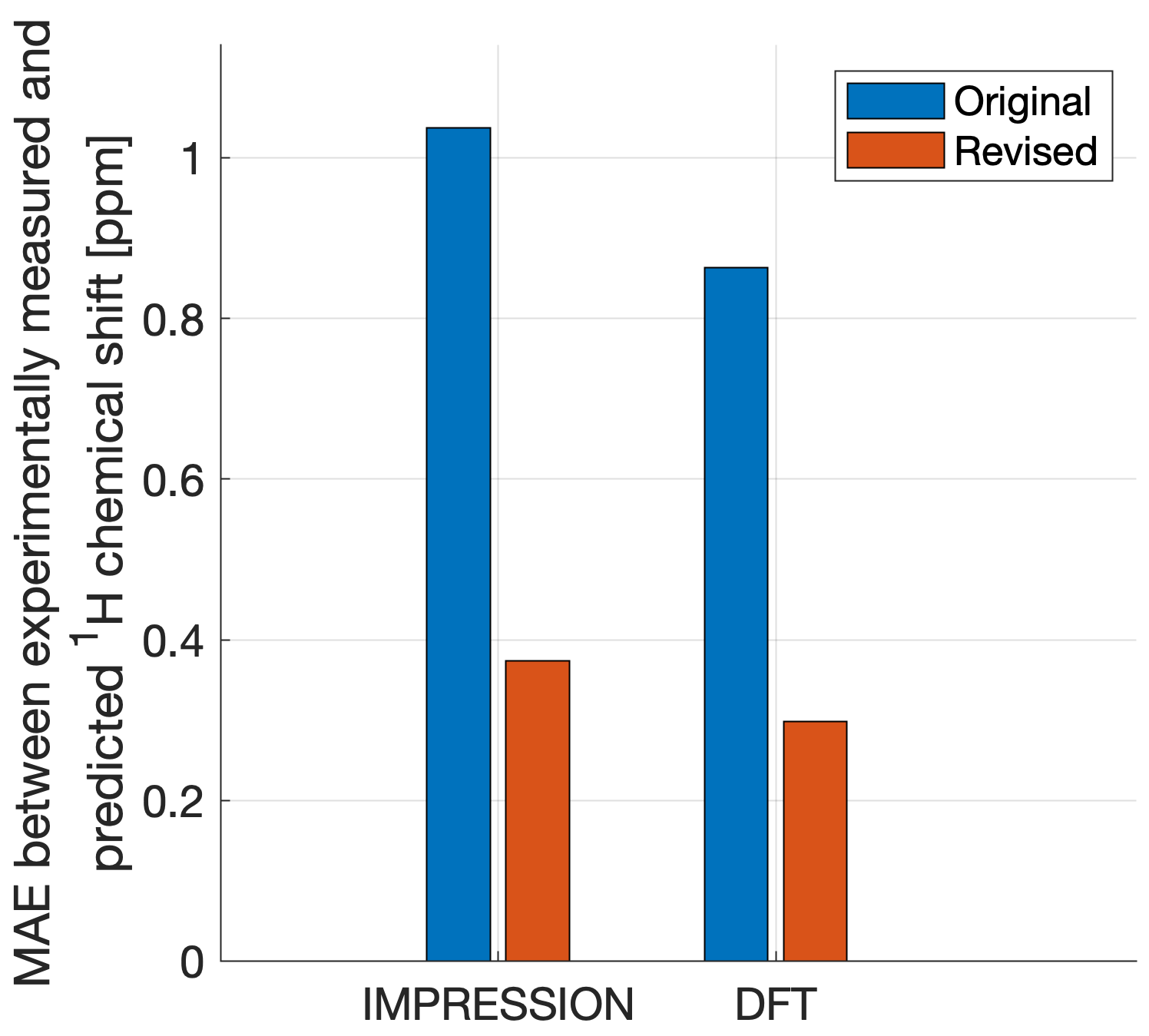}}
    \subfloat[]{\includegraphics[width=0.3\linewidth]{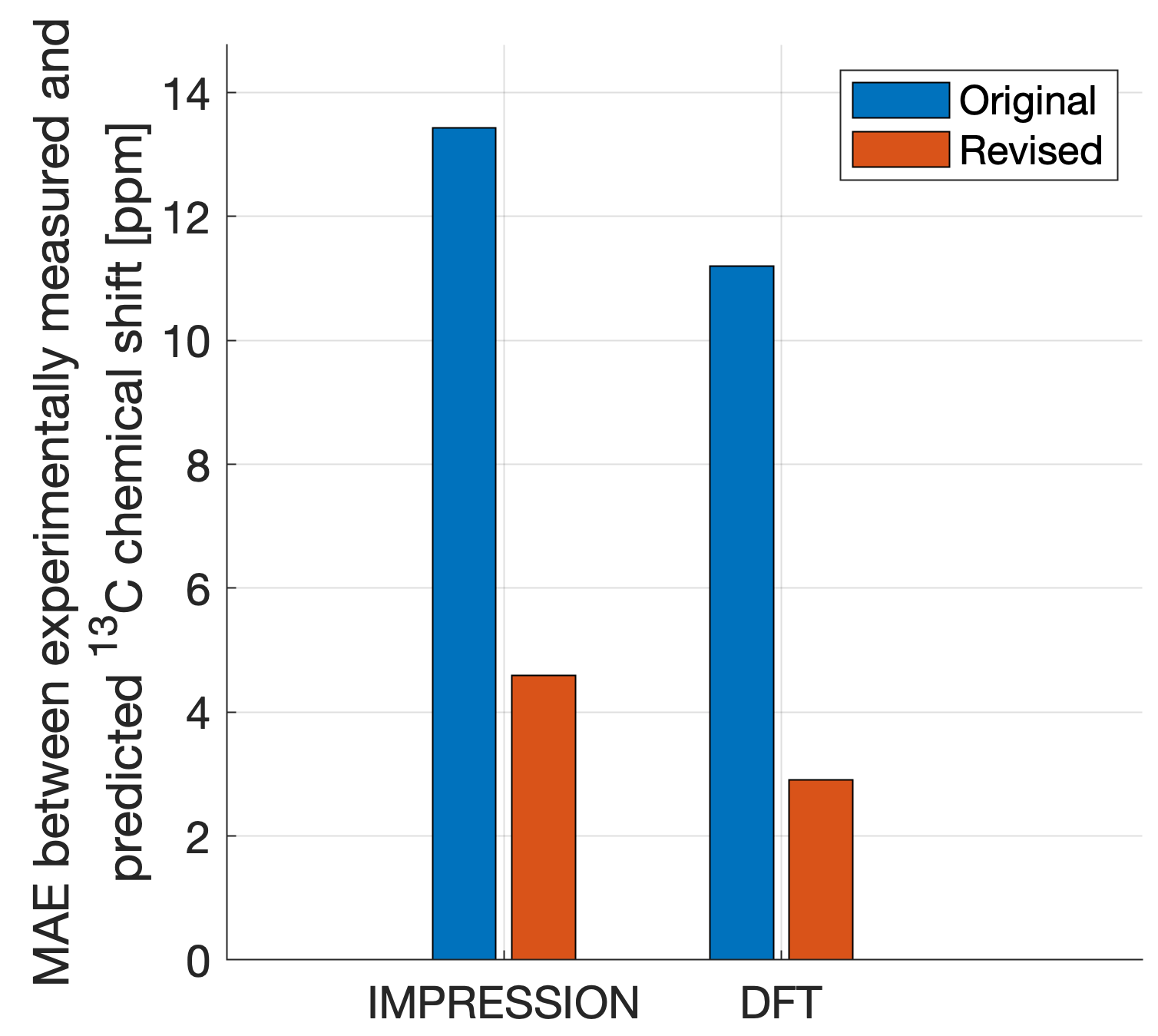}}
    \subfloat[]{\includegraphics[width=0.3\linewidth]{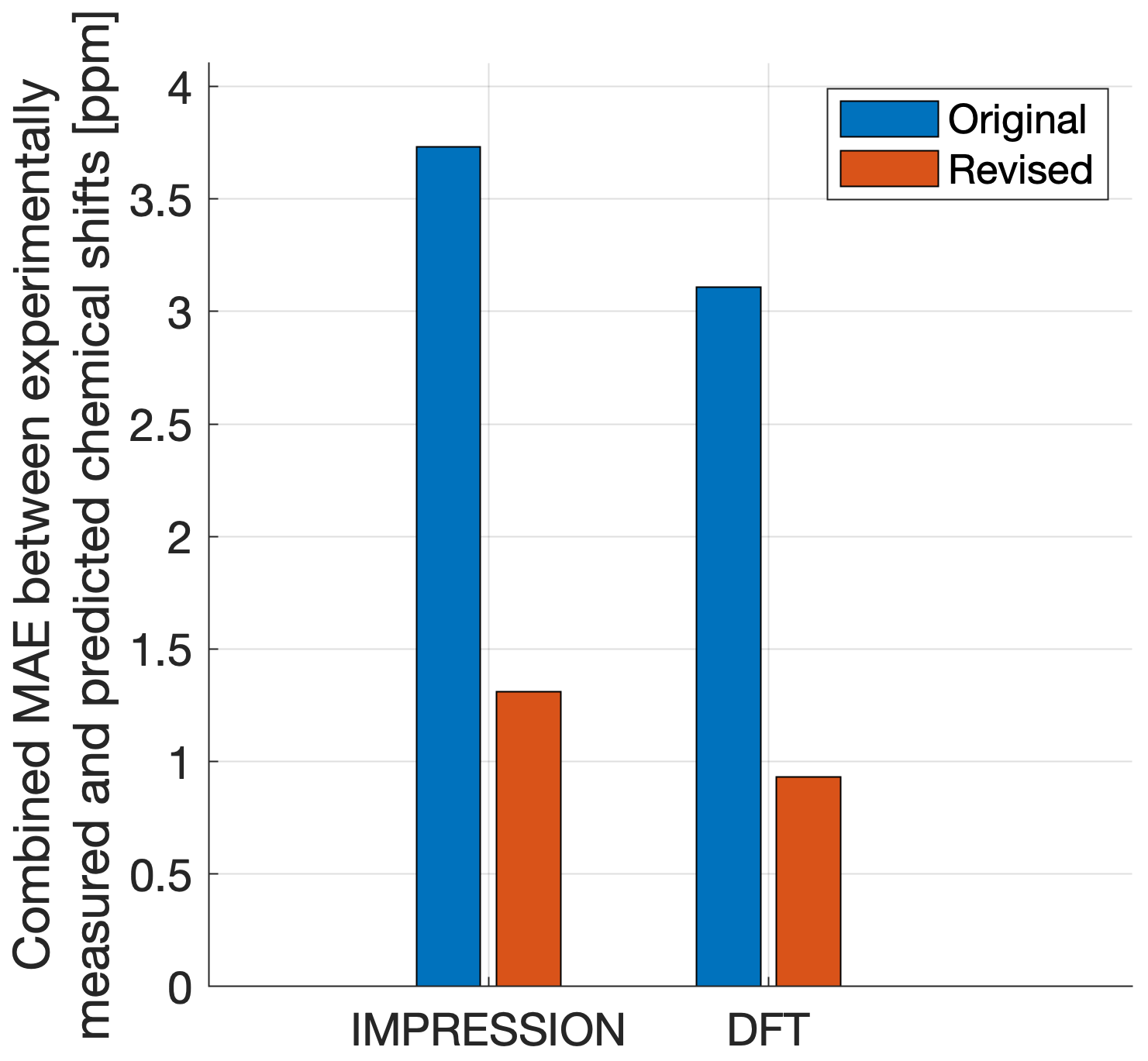}}
    
    \caption{Change in the fit between prediction and experiment for both DFT and IMPRESSION for Caespitenone. a) $\delta ^1\mathrm{H}$. b) $\delta ^{13}\mathrm{C}$. c) Geometric mean across both parameters.}
    \label{fig:caespitenone_bars}
\end{figure}

\FloatBarrier
\clearpage
\subsection{Secoafricane}
The reported values in the original work show a significant improvement in fit between experiment and calculation for both $^{1}H$ and $^{13}\mathrm{C}$ chemical shifts. The results from our DFT method show a smaller but still significant improvement in fit for both parameters, and IMPRESSION mimics these results, but with a smaller change in MAE for the $^{13}\mathrm{C}$ comparison.
\begin{figure}[h]
    \centering
    \includegraphics[width=0.5\linewidth]{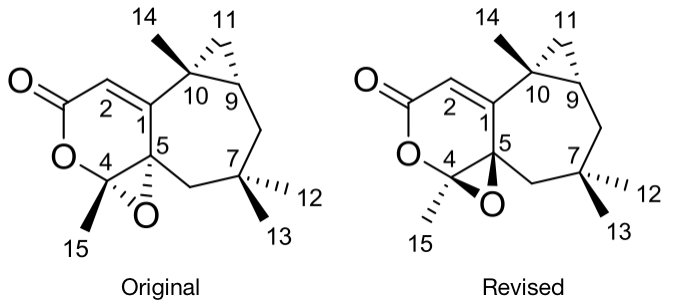}
    \caption{Original and revised structures for Secoafricane.}
    \label{fig:secoafricane}
\end{figure}
\begin{figure}[h]
    \centering
    \subfloat[]{\includegraphics[width=0.3\linewidth]{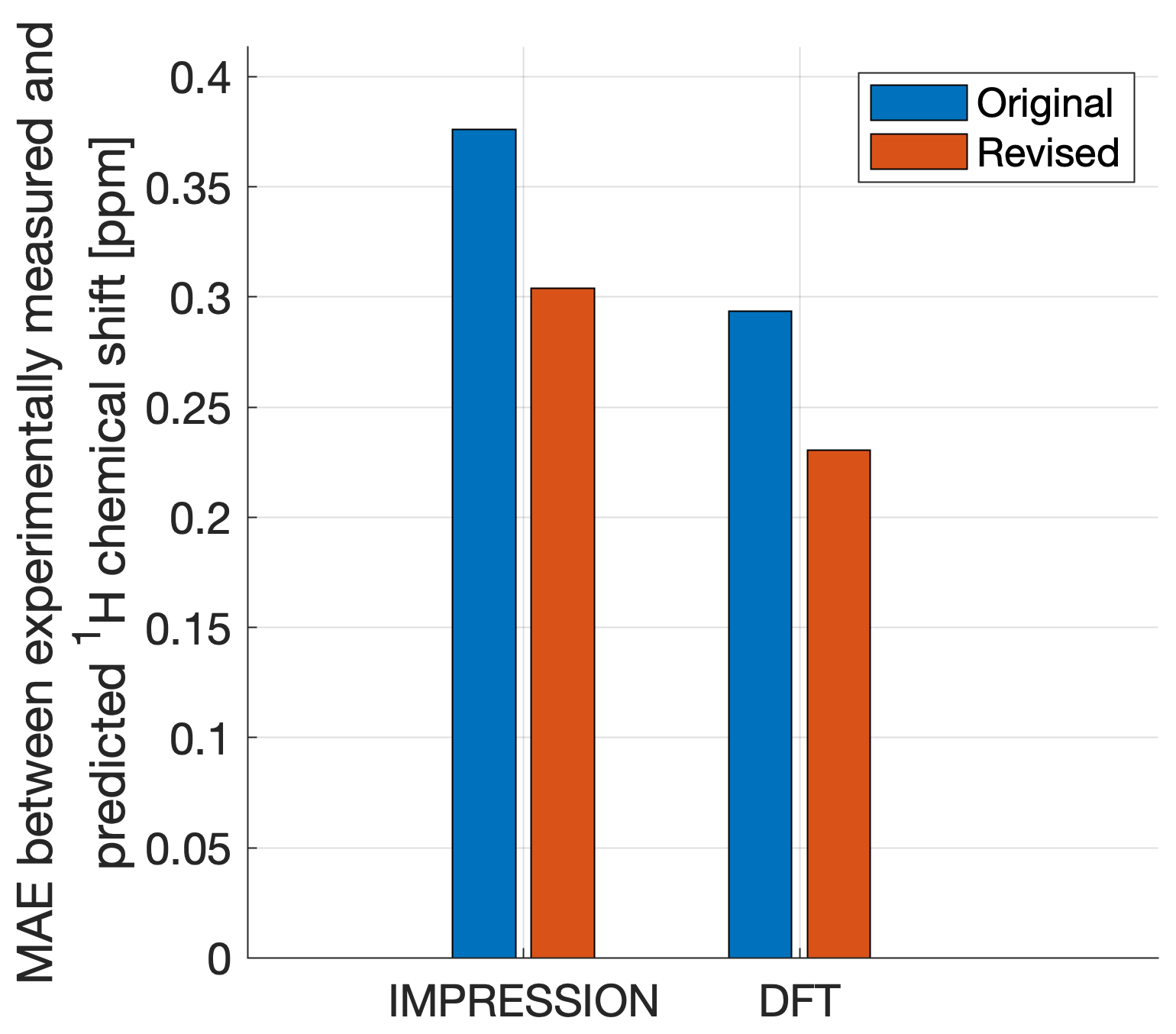}}
    \subfloat[]{\includegraphics[width=0.3\linewidth]{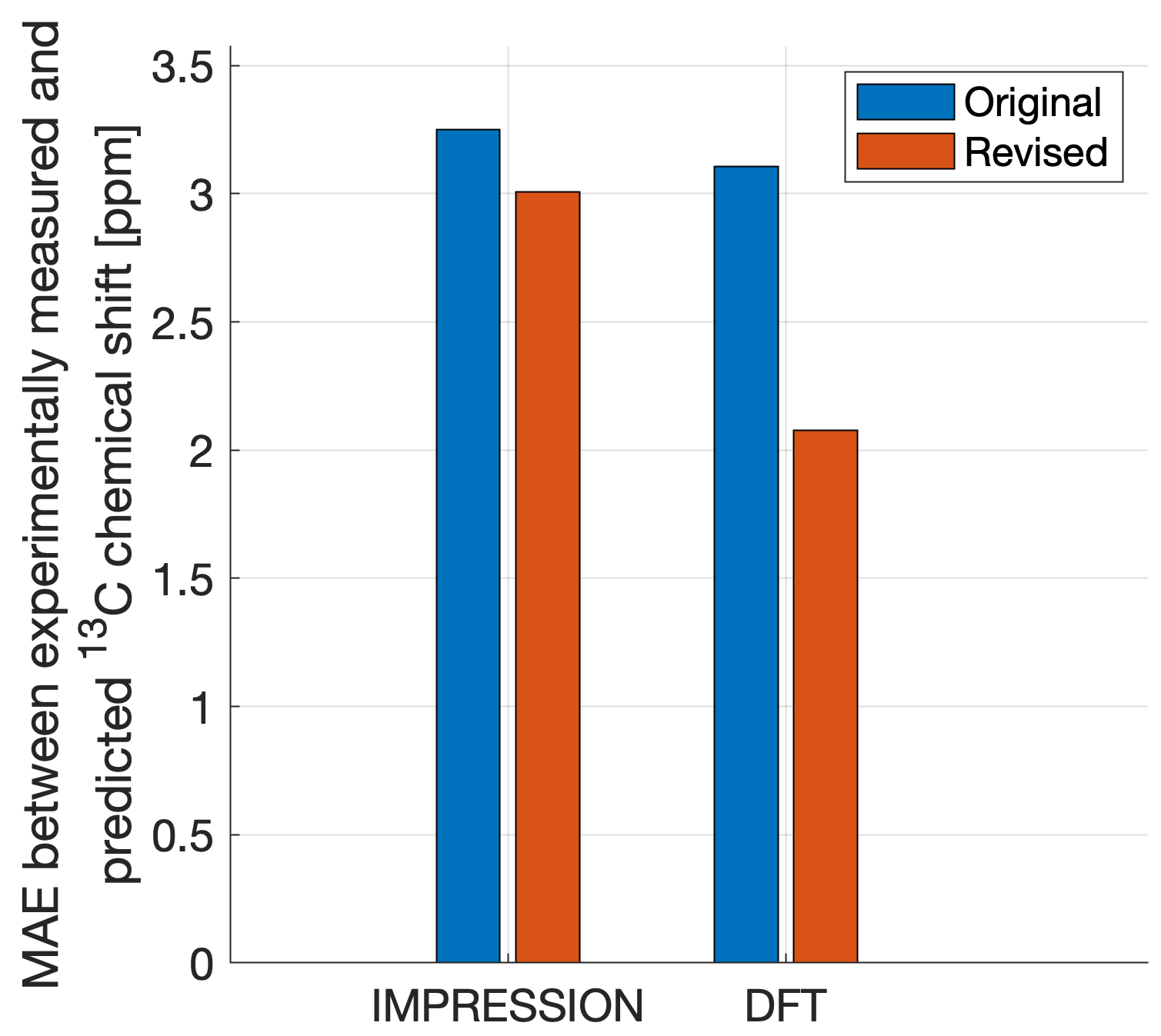}}
    \subfloat[]{\includegraphics[width=0.3\linewidth]{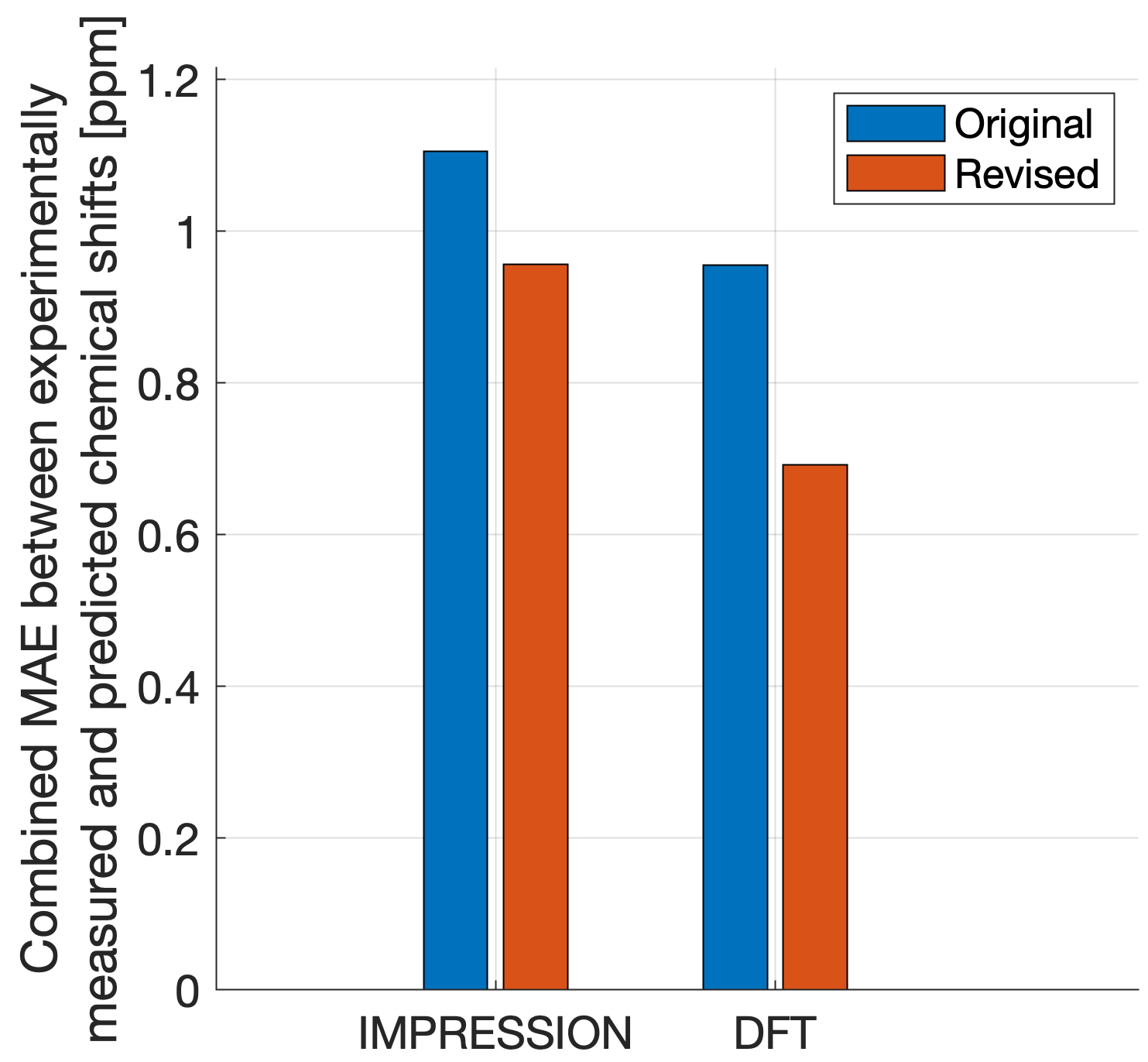}}
    
    \caption{Change in the fit between prediction and experiment for both DFT and IMPRESSION for Secoafricane. a) $\delta ^1\mathrm{H}$. b) $\delta ^{13}\mathrm{C}$. c) Geometric mean across both parameters.}
    \label{fig:secoafricane_bars}
\end{figure}

\FloatBarrier
\clearpage
\subsection{Grandilobalide B}
The literature results for Grandilobalide B show a large improvement in fit for $^{13}\mathrm{C}$ chemical shift, which is not reproduced in our results. The results for $^1\mathrm{H}$ chemical shifts are reproduced, in the literature a small decrease in the fit to experiment from 0.27ppm RMSE to 0.33ppm RMSE is reported. Both IMPRESSION and our DFT method show a small but significant reduction in fit for the revised structure.

\begin{figure}[h]
    \centering
    \includegraphics[width=0.5\linewidth]{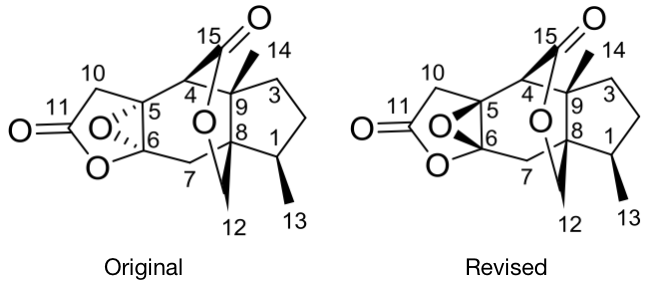}
    \caption{Original and revised structures for Grandilobalide B.}
    \label{fig:grandilobalide}
\end{figure}
\begin{figure}[h]
    \centering
    \subfloat[]{\includegraphics[width=0.3\linewidth]{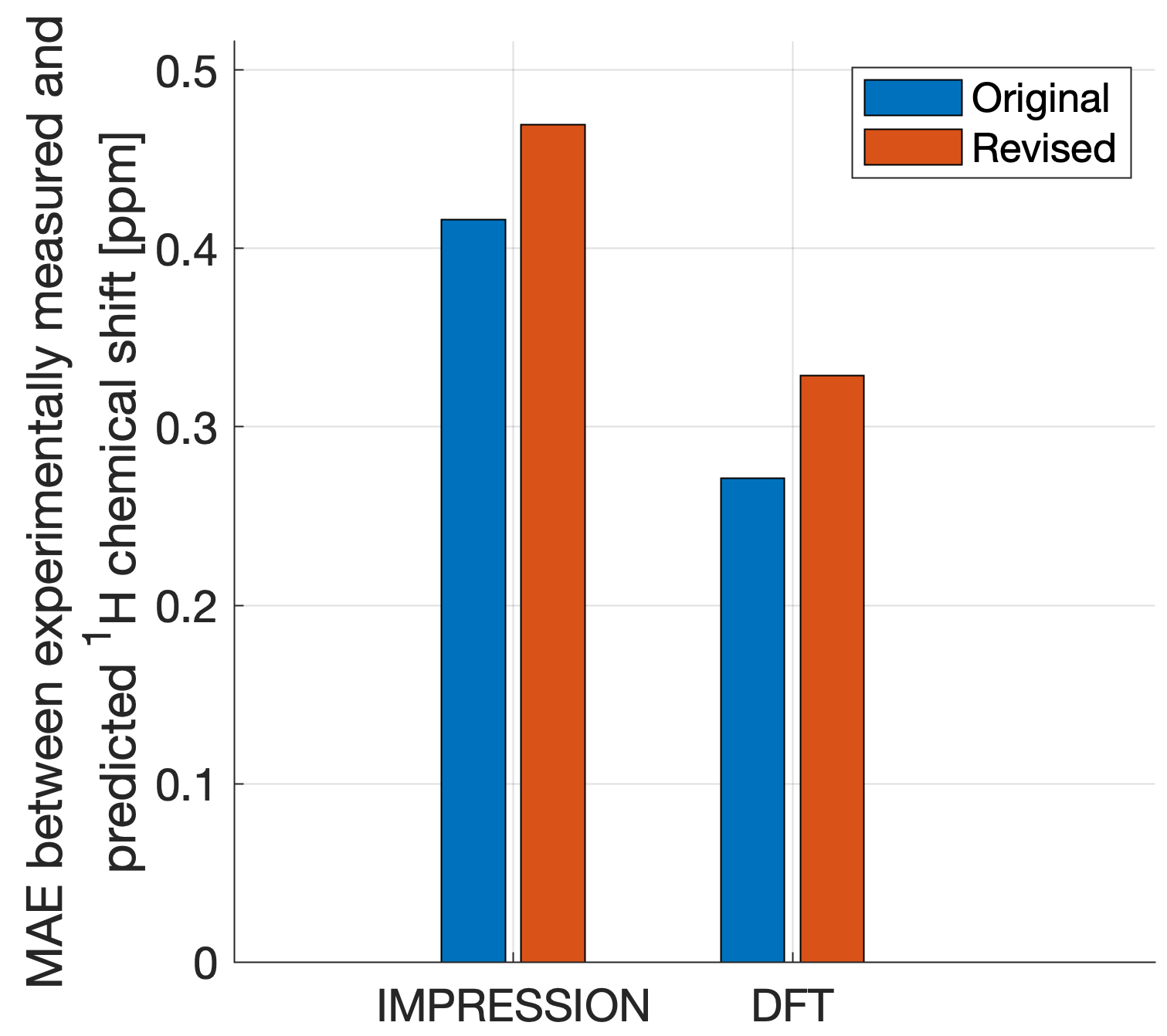}}
    \subfloat[]{\includegraphics[width=0.3\linewidth]{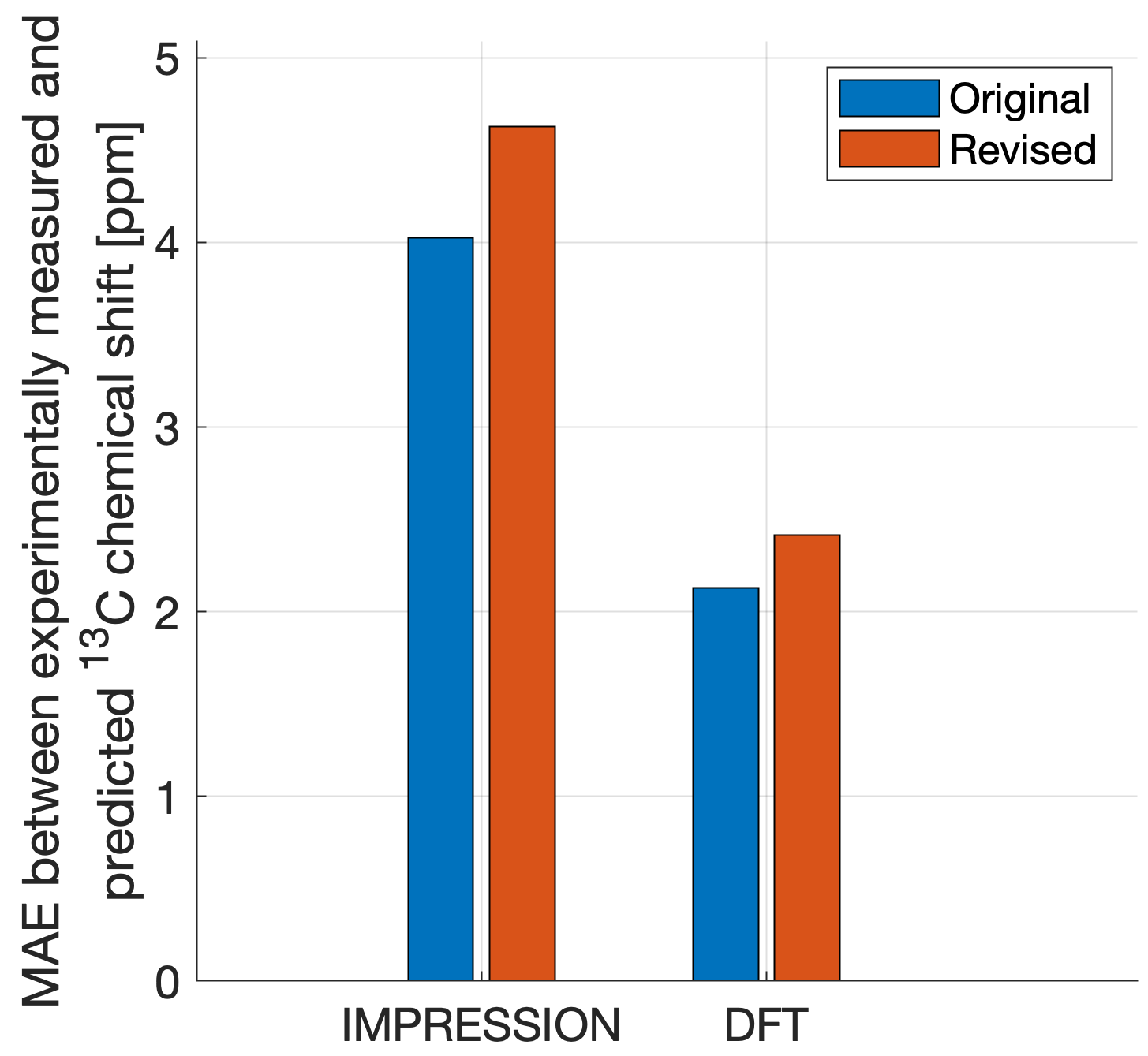}}
    \subfloat[]{\includegraphics[width=0.3\linewidth]{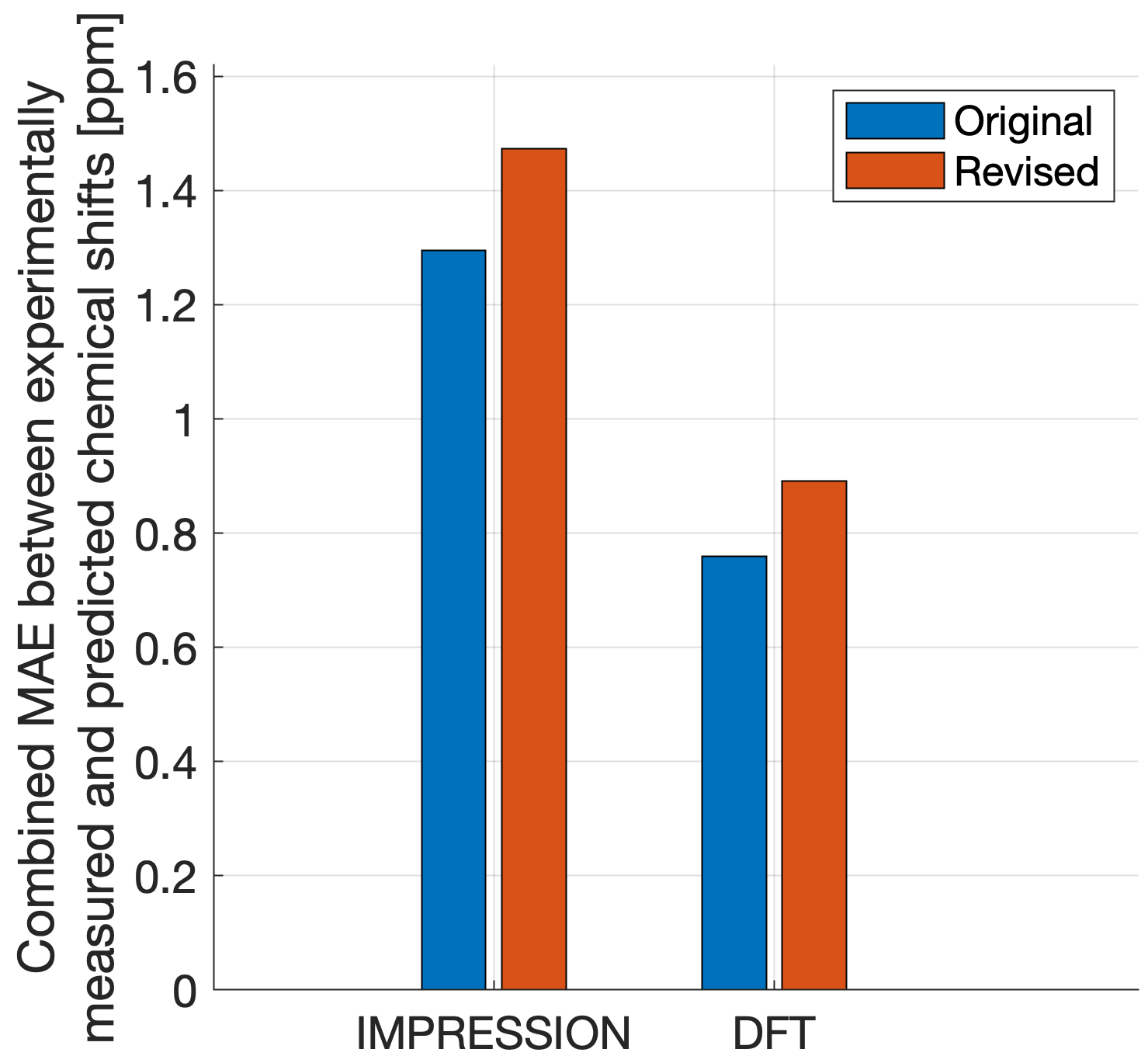}}
    
    \caption{Change in the fit between prediction and experiment for both DFT and IMPRESSION for Grandilobalide B. a) $\delta ^1\mathrm{H}$. b) $\delta ^{13}\mathrm{C}$. c) Geometric mean across both parameters.}
    \label{fig:grandilobalide_bars}
\end{figure}

\FloatBarrier
\clearpage
\subsection{Toluene dioxide}
In the original work, a large improvement in the fit to experiment for both $^{13}\mathrm{C}$ and $^1\mathrm{H}$ chemical shift was reported. The results from our DFT method were inconclusive for both parameters in this case. Pleasingly IMPRESSION mimics the DFT results, irrespective of the DFT methods fit to the reported results.
\begin{figure}[h]
    \centering
    \includegraphics[width=0.5\linewidth]{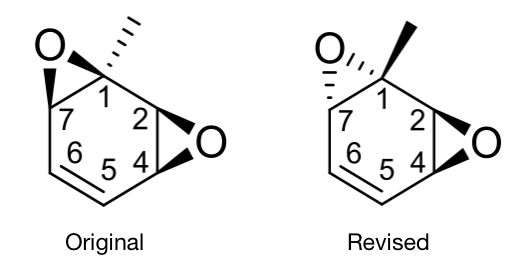}
    \caption{Original and revised structures for Toluene Dioxide}
    \label{fig:toluenedioxide}
\end{figure}
\begin{figure}[h]
    \centering
    \subfloat[]{\includegraphics[width=0.3\linewidth]{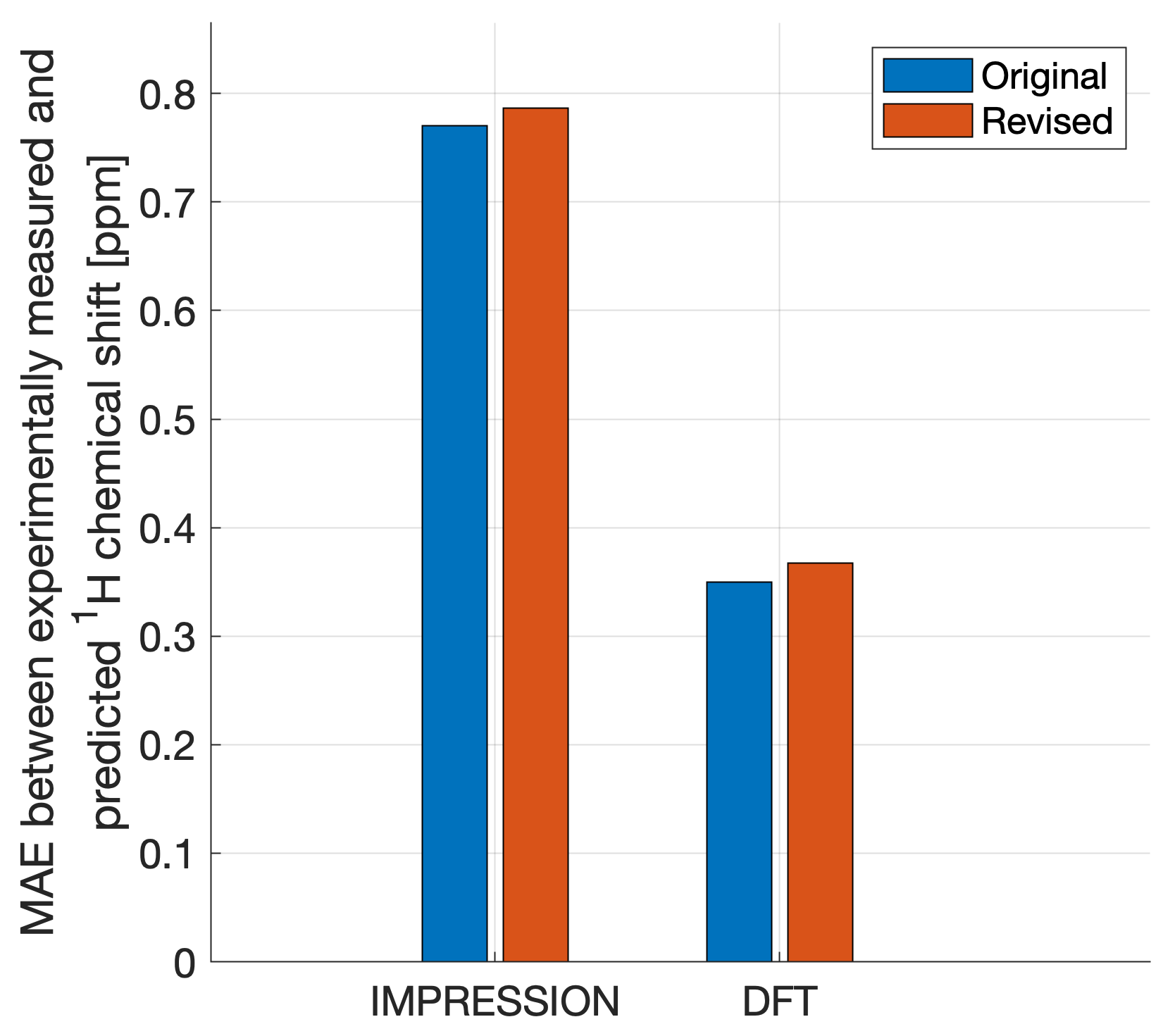}}
    \subfloat[]{\includegraphics[width=0.3\linewidth]{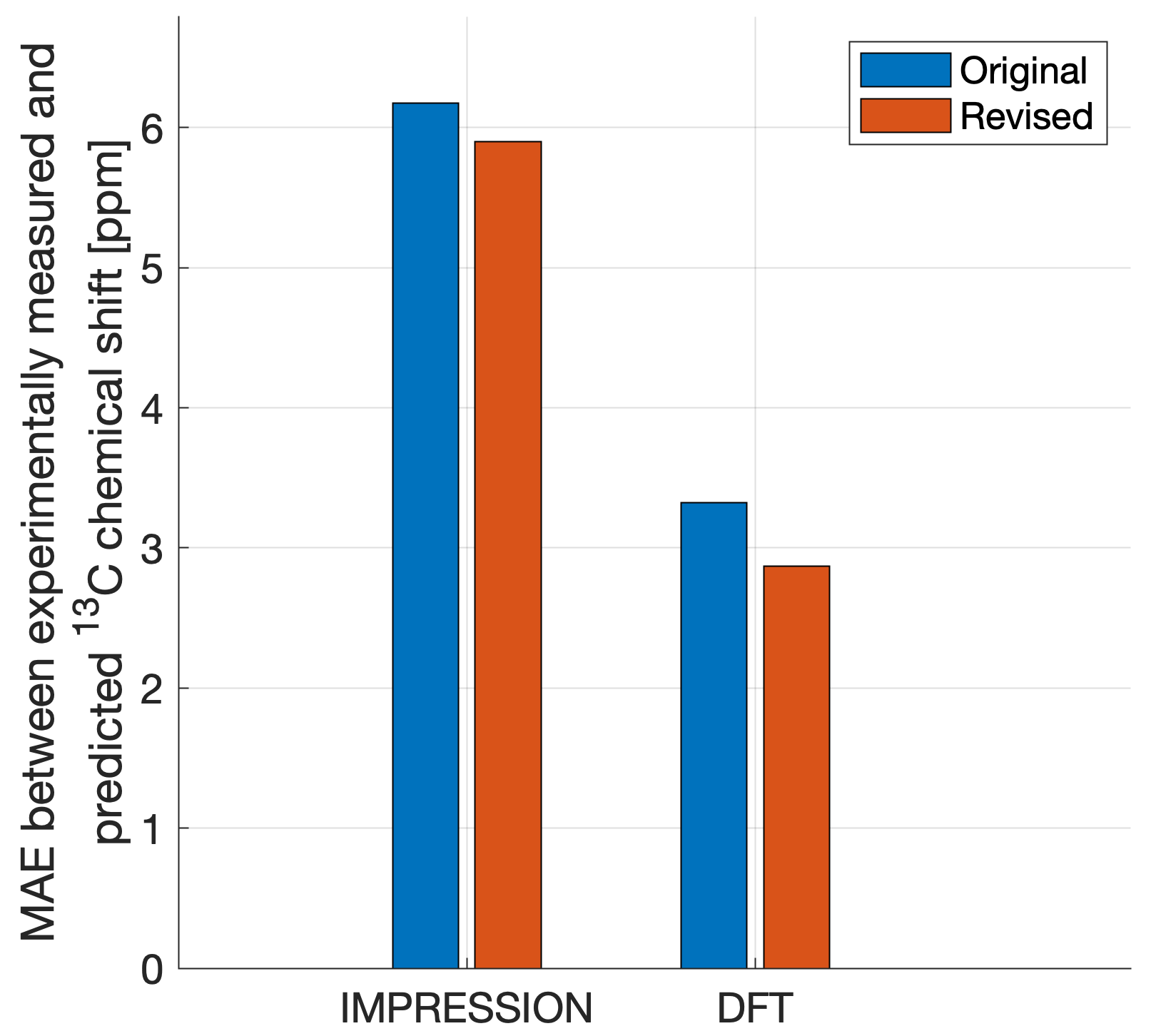}}
    \subfloat[]{\includegraphics[width=0.3\linewidth]{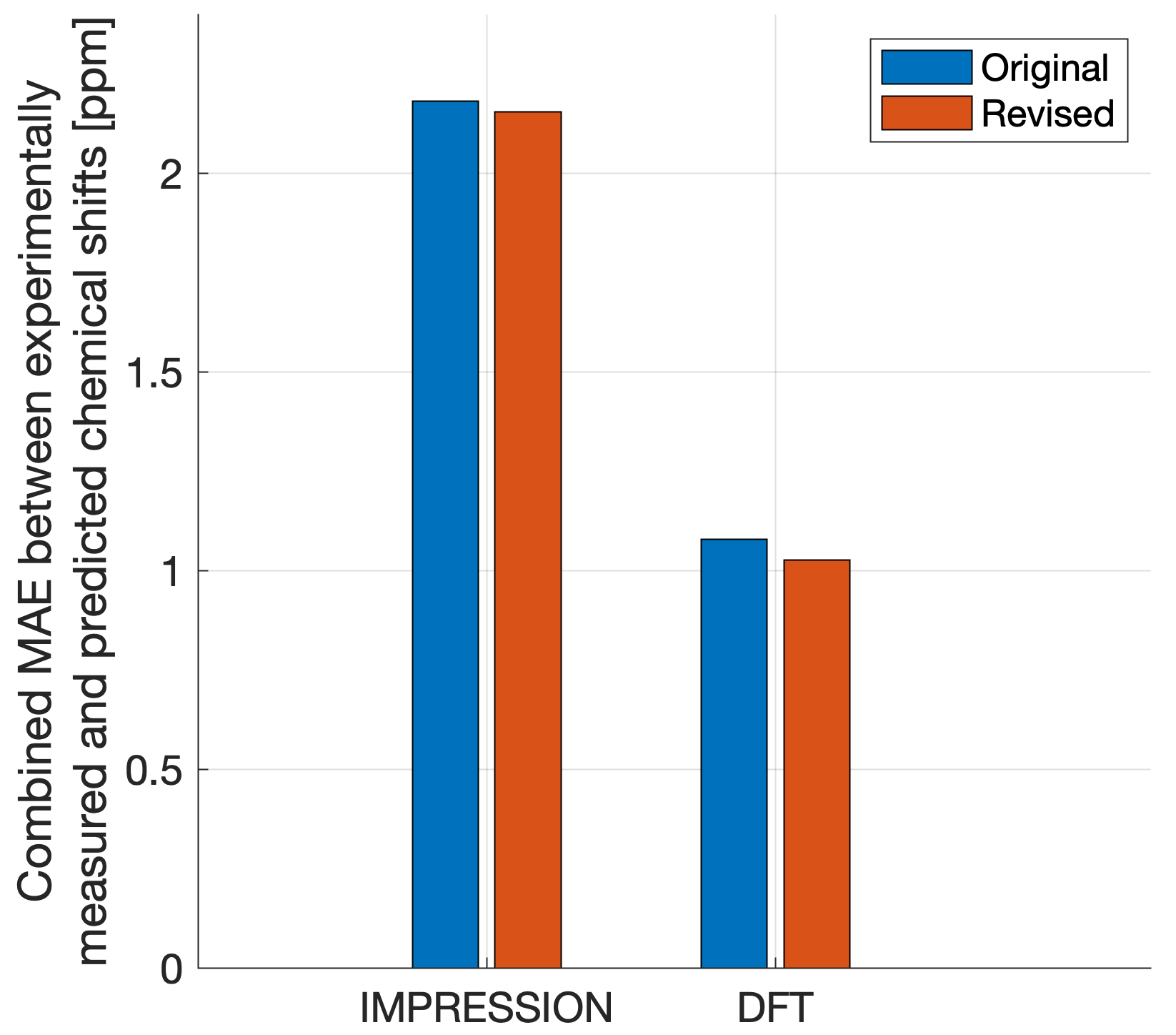}}
    
    \caption{Change in the fit between prediction and experiment for both DFT and IMPRESSION for Toluene Dioxide. a) $\delta ^1\mathrm{H}$. b) $\delta ^{13}\mathrm{C}$. c) Geometric mean across both parameters.}
    \label{fig:toluenedioxide_bars}
\end{figure}

\FloatBarrier
\clearpage
\section{Diastereotopic proton assignment in Strychnine}
\label{sec:strych_protons}

Further analysis was performed to see if the $^1\mathrm{\textit{J}_{CH}}$ IMPRESSION predictions could be used to assign the diastereotopic protons in strychnine. The $^1\mathrm{\textit{J}_{CH}}$ values for the 3 sets of diastereotopic protons for diastereomer 1 were compared across the three data sources: IMPRESSION predictions, DFT calculations and experimental measurements.

\begin{figure}[h]
    \centering
       \captionsetup[subfigure]{labelformat=empty}
    \subfloat[]{\includegraphics[width=0.9\linewidth]{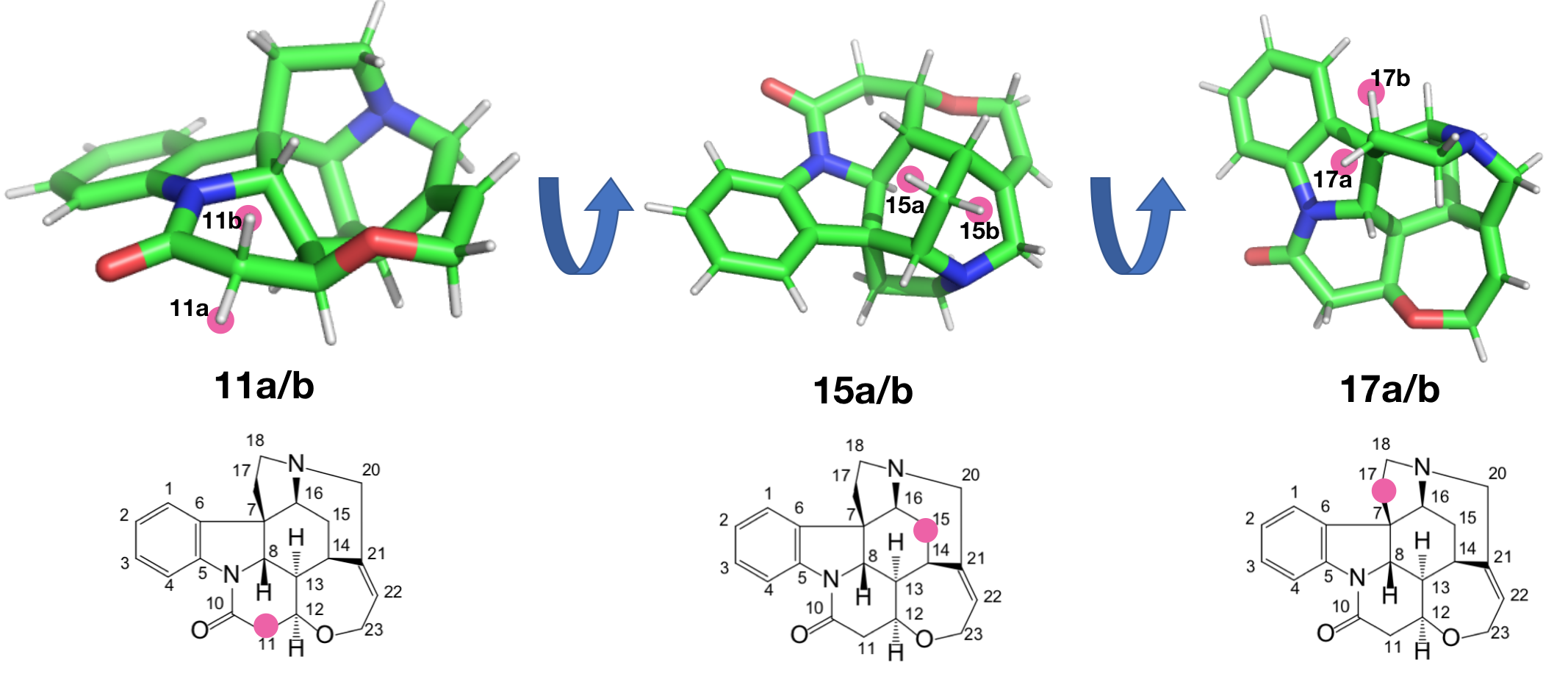}}\\
     \subfloat[]{\includegraphics[width=0.9\linewidth]{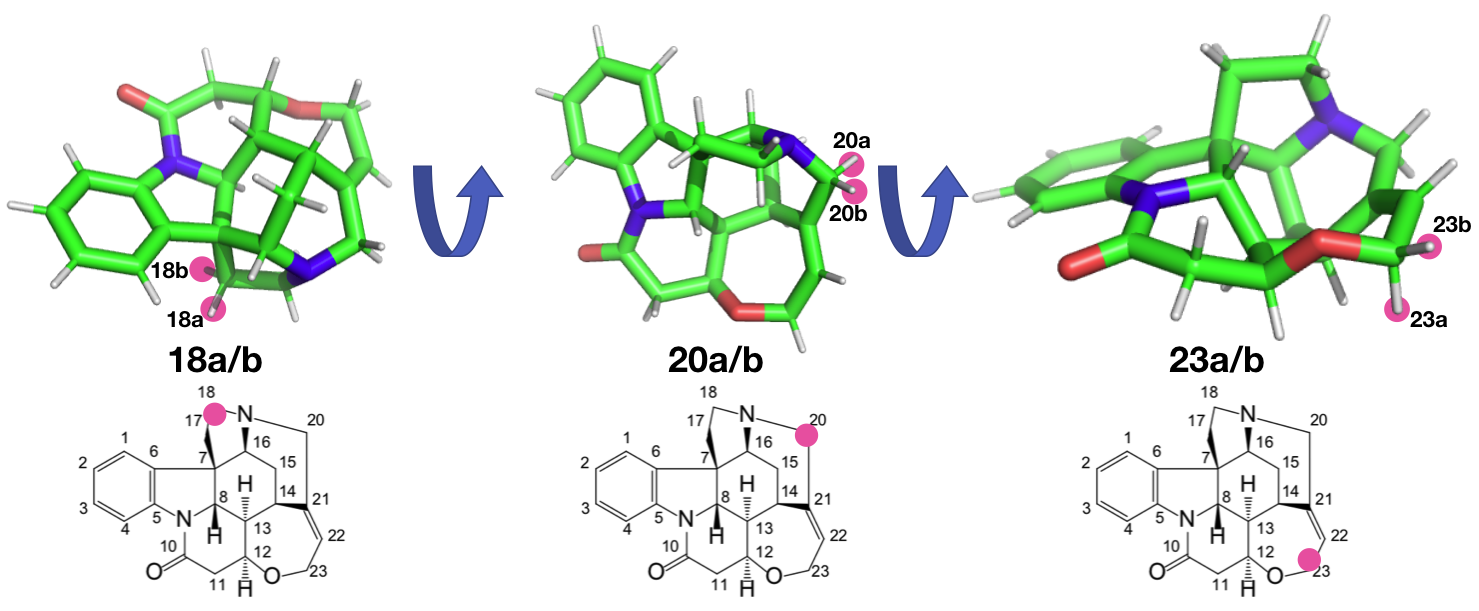}}
    \caption{Identification of the diastereotopic protons in Strychnine}
    \label{fig:strychnine_protons_diagram}
\end{figure}

\begin{figure}[h]
    \centering
    \subfloat[\label{fig:11ab}]{\includegraphics[width=0.3\linewidth]{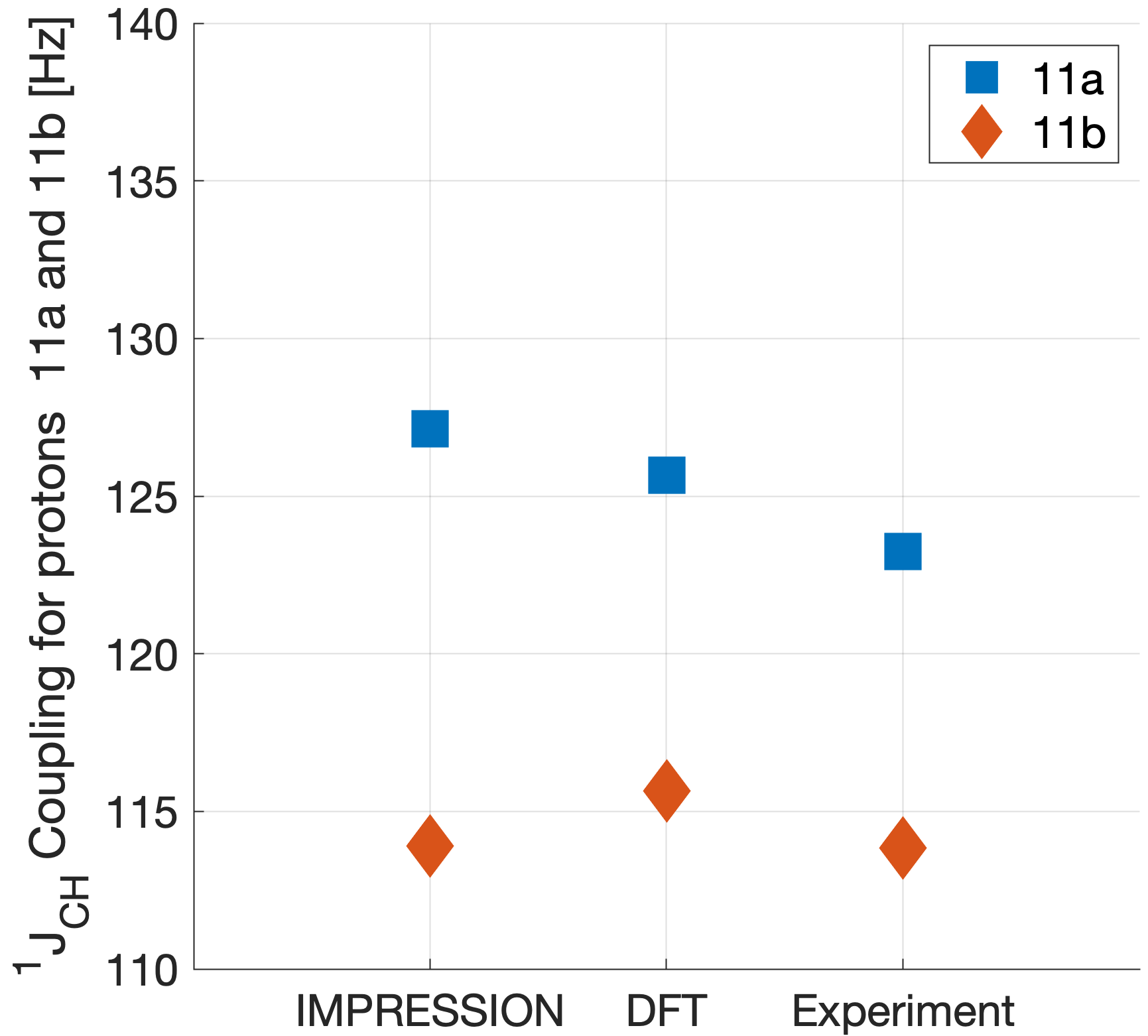}}
    \subfloat[\label{fig:18ab}]{\includegraphics[width=0.3\linewidth]{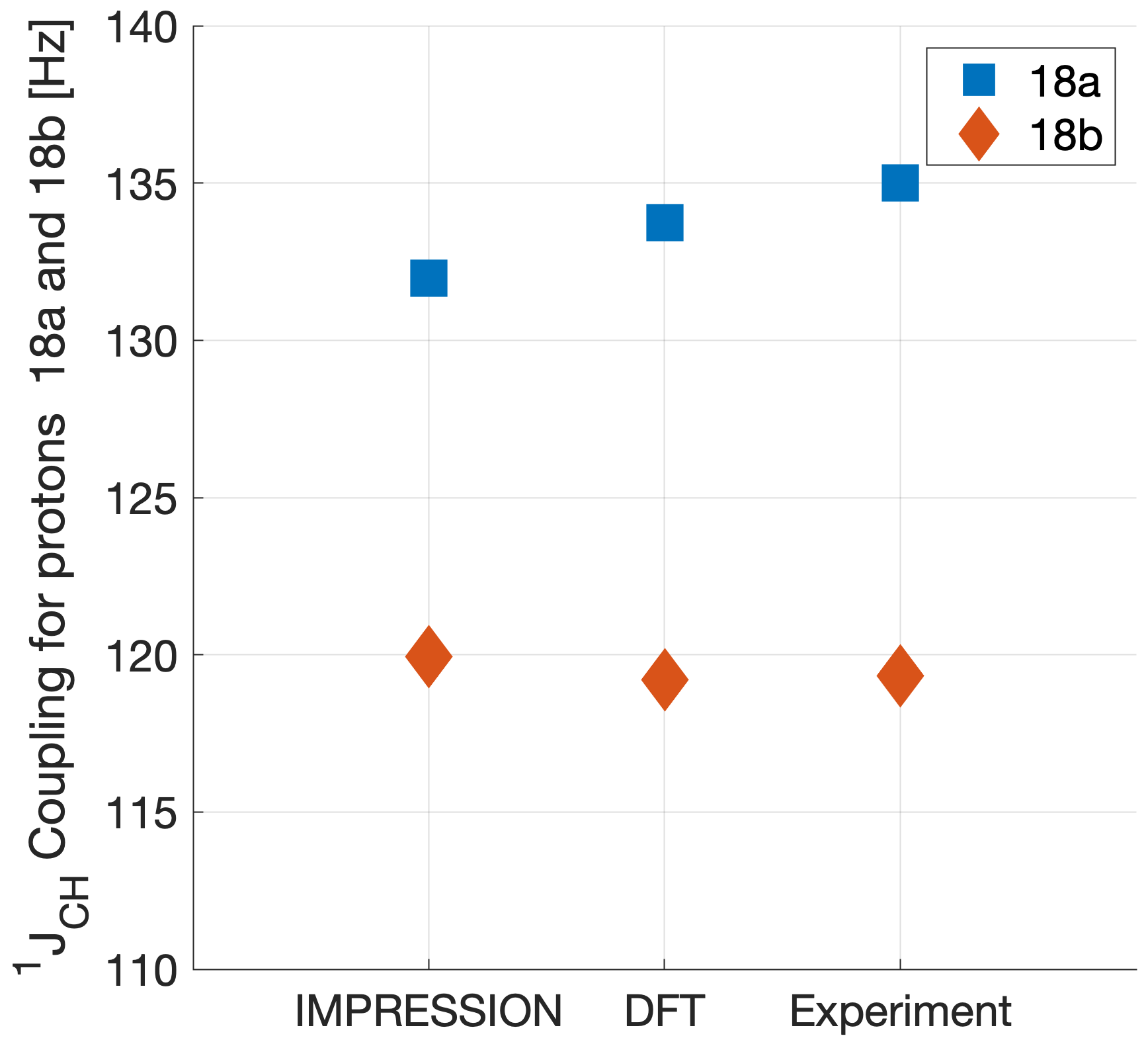}}
    \subfloat[\label{fig:23ab}]{\includegraphics[width=0.3\linewidth]{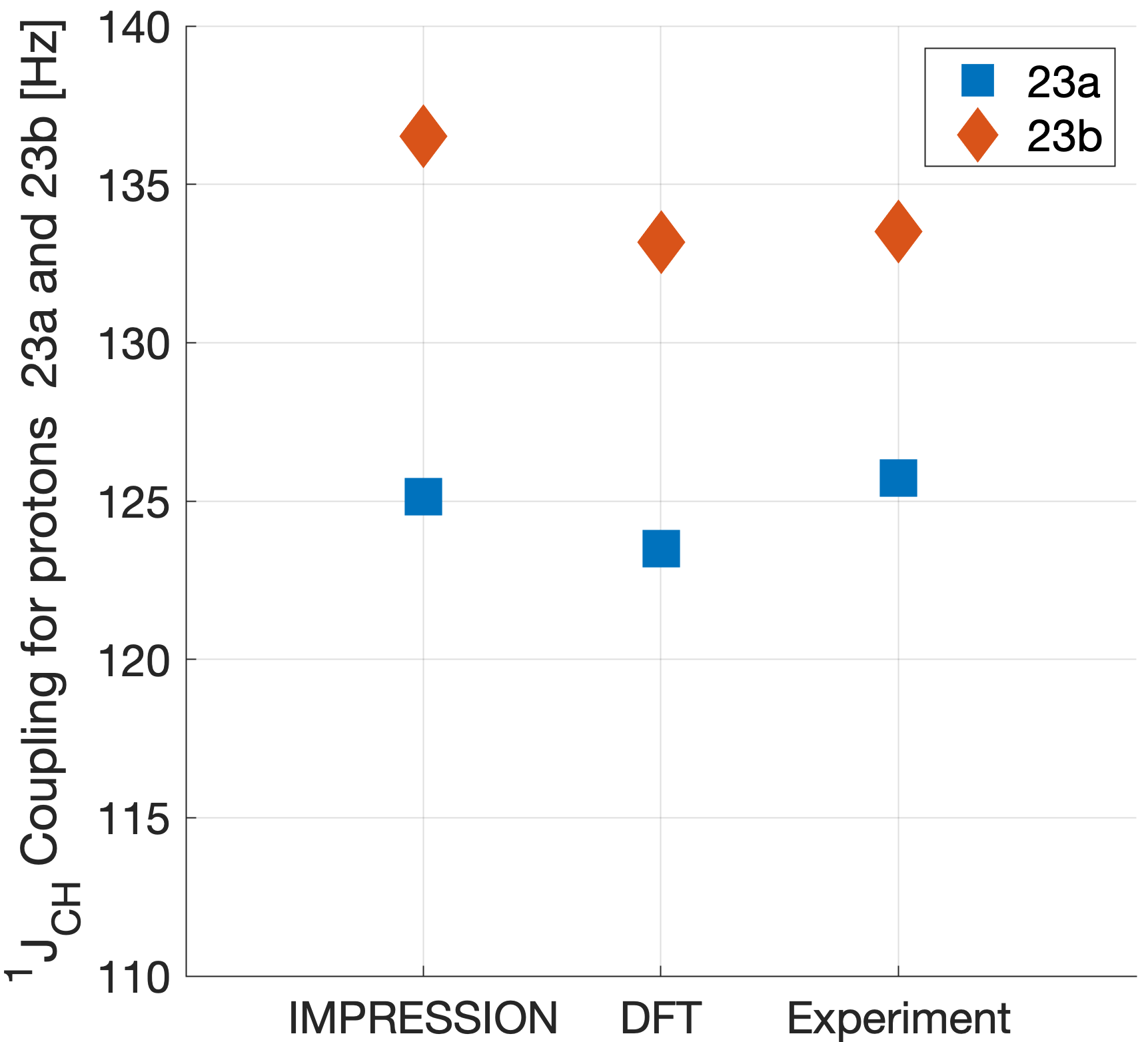}}
    \caption{Comparison of $^1\mathrm{\textit{J}_{CH}}$ values across all data sources for diastereotopic protons showing significant experimental difference. a) 11a/b. b)18a/b. c)23a/b.}
    \label{fig:protons1}
\end{figure}
\begin{figure}[h]
    \centering
    \subfloat[\label{fig:15ab}]{\includegraphics[width=0.3\linewidth]{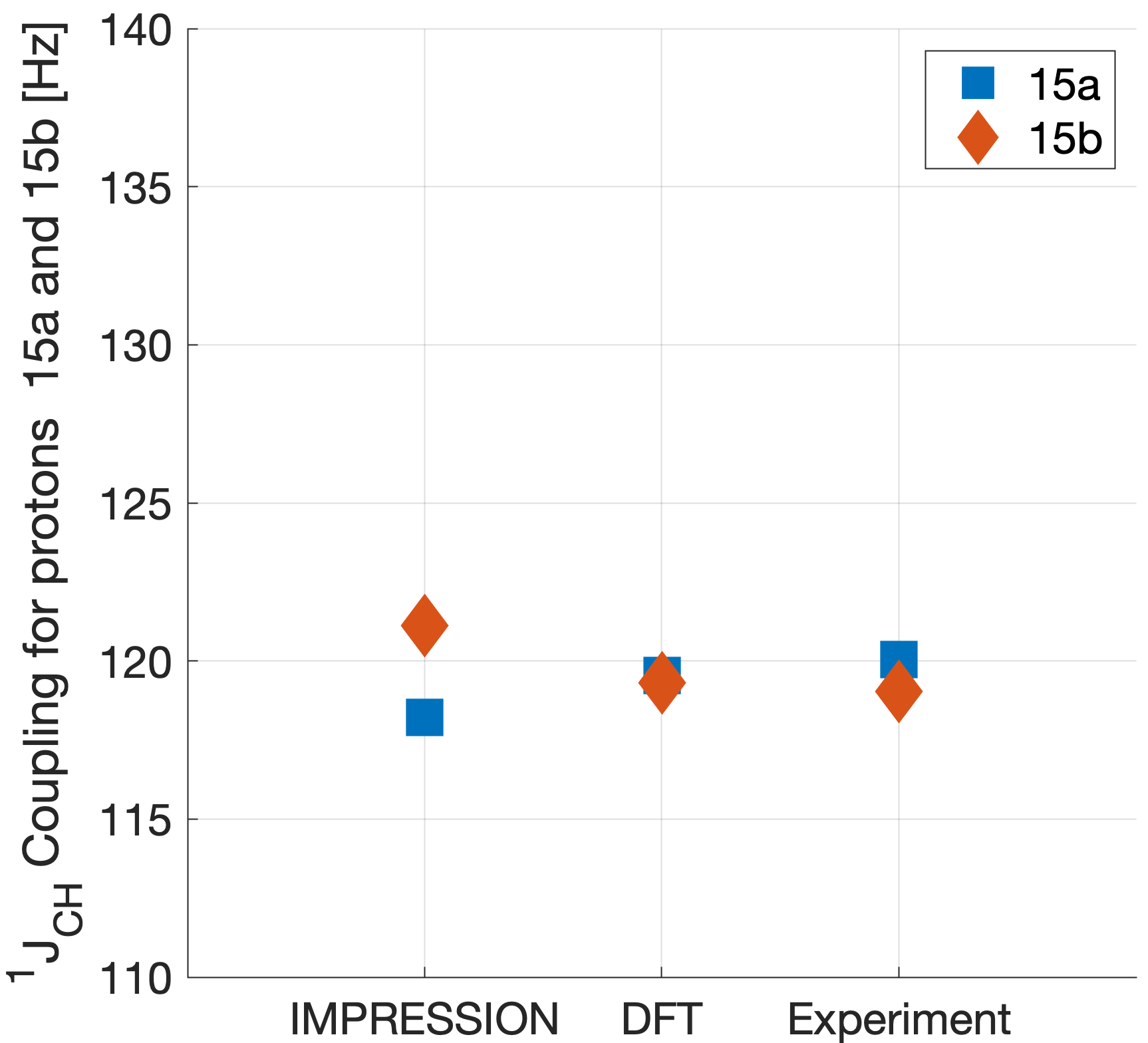}}
    \subfloat[\label{fig:17ab}]{\includegraphics[width=0.3\linewidth]{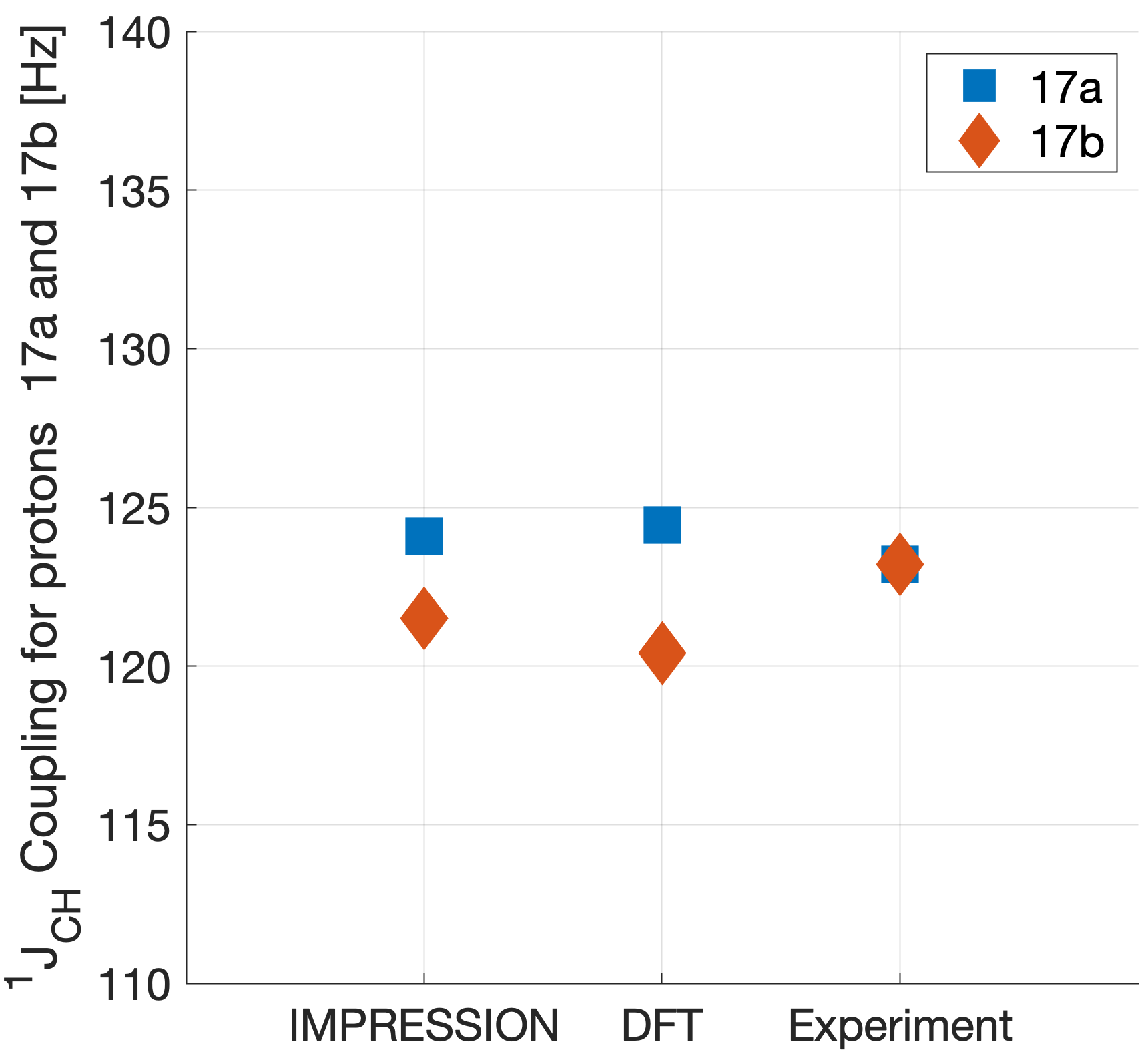}}
    \subfloat[\label{fig:20ab}]{\includegraphics[width=0.3\linewidth]{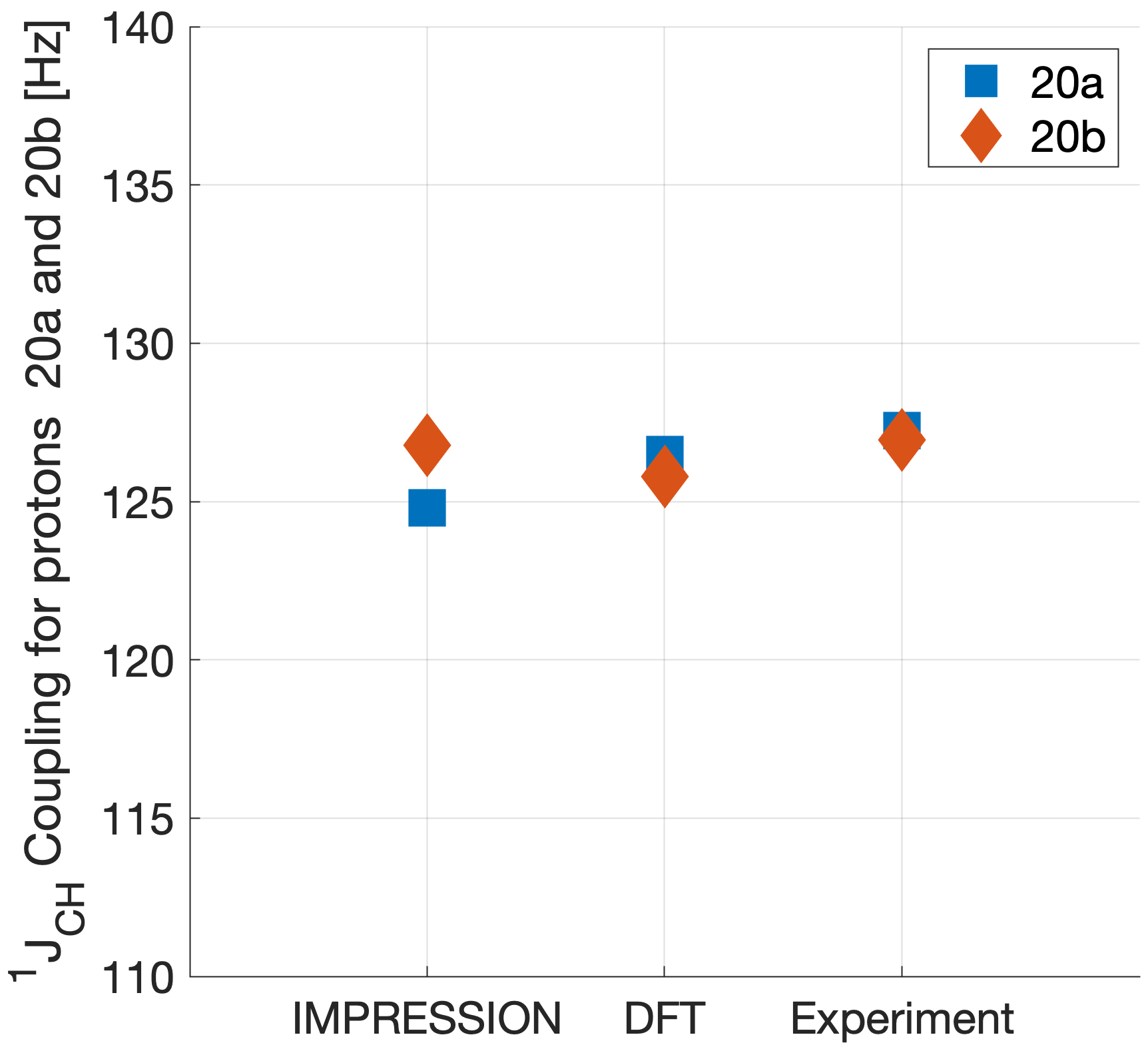}}
    \caption{Comparison of $^1\mathrm{\textit{J}_{CH}}$ values across all data sources for diastereotopic protons showing small experimental difference. a) 15a/b. b) 17a/b. c) 20a/b.}
    \label{fig:protons2}
\end{figure}

The results show that in cases where there is the three methylenes were there is any significant (>2Hz) difference in the experimental $^1\mathrm{\textit{J}_{CH}}$ values (figures \ref{fig:11ab}, \ref{fig:18ab}, \ref{fig:23ab}) the DFT method and IMPRESSION predictions can distinguish between the diastereotopic protons and correctly assign them. Where the difference in experimental values is small (Figure \ref{fig:protons2}, both DFT and IMPRESSION are not reliable for assignment.  

\FloatBarrier
\clearpage
\section{Strychnine diastereomers}
\label{sec:strych_SI}
MAE values for each of the parameters across each of the 14 strychnine structures compared to experiment:
\begin{center}
\begin{tabular}{|c|c|c|c|}
    \hline
    Structure & Parameters & MAE [IMP-EXP] & MAE [DFT-EXP] \\
    \hline
     1a & $\delta ^1\mathrm{H}$ & 0.28ppm & 0.25ppm \\
     1b & $\delta ^1\mathrm{H}$ & 0.73ppm & 0.56ppm \\
     2 & $\delta ^1\mathrm{H}$ & 0.23ppm & 0.19ppm \\
     3 & $\delta ^1\mathrm{H}$ & 0.37ppm & 0.35ppm \\
     4 & $\delta ^1\mathrm{H}$ & 0.69ppm & 0.52ppm \\
     5 & $\delta ^1\mathrm{H}$ & 0.44ppm & 0.32ppm \\
     6 & $\delta ^1\mathrm{H}$ & 0.42ppm & 0.35ppm \\
     7 & $\delta ^1\mathrm{H}$ & 0.77ppm & 0.54ppm \\
     8 & $\delta ^1\mathrm{H}$ & 0.71ppm & 0.53ppm \\
     9 & $\delta ^1\mathrm{H}$ & 0.70ppm & 0.66ppm \\
     10 & $\delta ^1\mathrm{H}$ & 0.48ppm & 0.49ppm \\
     11 & $\delta ^1\mathrm{H}$ & 0.55ppm & 0.49ppm \\
     12 & $\delta ^1\mathrm{H}$ & 0.57ppm & 0.36ppm \\
     13 & $\delta ^1\mathrm{H}$ & 0.73ppm & 0.45ppm \\
     \hline
\end{tabular}
\end{center}

\begin{center}
\begin{tabular}{|c|c|c|c|}
    \hline
    Structure & Parameters & MAE [IMP-EXP] & MAE [DFT-EXP] \\
    \hline
        1a & $\delta ^{13}\mathrm{C}$ &	2.26    &  	1.87 \\
        1b & $\delta ^{13}\mathrm{C}$ &	4.54    &  	3.84  \\   	
        2 & $\delta ^{13}\mathrm{C}$ &	2.75    &  	2.94  \\    	
        3 & $\delta ^{13}\mathrm{C}$ &	4.34    &  	3.98   \\   	
        4 & $\delta ^{13}\mathrm{C}$ &	7.69    &  	7.11\\      	
        5 & $\delta ^{13}\mathrm{C}$ &	3.44    &  	2.99  \\    	
        6 & $\delta ^{13}\mathrm{C}$ &	4.77    &  	3.63   \\   	
        7 & $\delta ^{13}\mathrm{C}$ &	7.77    &  	8.09 \\     	
        8 & $\delta ^{13}\mathrm{C}$ &	7.21    &  	7.41   \\   	
        9 & $\delta ^{13}\mathrm{C}$ &	8.47    &  	7.73     \\ 	
        10 & $\delta ^{13}\mathrm{C}$ &	3.98    &  	3.98      \\	
        11 & $\delta ^{13}\mathrm{C}$ &	4.80    &  	4.22      	\\
        12 & $\delta ^{13}\mathrm{C}$ &	5.05    &  	3.79      	\\
        13 & $\delta ^{13}\mathrm{C}$ &	5.53    &  	3.84      	\\
    \hline
     \hline
\end{tabular}
\end{center}

\begin{center}
\begin{tabular}{|c|c|c|c|}
    \hline
    Structure & Parameters & MAE [IMP-EXP] & MAE [DFT-EXP] \\
    \hline
     1a & $^1\mathrm{\textit{J}_{CH}}$ & 1.83Hz & 1.29Hz \\
     1b & $^1\mathrm{\textit{J}_{CH}}$ & 4.11Hz & 4.06Hz \\
     2 & $^1\mathrm{\textit{J}_{CH}}$ & 2.98Hz & 2.22Hz \\
     3 & $^1\mathrm{\textit{J}_{CH}}$ & 3.72Hz & 3.64Hz \\
     4 & $^1\mathrm{\textit{J}_{CH}}$ & 5.34Hz & 4.94Hz \\
     5 & $^1\mathrm{\textit{J}_{CH}}$ & 3.53Hz & 2.98Hz \\
     6 & $^1\mathrm{\textit{J}_{CH}}$ & 2.26Hz & 1.83Hz \\
     7 & $^1\mathrm{\textit{J}_{CH}}$ & 6.86Hz & 4.45Hz \\
     8 & $^1\mathrm{\textit{J}_{CH}}$ & 5.08Hz & 4.73Hz \\
     9 & $^1\mathrm{\textit{J}_{CH}}$ & 8.12Hz & 5.24Hz \\
     10 & $^1\mathrm{\textit{J}_{CH}}$ & 3.21Hz & 3.04Hz \\
     11 & $^1\mathrm{\textit{J}_{CH}}$ & 2.78Hz & 3.26Hz \\
     12 & $^1\mathrm{\textit{J}_{CH}}$ & 4.34Hz & 4.26Hz \\
     13 & $^1\mathrm{\textit{J}_{CH}}$ & 3.32Hz & 3.58Hz \\
     \hline
\end{tabular}
\end{center}

\begin{center}
\begin{tabular}{|c|c|c|c|}
    \hline
    Structure & Parameters & MAE [IMP-EXP] & MAE [DFT-EXP] \\
    \hline
     1a &  $\delta ^1\mathrm{H}$ + $\delta ^{13}\mathrm{C}$ & 0.80ppm & 0.68ppm \\
     1b &$\delta ^1\mathrm{H}$ + $\delta ^{13}\mathrm{C}$ & 1.83ppm & 1.47ppm \\
     2 &$\delta ^1\mathrm{H}$ + $\delta ^{13}\mathrm{C}$ & 0.80ppm & 0.75ppm \\
     3 &$\delta ^1\mathrm{H}$ + $\delta ^{13}\mathrm{C}$ & 1.27ppm & 1.18ppm \\
     4 &$\delta ^1\mathrm{H}$ + $\delta ^{13}\mathrm{C}$ & 2.30ppm & 1.92ppm \\
     5 &$\delta ^1\mathrm{H}$ + $\delta ^{13}\mathrm{C}$ & 1.24ppm & 0.98ppm \\
     6 &$\delta ^1\mathrm{H}$ + $\delta ^{13}\mathrm{C}$ & 1.41ppm & 1.13ppm \\
     7 &$\delta ^1\mathrm{H}$ + $\delta ^{13}\mathrm{C}$ & 2.45ppm & 2.09ppm \\
     8 &$\delta ^1\mathrm{H}$ + $\delta ^{13}\mathrm{C}$ & 2.27ppm & 1.99ppm \\
     9 &$\delta ^1\mathrm{H}$ + $\delta ^{13}\mathrm{C}$ & 2.43ppm & 2.25ppm \\
     10 &$\delta ^1\mathrm{H}$ + $\delta ^{13}\mathrm{C}$ & 1.39ppm & 1.40ppm \\
     11 &$\delta ^1\mathrm{H}$ + $\delta ^{13}\mathrm{C}$ & 1.62ppm & 1.43ppm \\
     12 &$\delta ^1\mathrm{H}$ + $\delta ^{13}\mathrm{C}$ & 1.58ppm & 1.17ppm \\
     13 &$\delta ^1\mathrm{H}$ + $\delta ^{13}\mathrm{C}$ & 1.78ppm & 1.32ppm \\
     \hline
\end{tabular}
\end{center}

\begin{center}
\begin{tabular}{|c|c|c|c|}
    \hline
    Structure & Parameters & MAE [IMP-EXP] & MAE [DFT-EXP] \\
    \hline
     1a &$\delta ^1\mathrm{H}$ + $\delta ^{13}\mathrm{C}$ + $^1\mathrm{\textit{J}_{CH}}$ & 1.05 & 0.84 \\
     1b &$\delta ^1\mathrm{H}$ + $\delta ^{13}\mathrm{C}$ + $^1\mathrm{\textit{J}_{CH}}$ & 2.39 & 2.06 \\
     2 & $\delta ^1\mathrm{H}$ + $\delta ^{13}\mathrm{C}$ + $^1\mathrm{\textit{J}_{CH}}$ & 1.24 & 1.08 \\
     3 & $\delta ^1\mathrm{H}$ + $\delta ^{13}\mathrm{C}$ + $^1\mathrm{\textit{J}_{CH}}$ & 1.82 & 1.72 \\
     4 & $\delta ^1\mathrm{H}$ + $\delta ^{13}\mathrm{C}$ + $^1\mathrm{\textit{J}_{CH}}$ & 3.04 & 2.63 \\
     5 & $\delta ^1\mathrm{H}$ + $\delta ^{13}\mathrm{C}$ + $^1\mathrm{\textit{J}_{CH}}$ & 1.75 & 1.42 \\
     6 & $\delta ^1\mathrm{H}$ + $\delta ^{13}\mathrm{C}$ + $^1\mathrm{\textit{J}_{CH}}$ & 1.65 & 1.33 \\
     7 & $\delta ^1\mathrm{H}$ + $\delta ^{13}\mathrm{C}$ + $^1\mathrm{\textit{J}_{CH}}$ & 3.45 & 2.69 \\
     8 & $\delta ^1\mathrm{H}$ + $\delta ^{13}\mathrm{C}$ + $^1\mathrm{\textit{J}_{CH}}$ & 2.97 & 2.65 \\
     9 & $\delta ^1\mathrm{H}$ + $\delta ^{13}\mathrm{C}$ + $^1\mathrm{\textit{J}_{CH}}$ & 3.64 & 2.98 \\
     10 &$\delta ^1\mathrm{H}$ + $\delta ^{13}\mathrm{C}$ + $^1\mathrm{\textit{J}_{CH}}$ & 1.84 & 1.81 \\
     11 &$\delta ^1\mathrm{H}$ + $\delta ^{13}\mathrm{C}$ + $^1\mathrm{\textit{J}_{CH}}$ & 1.94 & 1.89 \\
     12 &$\delta ^1\mathrm{H}$ + $\delta ^{13}\mathrm{C}$ + $^1\mathrm{\textit{J}_{CH}}$ & 2.21 & 1.80 \\
     13 &$\delta ^1\mathrm{H}$ + $\delta ^{13}\mathrm{C}$ + $^1\mathrm{\textit{J}_{CH}}$ & 2.19 & 1.84 \\
     \hline
\end{tabular}
\end{center}

\begin{figure}[!htb]
    \centering
    \includegraphics[width=0.8\linewidth]{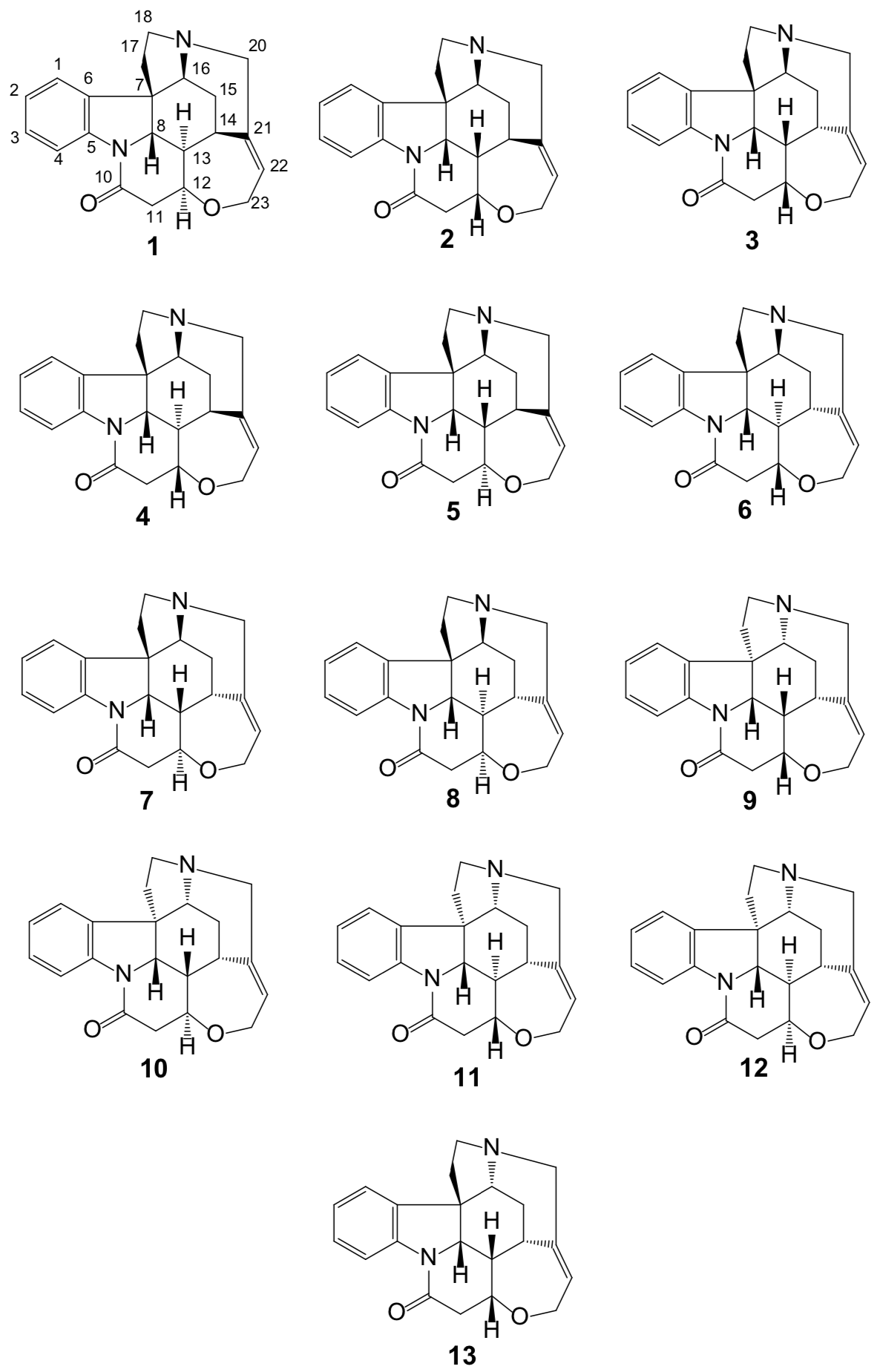}
    \caption{The 13 Energetically viable Strychnine diastereomers used for the IMPRESSION validation\cite{buevich2019enhancing}}
    \label{fig:strychnine_diasteromers}
\end{figure}

\newpage
\FloatBarrier
\clearpage
\section{Large errors}
\label{sec:big_errors}
The largest 5 individual errors between DFT and machine learning for the test set are shown here, for each of the parameters $\delta ^{13}\mathrm{C}$, $\delta ^1\mathrm{H}$, and $^1\mathrm{\textit{J}_{CH}}$.

\subsection{$^1\mathrm{H}$ chemical shifts}
\begin{center}
\begin{tabular}{|c|c|c|c|c|c|}
    \hline
    Mol ID & Atom ID & Error [ppm] & DFT [ppm] & ML [ppm] & Variance [ppm] \\
    \hline
    YEHWUD & 36 & 11.22 & -4.27 & 6.96 & 0.63 \\
    BEDFUM & 6 & 3.22 & 3.21 & 6.42 & 0.18 \\
    IQIKOI & 21 & 2.15 & 5.34 & 7.50 & 0.0023 \\
    AROKUN & 19 & 2.01 & 6.83 & 8.84 & 0.0025 \\
    WAWQUH & 37 & 1.94 & 0.76 & 2.70 & 0.014 \\
    \hline
\end{tabular}
\end{center}

\begin{figure}[h]
    \centering
    \subfloat[YEHWUD\label{fig:big_YEHWUDhcs}]{\includegraphics[width=0.3\linewidth]{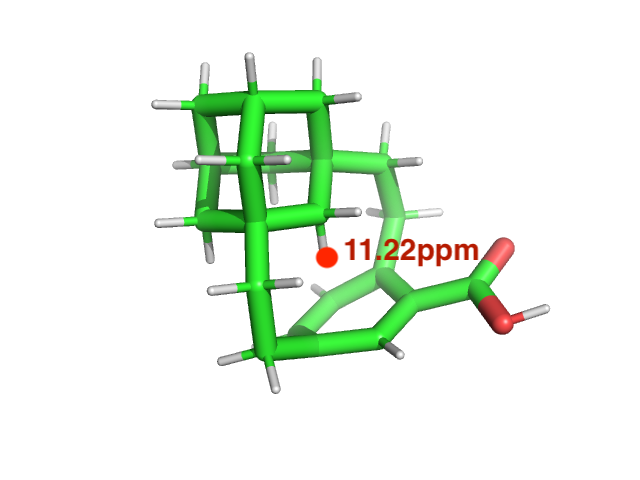}}
    \subfloat[BEDFUM\label{fig:big_BEDFUM}]{\includegraphics[width=0.3\linewidth]{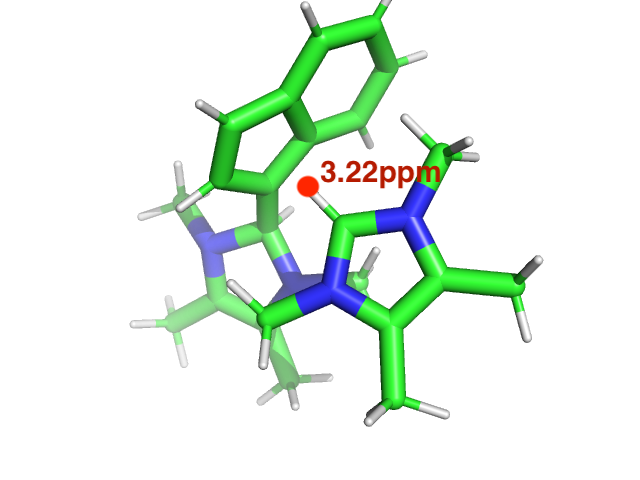}}
    \subfloat[IQIKOI\label{fig:big_IQIKOI}]{\includegraphics[width=0.3\linewidth]{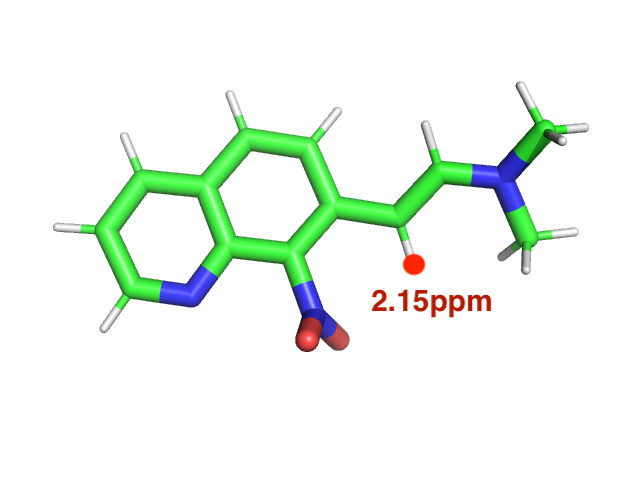}} \\
    \subfloat[AROKUN\label{fig:big_AROKUN}]{\includegraphics[width=0.3\linewidth]{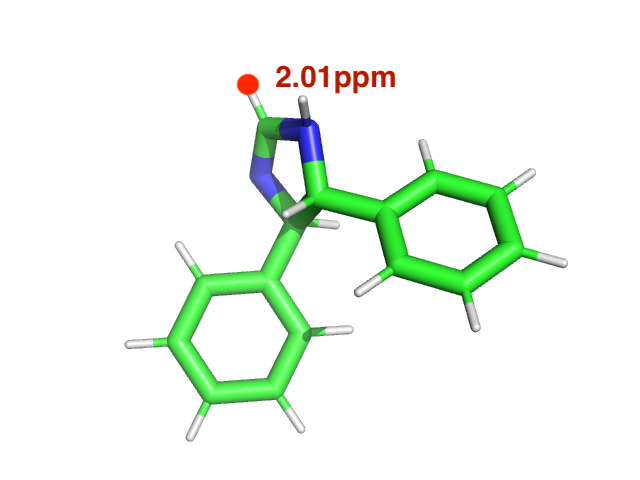}}
    \subfloat[WAWQUH\label{fig:big_WAWQUH}]{\includegraphics[width=0.3\linewidth]{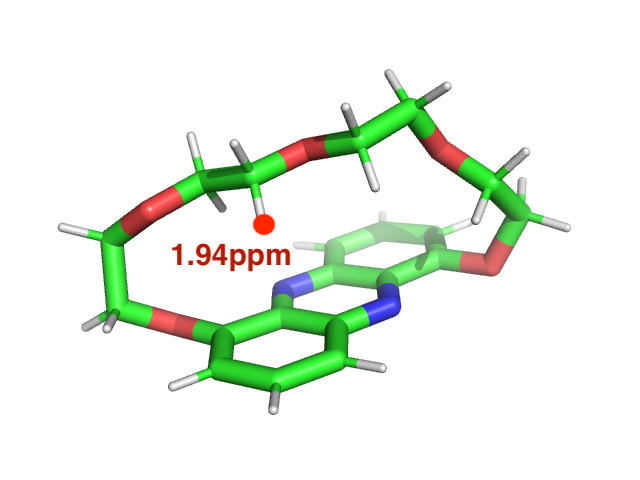}}
    \caption{Biggest errors in $^1\mathrm{H}$ prediction.}
    \label{fig:big_errs_hcs}
\end{figure}

\newpage
\FloatBarrier
\subsection{$^{13}\mathrm{C}$ chemical shifts}

\begin{center}
\begin{tabular}{|c|c|c|c|c|c|}
    \hline
    Mol ID & Atom ID & Error [ppm] & DFT [ppm] & ML [ppm] & Variance [ppm] \\
    \hline
    DOVWAM & 4 & -63.33 & 217.35 & 154.02 & 353.15 \\
    QUFCEZ & 15 & 42.92 & 60.82 & 103.75 & 45.22 \\
    RACGEJ & 10 & -37.87 & 180.27 & 142.40 & 2.02 \\
    BEHWER & 5 & 35.31 & 115.95 & 151.26 & 2.89 \\
    QOMVUK & 1 & 32.90 & 92.98 & 125.37 & 8.41 \\
    \hline
\end{tabular}
\end{center}

\begin{figure}[h]
    \centering
    \subfloat[DOVWAM\label{fig:big_DOVWAM}]{\includegraphics[width=0.3\linewidth]{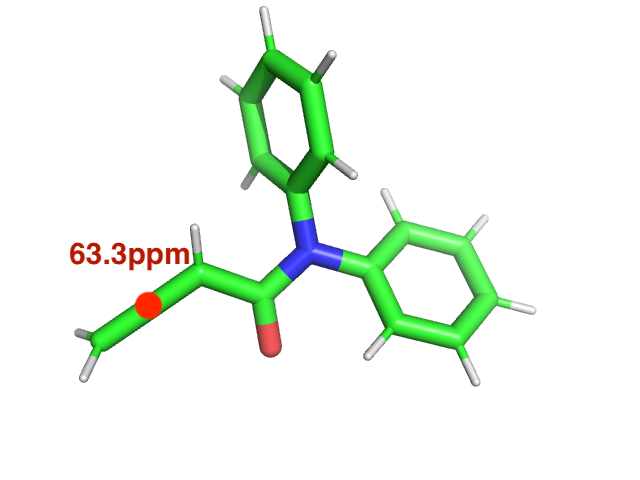}}
    \subfloat[QUFCEZ\label{fig:big_QUFCEZ}]{\includegraphics[width=0.3\linewidth]{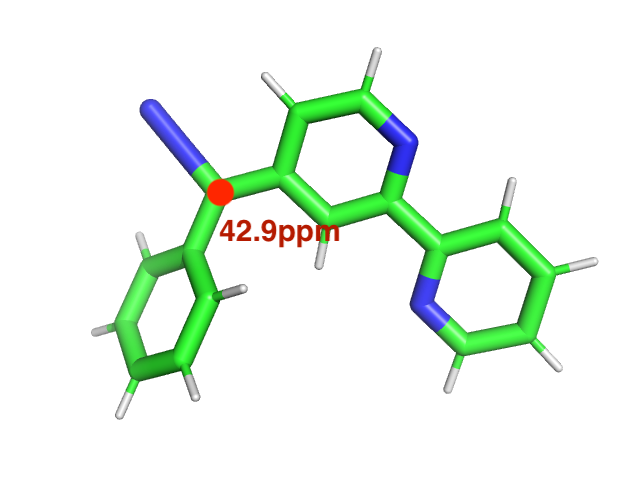}}
    \subfloat[RACGEJ\label{fig:big_RACGEJ}]{\includegraphics[width=0.3\linewidth]{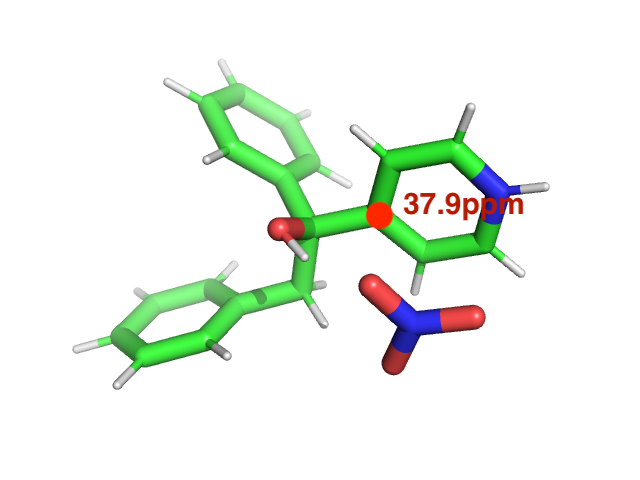}} \\
    \subfloat[BEHWER\label{fig:big_BEHWER}]{\includegraphics[width=0.3\linewidth]{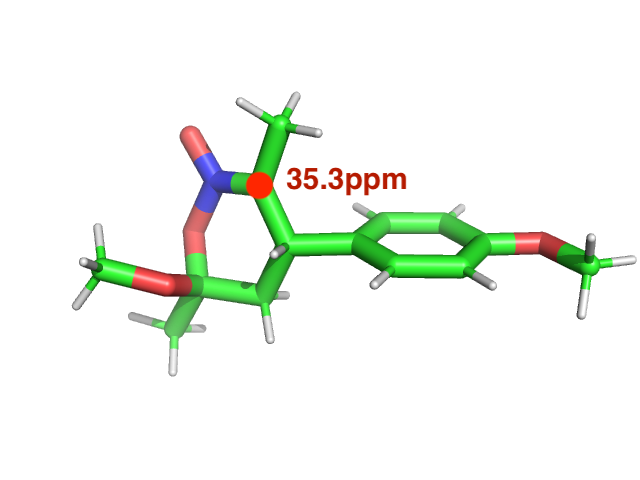}}
    \subfloat[QOMVUK\label{fig:big_QOMVUK}]{\includegraphics[width=0.3\linewidth]{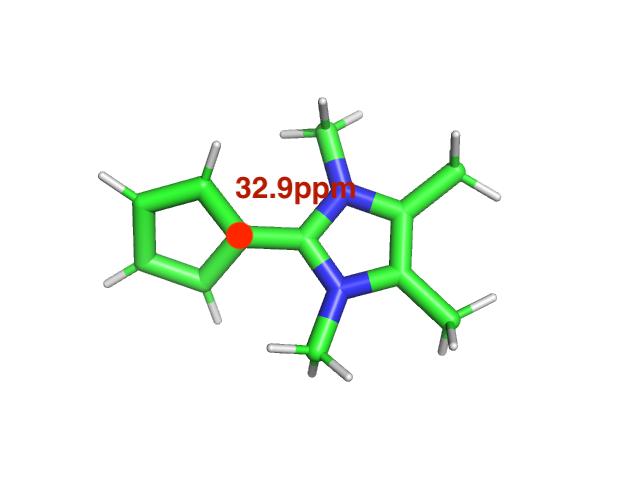}}
    \caption{Biggest errors in $^{13}\mathrm{C}$ prediction.}
    \label{fig:big_errs_ccs}
\end{figure}

\FloatBarrier
\newpage
\subsection{$^1\mathrm{\textit{J}_{CH}}$ coupling constant}

\begin{center}
\begin{tabular}{|c|c|c|c|c|c|c|}
    \hline
    Mol ID & $^1\mathrm{H}$ Atom ID & $^{13}\mathrm{C}$ Atom ID & Error [Hz] & DFT [Hz] & ML [Hz] & Variance [Hz] \\
    \hline
    YEHWUD & 10 & 36 & 24.63 & 116.31 & 140.94 & 365.70 \\
  JOTKIM01 & 50 & 51 & 24.40 & 194.51 & 218.91 & 8.52 \\
    ZEYLAS & 61 & 70 & -18.31 & 182.64 & 164.35 & 3.49 \\
    IDURIJ & 13 & 14 & -13.37 & 171.56 & 158.19 & 0.13 \\
     FEMXOK & 7 & 19 & 12.13 & 144.21 & 156.35 & 1.01 \\
    \hline
\end{tabular}
\end{center}

\begin{figure}[h]
    \centering
    \subfloat[YEHWUD\label{fig:big_YEHQUD1jch}]{\includegraphics[width=0.3\linewidth]{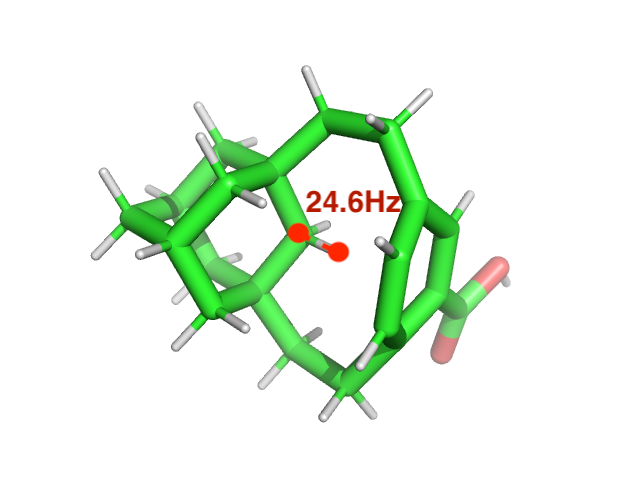}}
    \subfloat[JOTKIM01\label{fig:big_JOTKIM01}]{\includegraphics[width=0.3\linewidth]{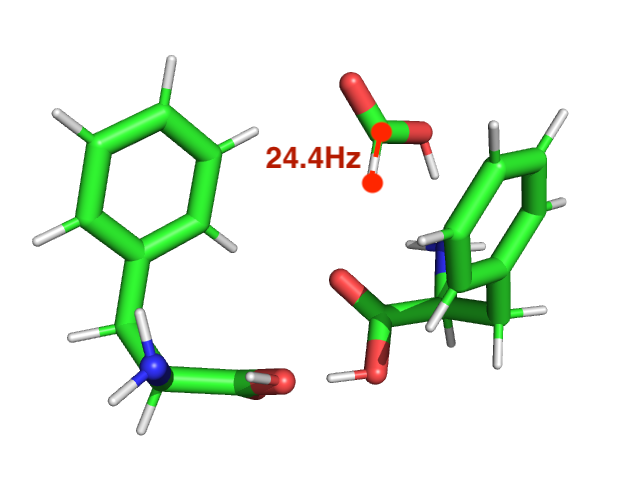}}
    \subfloat[ZEYLAS\label{fig:big_ZEYLAS}]{\includegraphics[width=0.3\linewidth]{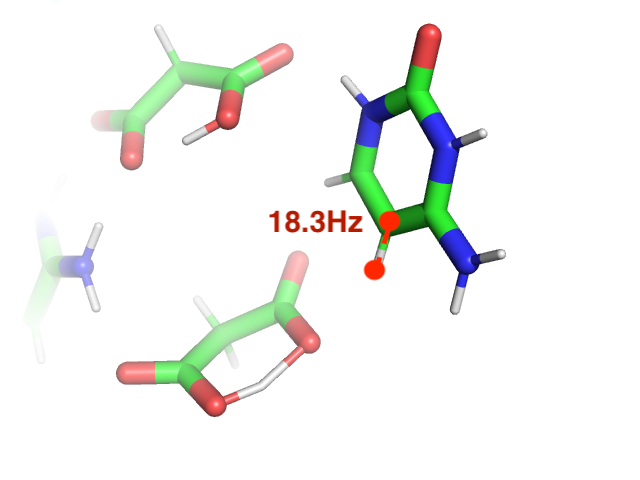}} \\
    \subfloat[IDURIJ\label{fig:big_IDURIJ}]{\includegraphics[width=0.3\linewidth]{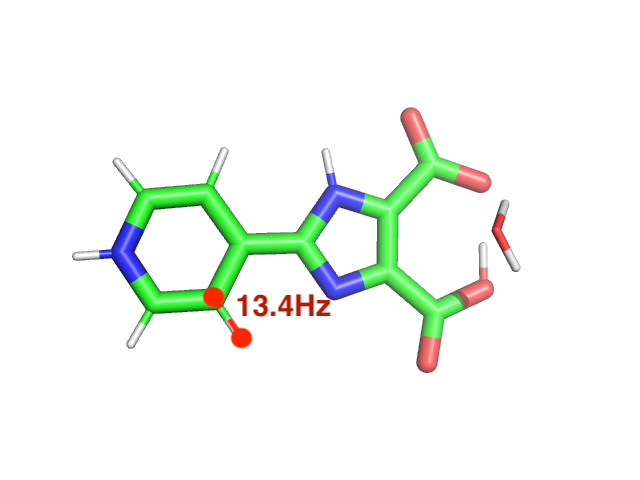}}
    \subfloat[FEMXOK\label{fig:big_FEMXOK}]{\includegraphics[width=0.3\linewidth]{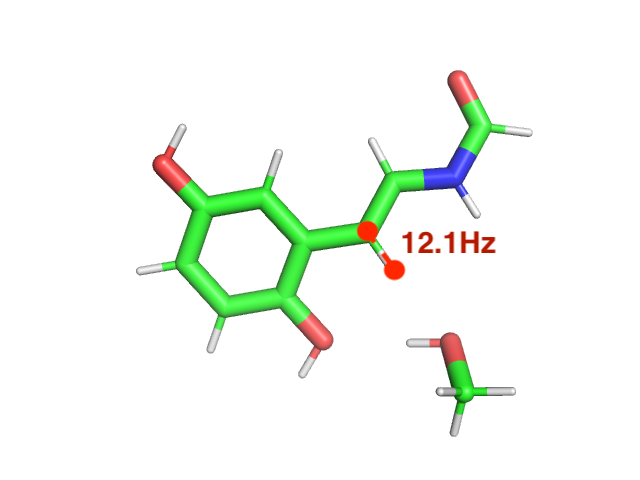}}
    \caption{Biggest errors in $^1\mathrm{\textit{J}_{CH}}$ prediction.}
    \label{fig:big_errs_1jch}
\end{figure}

\FloatBarrier
\newpage
\section{Gaussian input files}
Example input files for the Gaussian09 software are included here.

\begin{figure}[h]
    \centering
    \includegraphics[trim=0 13cm 0 2cm, clip, width=0.7\linewidth]{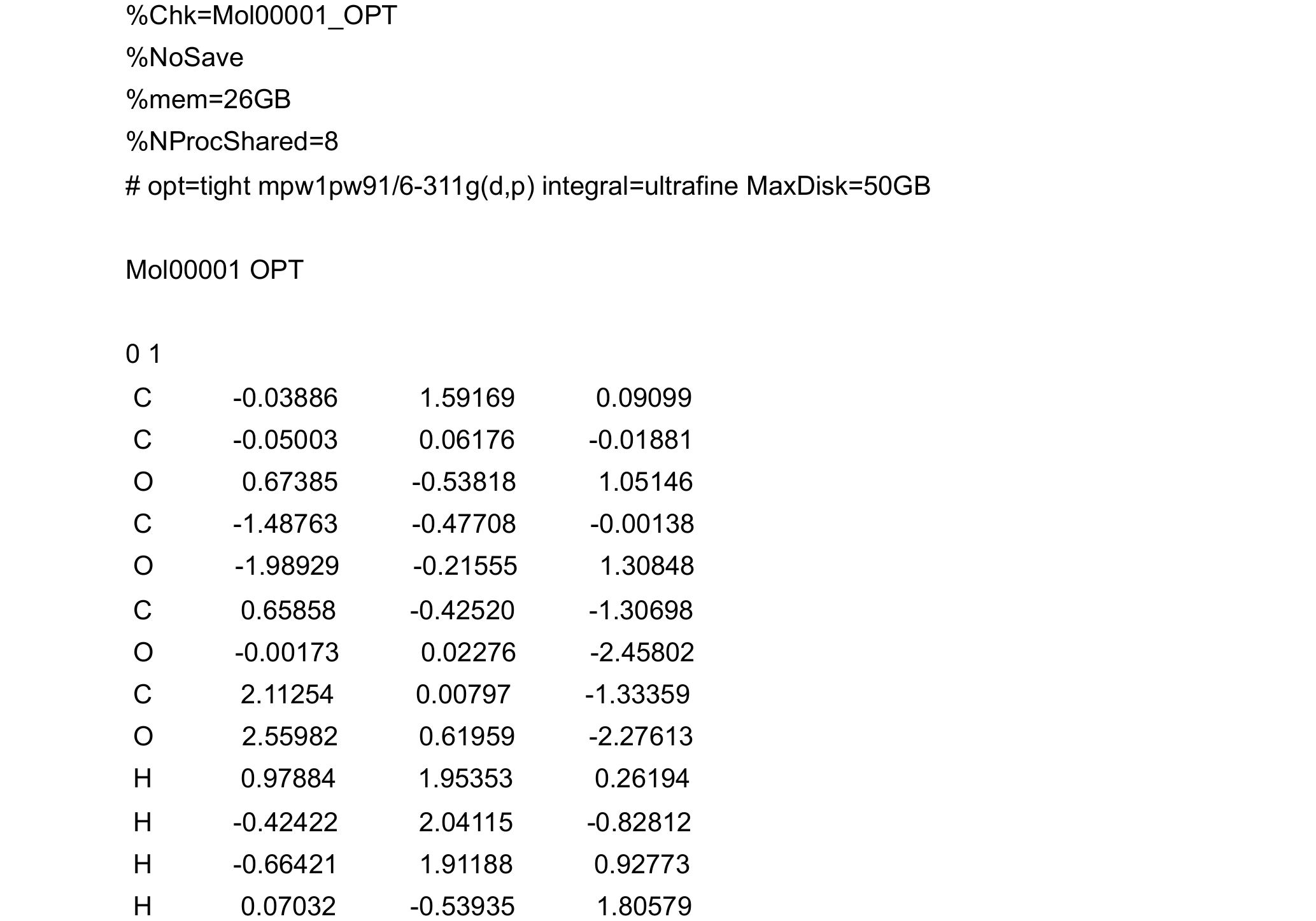}
    \caption{Gaussian Optimisation Input File}
    \label{fig:gauss_opt}
\end{figure}

\begin{figure}[h]
    \centering
    \includegraphics[trim=0 14cm 0 2cm,clip,width=0.7\linewidth]{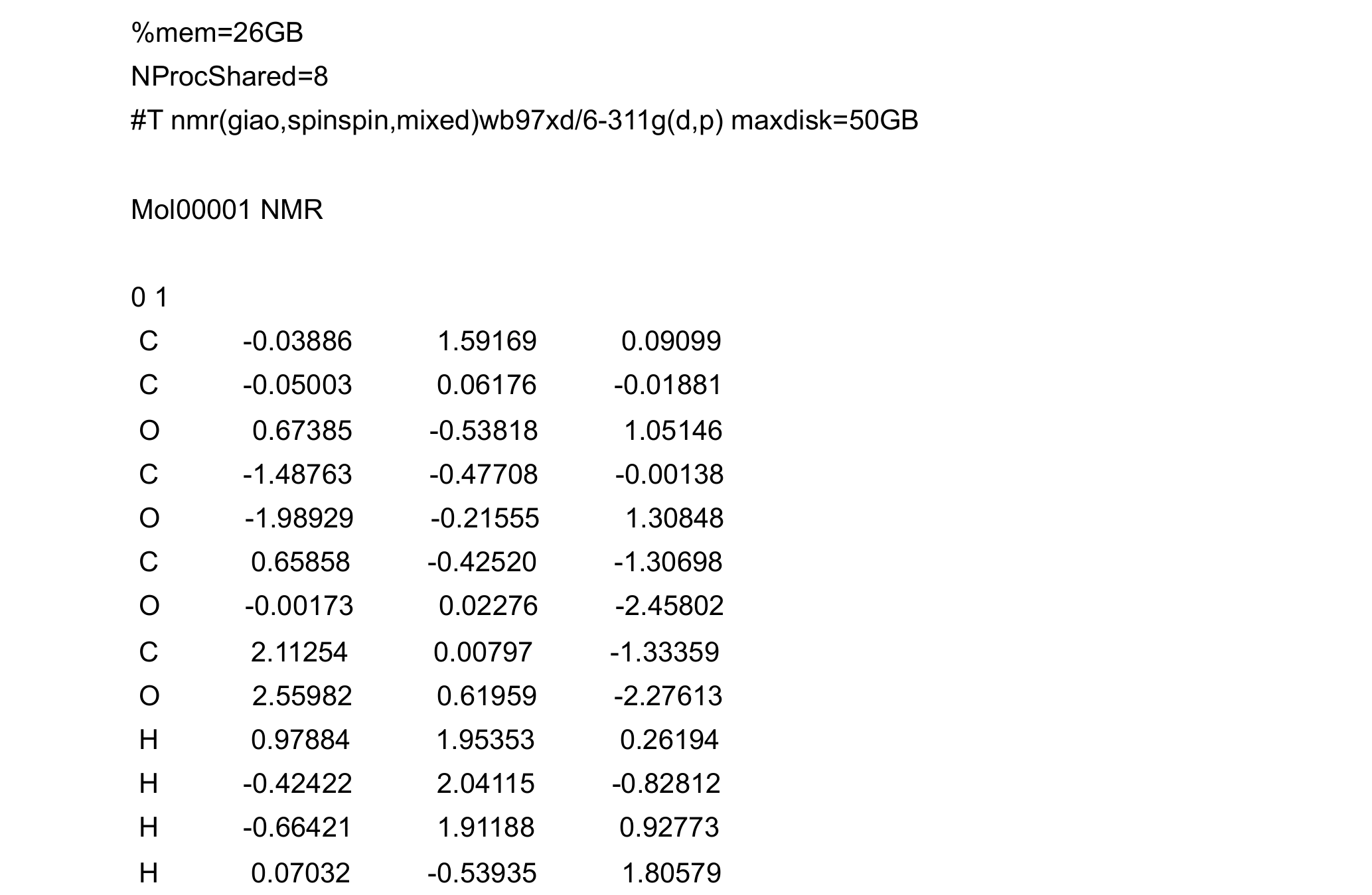}
    \caption{Gaussian NMR Calculation Input File}
    \label{fig:gauss_nmr}
\end{figure}

\FloatBarrier
\newpage
\section{CSD structures}
\begin{figure}
    \centering
    \includepdf[pages={1},pagecommand={},width=\textwidth]{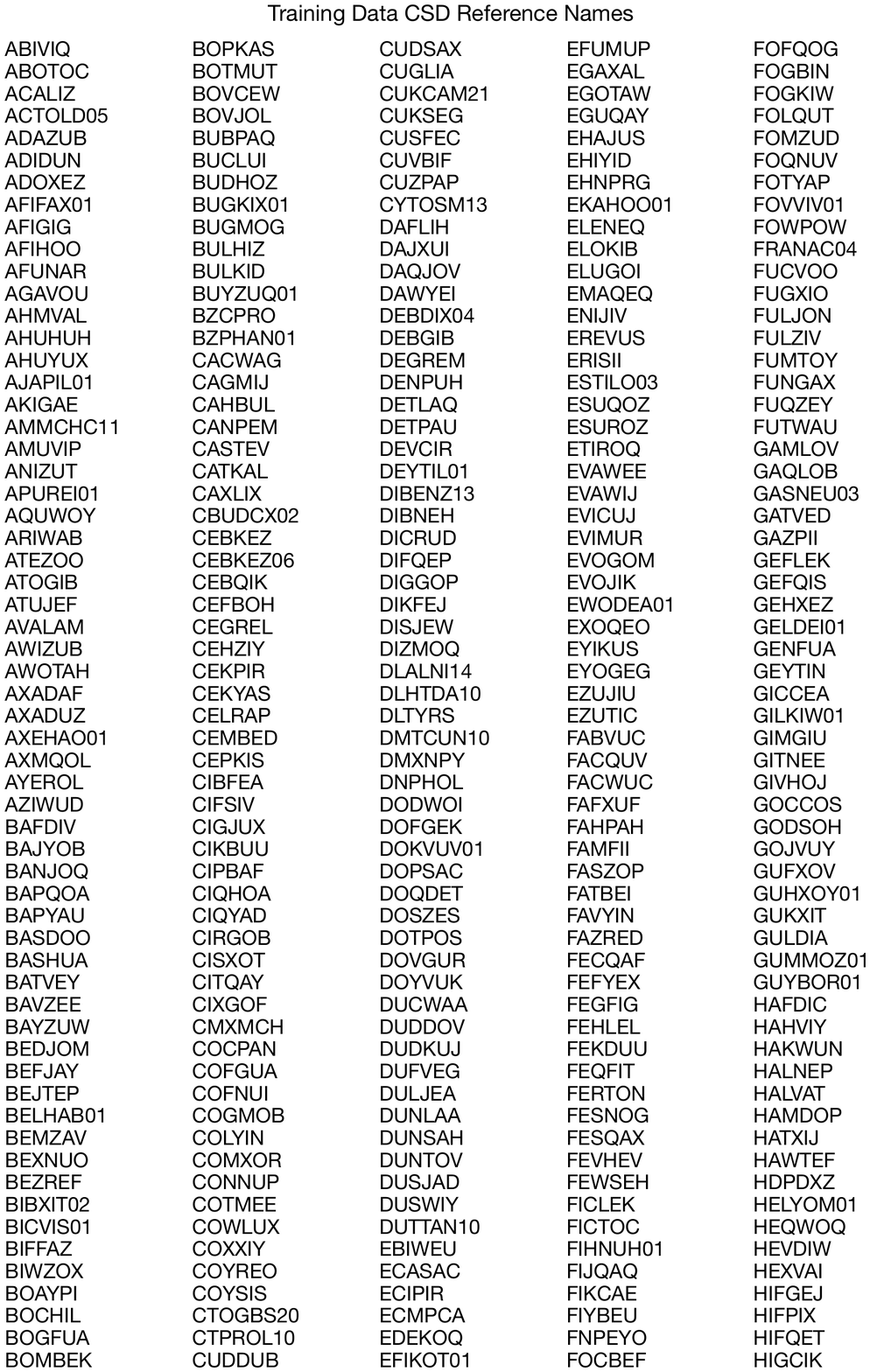}
\end{figure}

\includepdf[pages={2, 3},pagecommand={},width=\textwidth]{suppl_figures/data4_csdnames.pdf}
\includepdf[pages=-,pagecommand={},width=\textwidth]{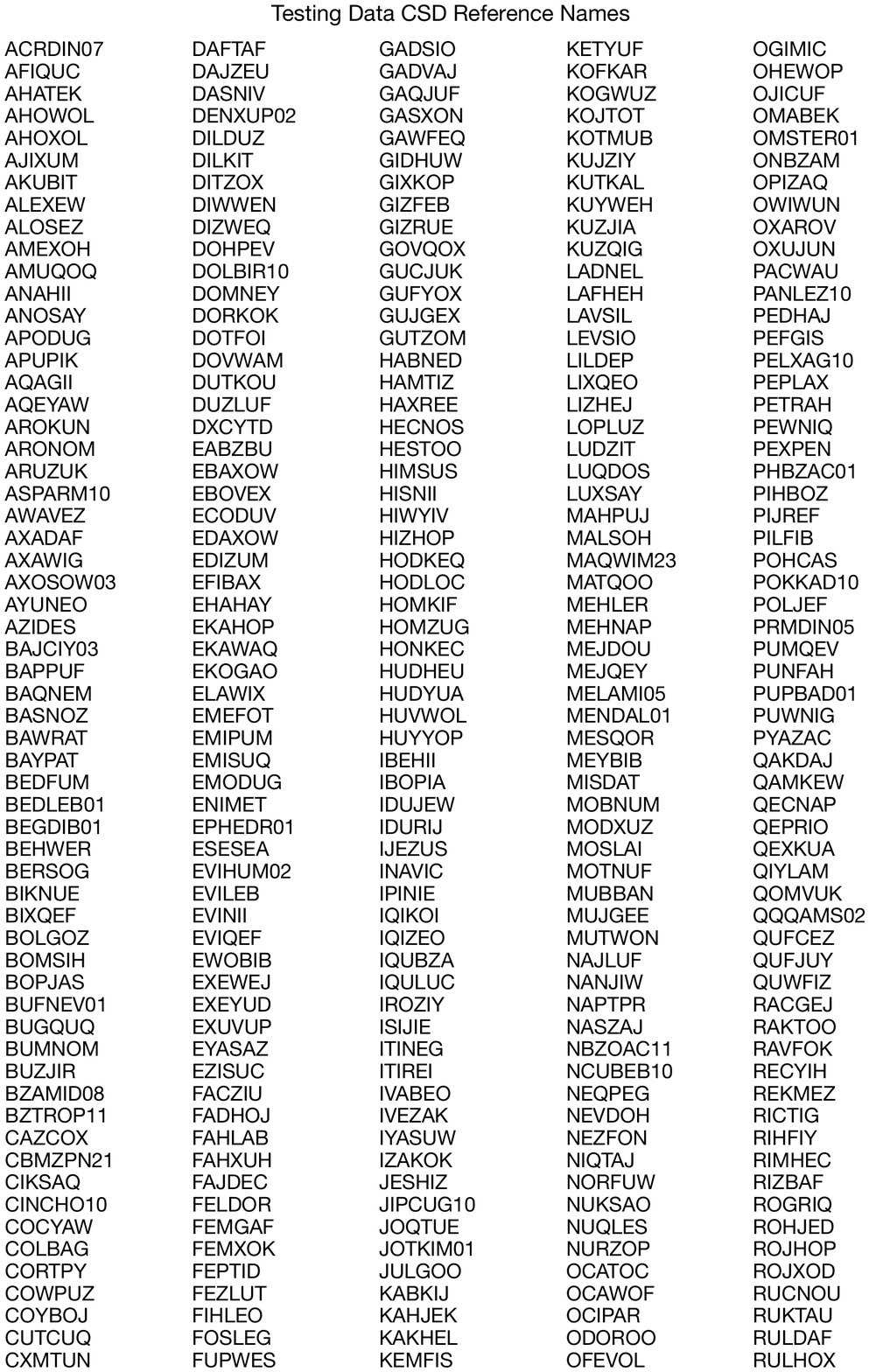}

\section{Full Gaussian reference} \label{sec:gaussian}
Gaussian 09, Revision D.01, M. J. Frisch, G. W. Trucks, H. B. Schlegel, G. E. Scuseria, M. A. Robb, J. R. Cheeseman, G. Scalmani, V. Barone, G. A. Petersson, H. Nakatsuji, X. Li, M. Caricato, A. Marenich, J. Bloino, B. G. Janesko, R. Gomperts, B. Mennucci, H. P. Hratchian, J. V. Ortiz, A. F. Izmaylov, J. L. Sonnenberg, D. Williams-Young, F. Ding, F. Lipparini, F. Egidi, J. Goings, B. Peng, A. Petrone, T. Henderson, D. Ranasinghe, V. G. Zakrzewski, J. Gao, N. Rega, G. Zheng, W. Liang, M. Hada, M. Ehara, K. Toyota, R. Fukuda, J. Hasegawa, M. Ishida, T. Nakajima, Y. Honda, O. Kitao, H. Nakai, T. Vreven, K. Throssell, J. A. Montgomery, Jr., J. E. Peralta, F. Ogliaro, M. Bearpark, J. J. Heyd, E. Brothers, K. N. Kudin, V. N. Staroverov, T. Keith, R. Kobayashi, J. Normand, K. Raghavachari, A. Rendell, J. C. Burant, S. S. Iyengar, J. Tomasi, M. Cossi, J. M. Millam, M. Klene, C. Adamo, R. Cammi, J. W. Ochterski, R. L. Martin, K. Morokuma, O. Farkas, J. B. Foresman, and D. J. Fox, Gaussian, Inc., Wallingford CT, 2016.

\bibliography{rsc} 
\bibliographystyle{rsc}